\documentclass[twocolumn]{aastex62}

\usepackage{multirow}
\usepackage{amsmath}
\usepackage{hyperref}

\newcommand{\ymean}{y_{R_j}}

\newcommand{\Test}[1]{\expandafter\hat#1}    

\graphicspath{{Figures/}}
\journalinfo{Accepted to AJ}
\shorttitle{NIRC2 vortex coronagraph performance}
\shortauthors{Xuan et al.}

\begin{document}

\title{\large Characterizing the performance of the NIRC2 vortex coronagraph at W.M. Keck Observatory}

\correspondingauthor{W. Jerry Xuan}
\email{wxob2015@mymail.pomona.edu}

\author[0000-0002-6618-1137]{W. Jerry Xuan}
\affil{Department of Astronomy, California Institute of Technology, 1200 E. California Blvd., Pasadena, CA 91125, USA}
\affil{Department of Physics and Astronomy, Pomona College, 333 N. College Way, Claremont, CA 91711}

\author[0000-0002-8895-4735]{Dimitri Mawet}
\affil{Department of Astronomy, California Institute of Technology, 1200 E. California Blvd., Pasadena, CA 91125, USA}
\affil{Jet Propulsion Laboratory, California Institute of Technology, 4800 Oak Grove Dr., Pasadena, CA 91109, USA}

\author[0000-0001-5172-4859]{Henry Ngo}
\affil{NRC Herzberg Astronomy and Astrophysics, 5071 West Saanich Road, Victoria, British Columbia, Canada}

\author[0000-0003-4769-1665]{Garreth Ruane}
\altaffiliation{NSF Astronomy and Astrophysics Postdoctoral Fellow}
\affil{Department of Astronomy, California Institute of Technology, 1200 E. California Blvd., Pasadena, CA 91125, USA}

\author[0000-0002-5407-2806]{Vanessa P. Bailey}
\affil{Jet Propulsion Laboratory, California Institute of Technology, 4800 Oak Grove Dr., Pasadena, CA 91109, USA}

\author[0000-0002-9173-0740]{\'{E}lodie Choquet}
\altaffiliation{Hubble Fellow}
\affil{Department of Astronomy, California Institute of Technology, 1200 E. California Blvd., Pasadena, CA 91125, USA}

\author[0000-0002-4006-6237]{Olivier Absil}
\altaffiliation{F.R.S.-FNRS Research Associate}
\affil{Space Sciences, Technologies, and Astrophysics Research (STAR) Institute, Universit\'{e} de Li\`{e}ge, Li\`{e}ge, Belgium}

\author{Carlos Alvarez}
\affil{W.M. Keck Observatory, Mamalahoa Hwy, Kamuela, HI 96743}

\author[0000-0002-6076-5967]{Marta Bryan}
\affil{Department of Astronomy, California Institute of Technology, 1200 E. California Blvd., Pasadena, CA 91125, USA}

\author{Therese Cook}
\affil{Department of Astronomy, California Institute of Technology, 1200 E. California Blvd., Pasadena, CA 91125, USA}
\affil{Department of Physics and Astronomy, University of California, Los Angeles, CA 90095}

\author{Bruno Femen\'{i}a Castell\'{a}}
\affil{W.M. Keck Observatory, Mamalahoa Hwy, Kamuela, HI 96743}

\author[0000-0003-2050-1710]{Carlos Gomez Gonzalez}
\affil{Universit\'{e}́ Grenoble Alpes, IPAG, F-38000 Grenoble, France}

\author{Elsa Huby}
\affil{LESIA, Observatoire de Paris, Universit\'{e} PSL, CNRS, Sorbonne Universit\'{e}, Univ. Paris Diderot, Sorbonne Paris Cit\'{e},\\5 place Jules Janssen, 92195 Meudon, France}

\author{Heather A. Knutson}
\affil{Division of Geological and Planetary Sciences, California Institute of Technology, 1200 E. California Blvd., Pasadena, CA 91125, USA}

\author{Keith Matthews}
\affil{Department of Astronomy, California Institute of Technology, 1200 E. California Blvd., Pasadena, CA 91125, USA}

\author{Sam Ragland}
\affil{W.M. Keck Observatory, Mamalahoa Hwy, Kamuela, HI 96743}

\author{Eugene Serabyn}
\affil{Jet Propulsion Laboratory, California Institute of Technology, 4800 Oak Grove Dr., Pasadena, CA 91109, USA}

\author{Zo\"{e} Zawol}
\affil{Department of Astronomy, California Institute of Technology, 1200 E. California Blvd., Pasadena, CA 91125, USA}

\begin{abstract}

The NIRC2 vortex coronagraph is an instrument on Keck II designed to directly image exoplanets and circumstellar disks at mid-infrared bands $L^\prime$ (3.4-4.1~$\mu$m) and $M_s$ (4.55-4.8~$\mu$m). We analyze imaging data and corresponding adaptive optics telemetry, observing conditions, and other metadata over a three year time period to characterize the performance of the instrument and predict the detection limits of future observations. We systematically process images from 359 observations of 304 unique stars to subtract residual starlight (i.e., the coronagraphic point spread function) of the target star using two methods: angular differential imaging (ADI) and reference star differential imaging (RDI). We find that for the typical parallactic angle (PA) rotation of our dataset ($\sim$10$^{\circ}$), RDI provides gains over ADI for angular separations smaller than 0.25\arcsec. Furthermore, we find a power-law relation between the angular separation from the host star and the minimum PA rotation required for ADI to outperform RDI, with a power-law index of -1.18$\pm$0.08. Finally, we use random forest models to estimate ADI and RDI post-processed detection limits a priori. These models, which we provide publicly on a website, explain 70\%-80\% of the variance in ADI detection limits and 30\%-50\% of the variance in RDI detection limits. Averaged over a range of angular separations, our models predict both ADI and RDI contrast to within a factor of 2. These results illuminate important factors in high-contrast imaging observations with the NIRC2 vortex coronagraph, help improve observing strategies, and inform future upgrades to the hardware.

\end{abstract}


\section{Introduction} \label{sec:intro}
High-contrast imaging in the infrared provides unique sensitivity to thermal emission from young giant planets and protoplanets as well as scattered light from circumstellar disks. Recent direct imaging surveys of nearby stars have discovered new giant exoplanets \citep[e.g.][]{Marois2008,Macintosh2015,Chauvin2017} and constrained their occurrence rates \citep[see review by][]{Bowler2016}.

Current ground-based high-contrast imagers, such as GPI~\citep{Macintosh2014}, SPHERE~\citep{Vigan2016}, SCExAO~\citep{Jovanovic2015}, P1640~\citep{Hinkley2011}, and NIRC2~\citep[e.g.][]{Marois2008,Bowler2012,Serabyn2017} combine adaptive optics and coronagraphs to detect faint planets at small angular separations from the host star. A standard measure of their performance is the limiting planet-to-star flux ratio, or contrast, as a function of angular separation from the primary star. The fundamental goal of high-contrast imaging is to improve contrast limits with advanced hardware and software, thereby expanding the parameter space of planets that could be uncovered behind the glare of their host stars.

There are many factors that may affect the contrast achieved from an instrument on a given observing night. Some of these factors are environmental, such as properties of the atmosphere and temperature of the optics. Other factors depend on the observing strategy and the timing of observations, which set the airmass during the observation, the amount of time per target, and the amount of parallactic angle (PA) rotation (the angle through which the field of view rotates during an observation set when observing in fixed pupil mode on an altitude-azimuth telescope). The magnitude of the target star and the choice of post-processing algorithm also influence the detection limits. Large surveys of many targets with a single instrument reveal correlations between these different factors and the resulting detection limits, helping determine the bottlenecks of instrument performance and improve observing strategies. They can also help identify areas of potential improvement for future instrument upgrades. 

Studies of GPI and SPHERE survey data \citep{Poyneer2016, Bailey2016, Milli2017} have demonstrated direct relationships between the science image contrast and factors such as adaptive optics (AO) system performance and environmental effects. For instance, the pernicious low-wind effect caused by heat exchange between the telescope spiders and the surrounding air has been known to degrade the quality of the PSF in a way unseen by the AO system \citep{Sauvage2016}. In addition, analysis of GPI data has shown that temperature disequilibrium between the instrument, telescope, and dome also degrades AO residual wavefront error and science image contrast, due to additional induced dome seeing \citep{Tallis2018}. Furthermore, some of these studies have shown relationships between contrast and atmospheric seeing. While the amplitude of the seeing in arcseconds is not well-correlated with contrast, the coherence time of the seeing ($\tau_0$, related to the speed of the turbulent wind layer) is found to be one of the strongest predictors of image quality \citep{Bailey2016}. As long as the seeing amplitude is within the capture range of the deformable mirror, the turbulence can be corrected. However, if the turbulence evolves on timescales faster than the AO system correction bandwidth, performance will be degraded.

In this paper, we present our characterization of the performance of the Keck/NIRC2 high-contrast imager. In 2015, a vector vortex coronagraph \citep{Mawet2005} optimized for $L^\prime$ and $M_s$ imaging was installed on NIRC2 \citep{Serabyn2017}. NIRC2 (PI: Keith Matthews) is a near-infrared camera installed behind Keck II's AO system \citep{Wizinowich2000}, which is equipped with a low-order deformable mirror and a visible light Shack-Hartmann wavefront sensor (WFS). Since 2015, we have implemented a streamlined workflow for NIRC2 vortex data, including an automated pipeline, a database, and a web server. After systematically reprocessing all of our vortex coronagraph data with this new infrastructure, we obtained a homogeneous and complete dataset suitable for a robust statistical study.

In our pipeline, principal component analysis (PCA) is used to model the coronagraphic PSF \citep{Soummer2012}. We study the imaging sensitivity of the PCA post-processing technique for two observing strategies, angular differential imaging \citep[ADI,][]{Marois2005} and reference differential imaging \citep[RDI, e.g.][]{Lafrenière2009,Soummer2011,Gerard2016,Ruane2017}. Both techniques perform PSF subtraction by synthesizing a model PSF from a library of images: ADI models the stellar PSF from the target data itself, taking advantage of the PA rotation of the field over the course of the observing sequence. On the other hand, RDI uses images of other similar targets to build a model stellar PSF. Since ADI is limited by self-subtraction at small PA rotations, we expect that RDI should yield deeper contrasts below a certain threshold \citep{Ruane2017}. Using data processed in a uniform manner, we compare the contrast from ADI and RDI as a function of PA rotation. We also investigate contrast as a function  of other variables such as stellar magnitude, $\tau_0$ over WFS integration time, seeing, and temperature differentials between different parts of the instrument.

Lastly, we present predictive models to accurately estimate contrast at a broad range of separations for both ADI and RDI. These models not only improve the efficiency of future NIRC2 vortex observations by optimizing observing strategy, but also provide quantitative measurements of how important each variable is in determining contrast.

\begin{figure*}[t]
  \centering
\plottwo{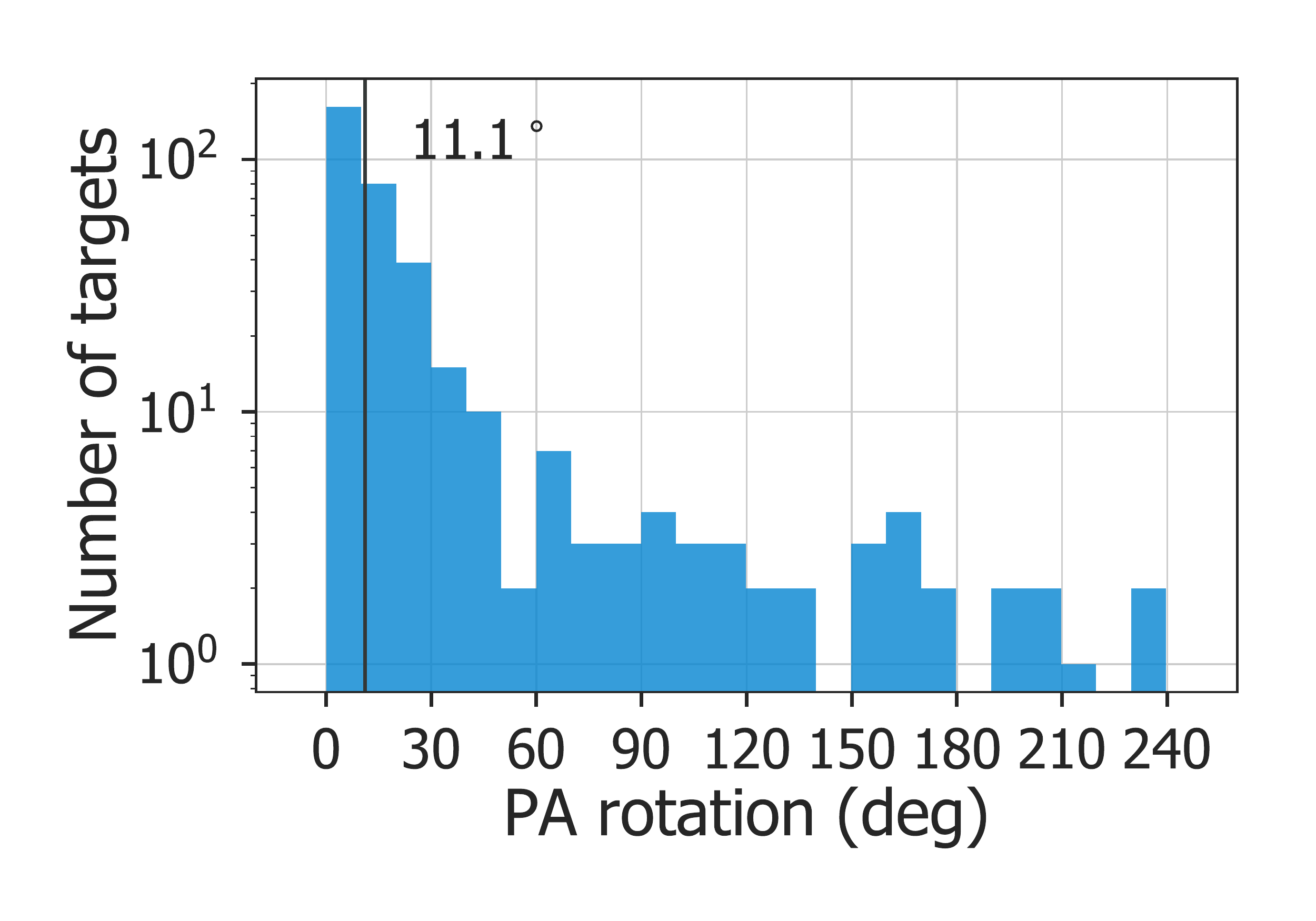}{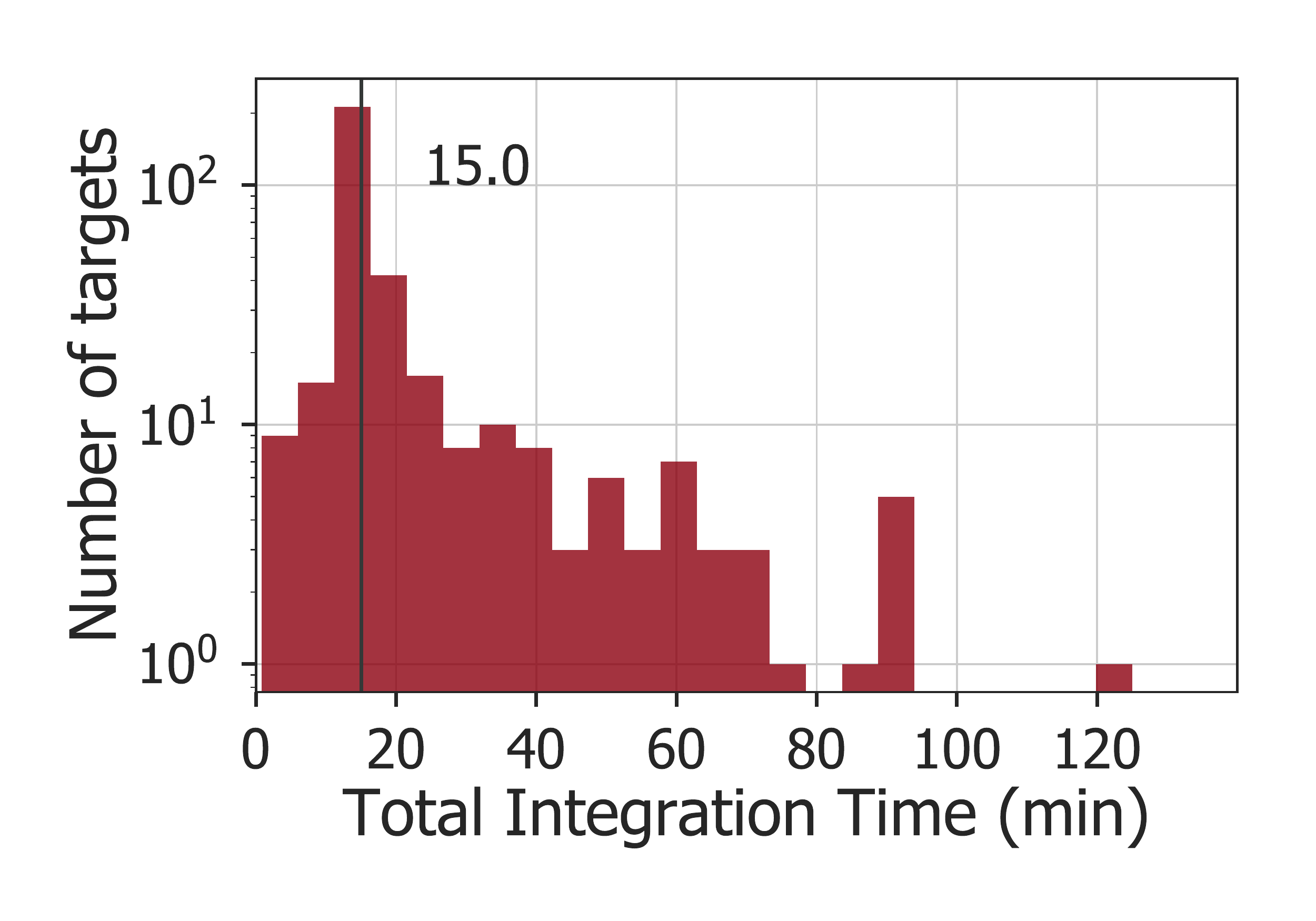}
  \caption{Left panel: Distribution of PA rotation of the targets in our sample set. The median PA rotation (solid vertical line) is at 11.1$^{\circ}$. Right panel: Distribution of total integration time. The median (solid vertical line) is at 15.0 minutes. \label{fig:PA_rot_totitime}}
\end{figure*}

We organize this paper into the following sections. Section~\ref{sec:re-reduction} provides an overview of the observations, the dataset, and the processing steps that lead to the limiting planet-to-star flux ratio. Section~\ref{sec:analysis} describes the effect of a few prominent variables on image quality, and compares the two techniques ADI and RDI. The random forest models are presented in Section~\ref{sec:stat-models}, where the relative importance of variables is measured, and the prediction accuracy is examined. Finally, we conclude in Section~\ref{sec:conclusion}.

\section{Observations, Data, and Systematic Re-processing}\label{sec:re-reduction}
Our study's sample is a set of 359 observations of 304 unique targets observed from 2015 December 26 to 2018 January 5. These observations were taken with the vortex coronagraph installed in NIRC2 \citep{Serabyn2017}, using the \texttt{QACITS} automatic, real-time coronagraphic PSF centering algorithm~\citep{Huby2015,Huby2017}. The typical centering accuracy provided by \texttt{QACITS} is 2.4 mas rms \citep{Huby2017}, or $\simeq$0.025$\lambda/D$ rms in $L^\prime$ band. In comparison, the pixel scale of the NIRC2 vortex is 9.942 mas per pixel \citep{Service2016}. The full dataset contains images taken in the $L^\prime$ (central wavelength of $3.776\,\mu$m) and $M_s$ (central wavelength of $4.670\,\mu$m) bandpasses. In this paper, we include only the targets observed in $L^\prime$ band in our sample set, which accounts for over 98\% of the data. We note that around 2/3 of our sample is composed of stars from surveys designed to use RDI and have limited PA rotation. For our sample set, the median and mean PA rotations are 11.1$^{\circ}$ and 26.0$^{\circ}$, respectively. The median and mean total integration times are 15 mins and 22 mins, respectively (see Fig.~\ref{fig:PA_rot_totitime}).

The typical observing sequence consists of one image of the star without the coronagraph to characterize the unocculted PSF, one sky frame of a blank field 10\arcsec\ away from the target, and then 10-30 science frames with the star centered on the vortex representing $\sim$10-60 seconds of integration time for each frame. For longer observations or in rapidly changing conditions, the full sequence is repeated every 10-30 minutes to sample potential variations in the unocculted PSF and sky background. All observations were taken with the telescope's field rotator set to track the telescope pupil in order to exploit the natural rotation of the sky.

We uniformly reprocess our sample set with a pipeline that automatically downloads, sorts, and processes data relying on the functionality of the Vortex Image Processing (\texttt{VIP}) software package~\citep[][]{GomezGonzalez2017} as well as custom programs.

In the pipeline, we perform a series of pre-processing steps. First, we apply a flat field correction to both science and sky background frames. The flat field image is the median of 5-10 images of the blank sky taken with the vortex mask removed near the end of the night, or another night close in time if same-night flats are not available.

Then, we use \texttt{VIP} to remove bad pixels from both science and sky frames. For the mean science and sky frames, bad pixels are identified as those with values greater than 3 standard deviations above the median of a $5\times5$ box centered on that pixel. NIRC2 also has hot pixels and dead pixels, which are identified in a similar way in the dark frames and flat field frames, respectively. We replace the value of bad, hot, and dead pixels with the median of the neighboring pixels in the $5\times5$ box. However, we avoid bad pixel correction in a circular region of diameter equal to the FWMH, centered on the star. Next, we use a PCA-based algorithm from \texttt{VIP} to subtract the sky from our science frames, identifying and removing the number of sky principal components equal to the number of sky frames for each target. Finally, these sky-subtracted images are registered to the target star's position and de-rotated according to the parallactic angle recorded by the instrument so that north points up and east points left. To register the images, we identify the star's position in each frame by aligning the speckle pattern with the median frame using a cross-correlation. Specifically, \texttt{VIP} uses the ``register\_translation'' method from the \texttt{scikit-image} package, which implements the algorithm developed by \citet{Guizar-Sicairos2008}.

Then, for each target, we subtract the PSF of the target star using a full-frame PCA-based approach. For ADI, the PSF library is built with all science images from the target. For RDI, we use designated reference stars if they are available. These are reference stars imaged before and after the target star on the same night, and are usually late-type stars that are unlikely to contain young planets emitting strongly in thermal wavelengths \citep[see e.g.][]{Ruane2017}. Most of our targets do not have designated reference targets, in which case we use all other targets from the same night and the same observing program to build the reference PSF library (similar to the self-referencing strategy with the Hubble Space Telescope data from \citet{Lafrenière2009} and \citet{Soummer2011} in the sense that these two studies also used multiple reference stars that were observed as part of the same observing program, with similar observing times and noise levels). The size of the reference PSF library thus depends on how many other stars were observed on a given night and how many frames were acquired on each star. Accordingly, we include the size of the reference PSF library as a variable in our study. For our sample set, the median reference library size is 150 frames. We use a set of different frame sizes for ADI and RDI in order to optimize the detection limits at different separations from the host star. In addition, we use numerical masks to improve PSF subtraction with PCA. The mask is centered on the star and set to a different radius for each frame size, as summarized in Table~\ref{tab:post_params}.

\begin{deluxetable}{cccc}
\tablecolumns{4}
\tabletypesize{\scriptsize}
\tablewidth{\textwidth}
\tablecaption{Post-processing parameters \label{tab:post_params}}
\tablehead{\colhead{PCA} & \colhead{Frame size} & \colhead{Mask radius/Inner radius} & \colhead{Outer radius}\vspace{-8pt} \\
\colhead{algorithm} & \colhead{(arcsec)} & \colhead{(arcsec)} & \colhead{(arcsec)}}
\startdata
ADI & 1.0 & 0.08 & 0.5 \\
    & 1.5 & 0.08 & 0.75 \\
    & 2.0 & 0.16 & 1.0 \\
    & 3.5 & 0.16 & 1.75 \\
    & 5.5 & 0.40 & 2.75 \\
RDI & 1.0 & 0.08 & 0.5 \\
    & 1.5 & 0.08 & 0.75
\enddata
\tablecomments{Our study examines post-processing results in five analysis regions for ADI and two regions for RDI. The inner ADI regions are computed for a direct comparison with RDI. Images are cropped to a square frame of the given size, centered on the target, prior to processing. A circular region of the given mask radius centered on the target is masked out, which sets the inner radius of the analysis regions. The outer radius of the analysis regions is equal to half of the frame size.}
\end{deluxetable}

We compute the detection limit as a function of angular separation for each target and each set of frame sizes. The ratio of this limit to the host star flux is called the contrast curve. Within each frame size, we compute several contrast curves using a range of principal components to model the PSF. In our pipeline, the contrast corresponding to one standard deviation was calculated using the \verb|contrast_curve| function in the \texttt{VIP}, which performs fake companion injection and retrieval to determine and compensate for signal losses owing to self-subtraction and over-subtraction effects. Planets with a signal-to-noise of 10 are injected and retrieval is attempted to estimate point source over-subtraction by PCA as a function of separation. These injected planet signals are placed 0.32\arcsec (four FWHM) apart, and we repeat this four times, shifting the position of the injected planet by 0.08\arcsec (one FWHM) each time, in order to compute the retrieval at every 0.08\arcsec. We do not include the effect of small sample statistics on contrast described in \citet{Mawet2014} in order to make our results directly comparable to past studies on GPI \citep{Bailey2016} and SPHERE \citep{Milli2017}. Furthermore, the inclusion of small sample statistics would not alter our results, since it would merely constitute a multiplicative factor to the contrast values (e.g., it would not change the power laws we find). It is important to note that contrast depends on the post-processing algorithm used, and our study refers only to the full-frame PCA subtraction algorithm.

At each stage of the pipeline, data products such as reduced images and contrast curve calculations are stored along with the target's metadata in a \texttt{Mongo} database. While the pipeline enables our characterization study by allowing us to uniformly process all the data in our sample set, the database is essential for making the comparisons between detection limits and relevant explanatory variables.

\section{NIRC2 Vortex Characterization}\label{sec:analysis}

\subsection{Response variable: optimal contrast}\label{sec:opt-contrast}
In our study, the response variable is 5$\sigma$ contrast, obtained by multiplying five to the 1$\sigma$ contrast level computed by the pipeline. Since each of our targets has a set of contrast curves for different frame sizes, mask sizes, and numbers of principal components, we extract an ``optimal contrast curve'' for each target with the following steps. First, we re-sample all of the contrast curves computed from all sets of post-processing parameters to 0.01\arcsec intervals in angular separation, interpolating between two values when necessary. Then, for each re-sampled separation, we compare all of the available contrast curves and choose the smallest contrast value as the ``optimal contrast'' for this separation. Lastly, we combine the optimal contrast value at every separation to form the combined optimal contrast curve for the target. Thus, the optimal contrast curve represents the ideal contrast one could achieve for a target under the conditions of the given observing window. We repeat this process twice to generate optimal ADI and RDI contrast curves separately.

\begin{figure}[t]
    \centering
    \includegraphics[width=\linewidth]{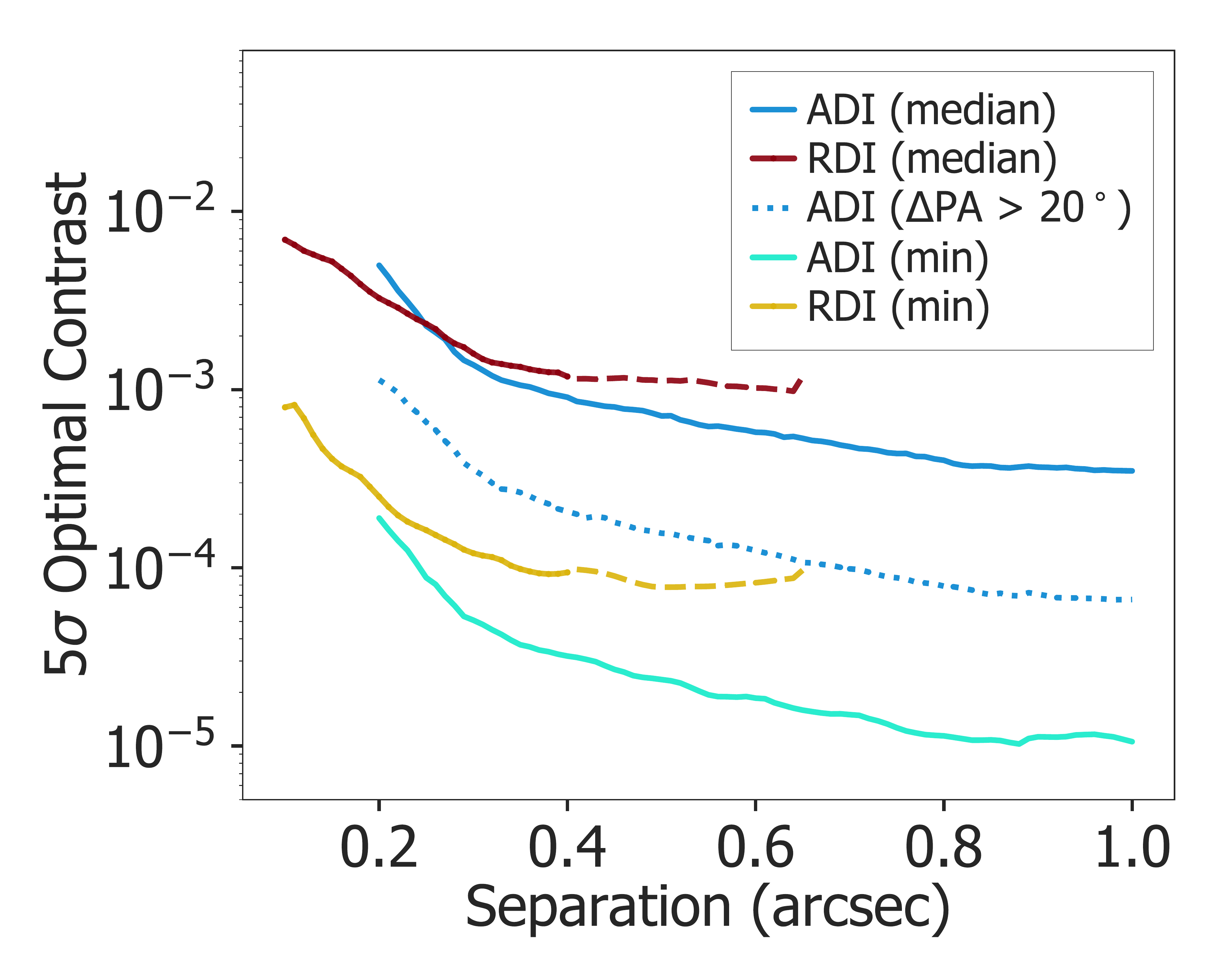}
    \caption{Median contrast curves (top three) and minimum contrast curves (bottom two) for ADI and RDI. For the median contrast, the intersection between ADI and RDI is at 0.25\arcsec. The dotted segments for the RDI curves represent the portion beyond 0.4\arcsec where only 75\% of targets are available for median calculation. This artificially pulls the RDI contrast beyond 0.4\arcsec upwards slightly. The dotted line in the middle represents ADI median contrast from targets that have PA rotation larger than 20$^{\circ}$.
    \label{fig:median_contrast}}
\end{figure}

Fig.~\ref{fig:median_contrast} shows the median and minimum of all optimal contrasts for ADI and RDI post-processed images. Because we rely on ADI to probe large separations, RDI data is not complete for separations beyond 0.4\arcsec. Specifically, only 75\% of targets in our sample set have RDI contrast between 0.4\arcsec and 0.65\arcsec. However, these larger separations are not as important for RDI, since the top two curves in Fig.~\ref{fig:median_contrast} show that ADI performs better at separations larger than 0.25\arcsec on average. At closer separations, ADI performance degrades because of self-subtraction effects. Therefore, our ADI median contrast is pulled upwards by the large number of low PA rotation targets. The middle dotted curve shows the median ADI contrast for targets with PA rotations larger than 20$^{\circ}$ in order to demonstrate the performance on well-timed ADI sequences. The bottom two curves show the minimum of all optimal contrast curves, demonstrating the best contrast limits obtained with the NIRC2 vortex coronagraph in $L^\prime$ band.

\subsection{Explanatory variables}
Explanatory variables are factors that may influence the response variable. Unlike truly independent variables, explanatory variables may be correlated with each other as well as the response variable. We list the explanatory variables considered for our models in Table~\ref{tab:exp_var}, where we divide them into three categories: observing conditions, observation parameters, and stellar magnitudes. For variables that change during the course of an observation, such as airmass, we use the median value over the observation time frame of the target. We also measure the standard deviation of such dynamic variables over the observing sequence. For instance, seeing shows a median standard deviation of 0.12\arcsec and $\tau_{0}$ shows a median standard deviation of 0.30ms across all observations in our sample set. We find that such measures of variability show no correlation with contrast, and hence do not study them further in this paper.

\begin{deluxetable}{ll}
\tablecolumns{2}
\tabletypesize{\scriptsize}
\tablewidth{\textwidth}
\tablecaption{Explanatory variables \label{tab:exp_var}}
\tablehead{\colhead{Variable} & \colhead{Source}}
\startdata
\cutinhead{Observing conditions}
\tableline
$\tau_{0}$ & AO Telemetry \\
Seeing & AO Telemetry \\
WFS Frame Rate & AO Telemetry \\
Airmass & Fits Header \\
Primary Mirror Temperature & Keck II sensors \\
AO Optical Bench Temperature & Keck II sensors \\
AO Acquisition Camera Enclosure Temperature & Keck II sensors \\
Dome Temperature & Keck II sensors \\
Dome Humidity & Keck II sensors \\
Wind Speed & Keck II sensors \\
Pressure & Keck Weather Station \\
\cutinhead{Observation parameters}
\tableline
PA Rotation & Fits Header \\
PSF x FWHM & Pipeline Product \\
PSF y FWHM & Pipeline Product \\
Total Science Integration Time & Fits Header \\
RDI Reference Library Size & Pipeline Product \\
\cutinhead{Stellar magnitudes}
\tableline
$R$ magnitude & UCAC4 \\ 
W1 magnitude & WISE All-Sky $\&$ AllWISE\\ 
\enddata
\tablecomments{A list of explanatory variables considered for our analysis. We obtain weather and temperature data from the Keck weather station and sensors in Keck II, seeing estimations and the atmospheric coherence time $\tau_0$ from AO telemetry data (see Sections~\ref{sec:tau0} and~\ref{sec:seeing}). Stellar magnitudes come from the UCAC4~\citep{Zacharias2012, Zacharias2013}, WISE All-Sky~\citep{Wright2010,Cutri2012}, and AllWISE~\citep{Cutri2014} catalogs, as indicated. RDI reference library size refers to the number of PSF frames included in the reference PSF library for a given target.}
\end{deluxetable}

\subsection{Noise regimes}\label{sec:noise-reg}
The achieved contrast can be broken into two regimes: background noise-limited and speckle noise-limited. The separation beyond which a target reaches the background limit depends on the total integration time and the magnitude of the target, and therefore differs between targets (larger separation for brighter targets). In the speckle noise-limited regime, the performance is limited by speckle noise from the residual PSF of the star. Compared to the background noise limit, we expect the speckle noise limit to be controlled by a wider range of factors.

\begin{figure}[t]
    \centering
    \includegraphics[width=\linewidth]{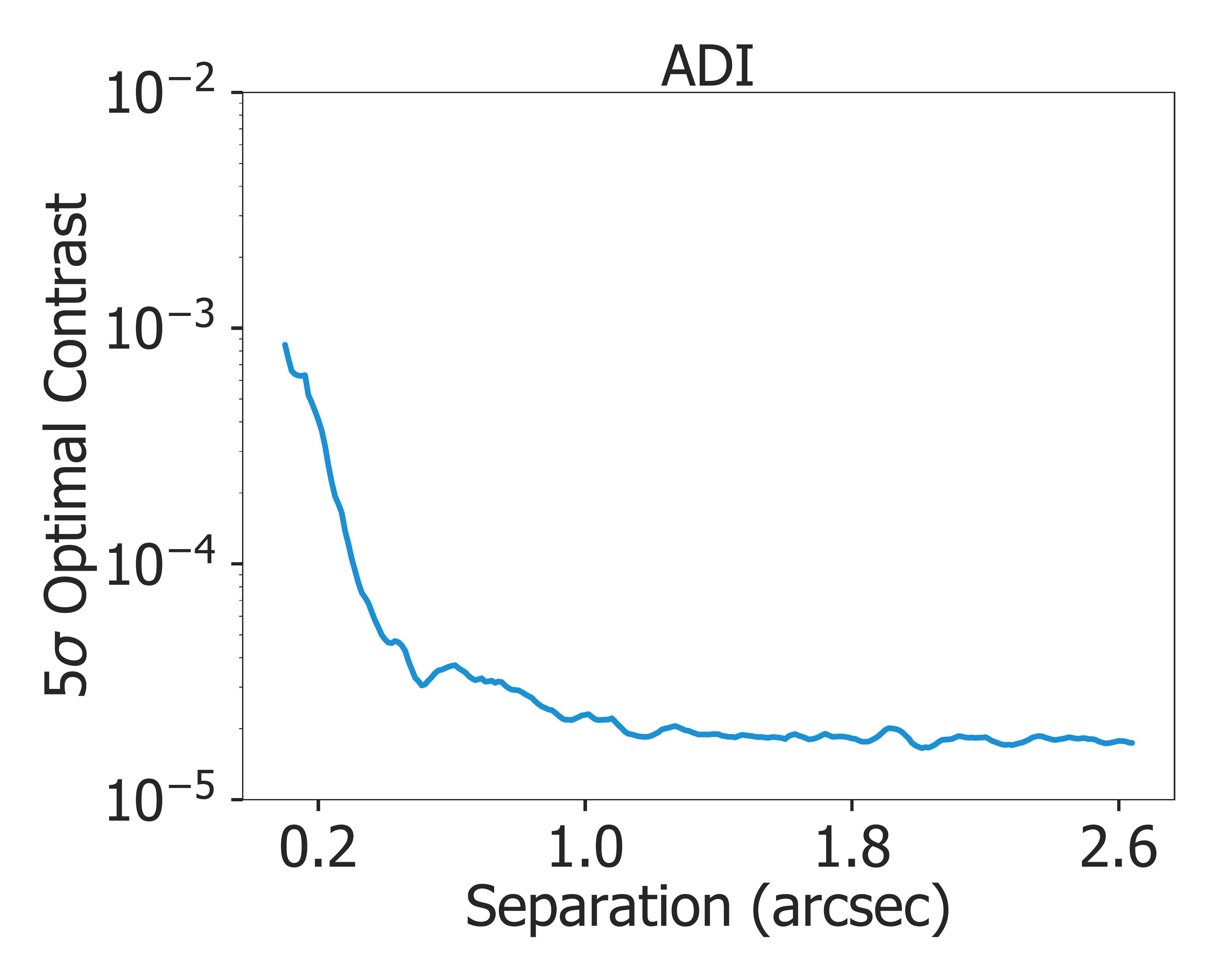}
    \caption{A sample ADI optimal contrast curve from a target in our dataset. The contrast plateaus at large separations from about 1.6\arcsec and is safely in the background-limited regime at 2.6\arcsec. 
    \label{fig:large_cc}}
\end{figure}

In Section~\ref{sec:PA}, where we compare ADI and RDI performance through the lens of PA rotation, we exclude contrast data that are background-limited. To do this, we empirically determine the background-limited contrast as the contrast at the largest separation ($\approx$2.6\arcsec), where we assume all targets to be in the background noise regime (very reasonable assumption for images in $L^\prime$ band; see Fig.~\ref{fig:large_cc} for an example). We then remove any optimal contrasts between 0.2\arcsec and 0.4\arcsec that are within a factor 3 of the optimal contrast at $\approx$2.6\arcsec. We find that under this threshold, 37\% of ADI contrasts and 20\% of RDI contrasts at 0.4\arcsec are background-limited. On the other hand, in all other subsections after Section~\ref{sec:PA}, we include both speckle-limited and background-limited contrasts in the study. This is because unlike in Section~\ref{sec:PA}, where we attempt to compare ADI and RDI in an unbiased manner and extract an empirical relation that may be generally applicable, in the subsequent sections we focus on summarizing the past performance of the NIRC2 vortex, and finding trends specific to this instrument. In addition, in the predictive models in Section~\ref{sec:stat-models}, we also include all data to allow the models to differentiate between the two noise limits. Both limits are present in real data so the inclusion of both is necessary for accurate predictions across a broad range of separations. 

\subsection{Relationships between contrast and explanatory variables}
We investigate the relationship between contrast and explanatory variables for both ADI and RDI, using univariate fits. In order to directly compare ADI and RDI in Section~\ref{sec:PA}, we use only speckle noise-limited data, as described in Section~\ref{sec:noise-reg}. For all other variables, we include all data in the sample set. We mainly focus on contrast at two separations: 0.2\arcsec and 0.4\arcsec. Where power-law fits are expected, we take the log of the explanatory variable. Where no theoretical expectation exists, we try plotting the variable in both linear scale and log scale with log contrast and search for an empirical relation. We compute the slopes using a least squares linear fit, and determine the uncertainty of the fit as the square root of the variance estimate for the slope parameter.

\begin{figure*}[t!]
    \centering
    \plottwo{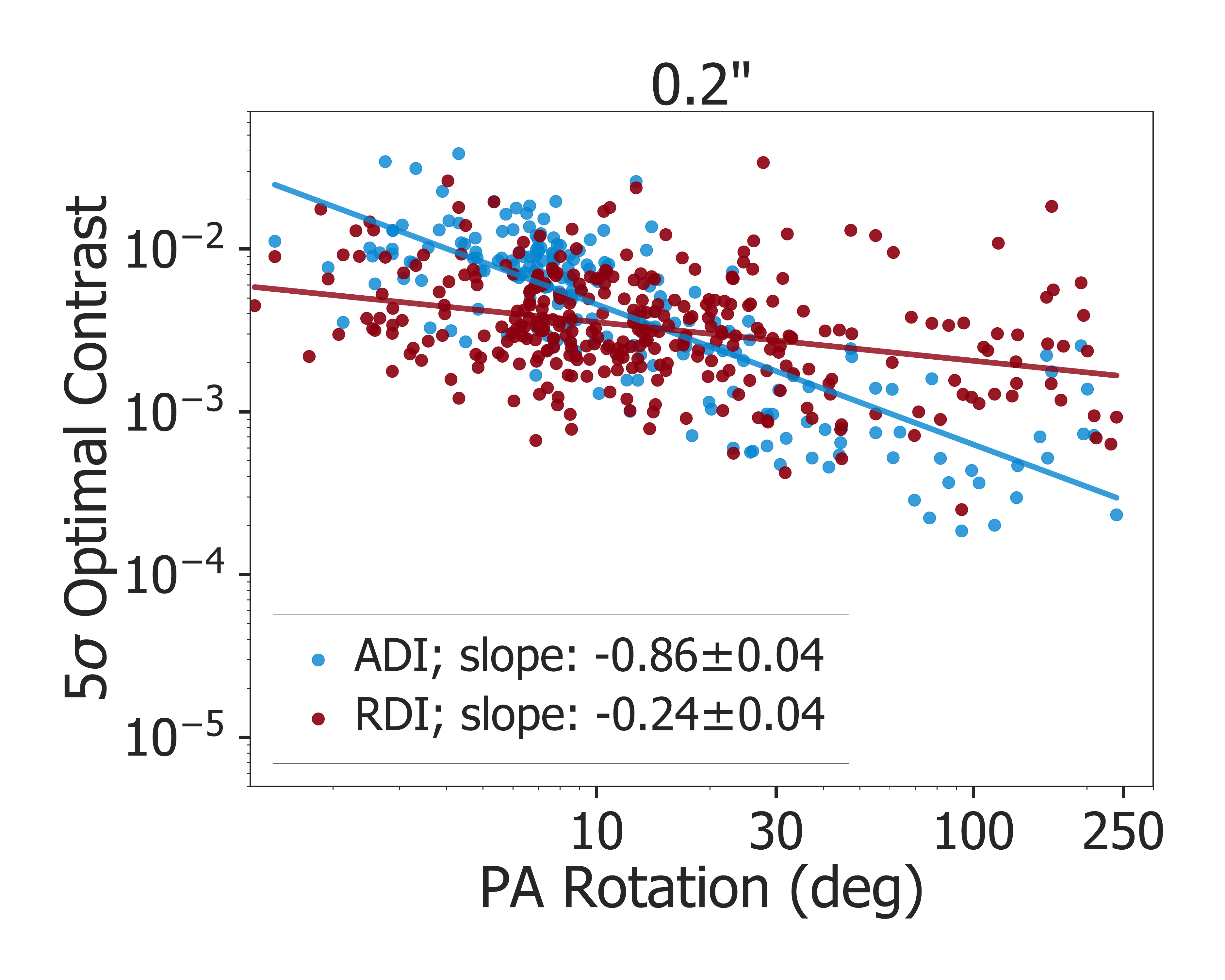}{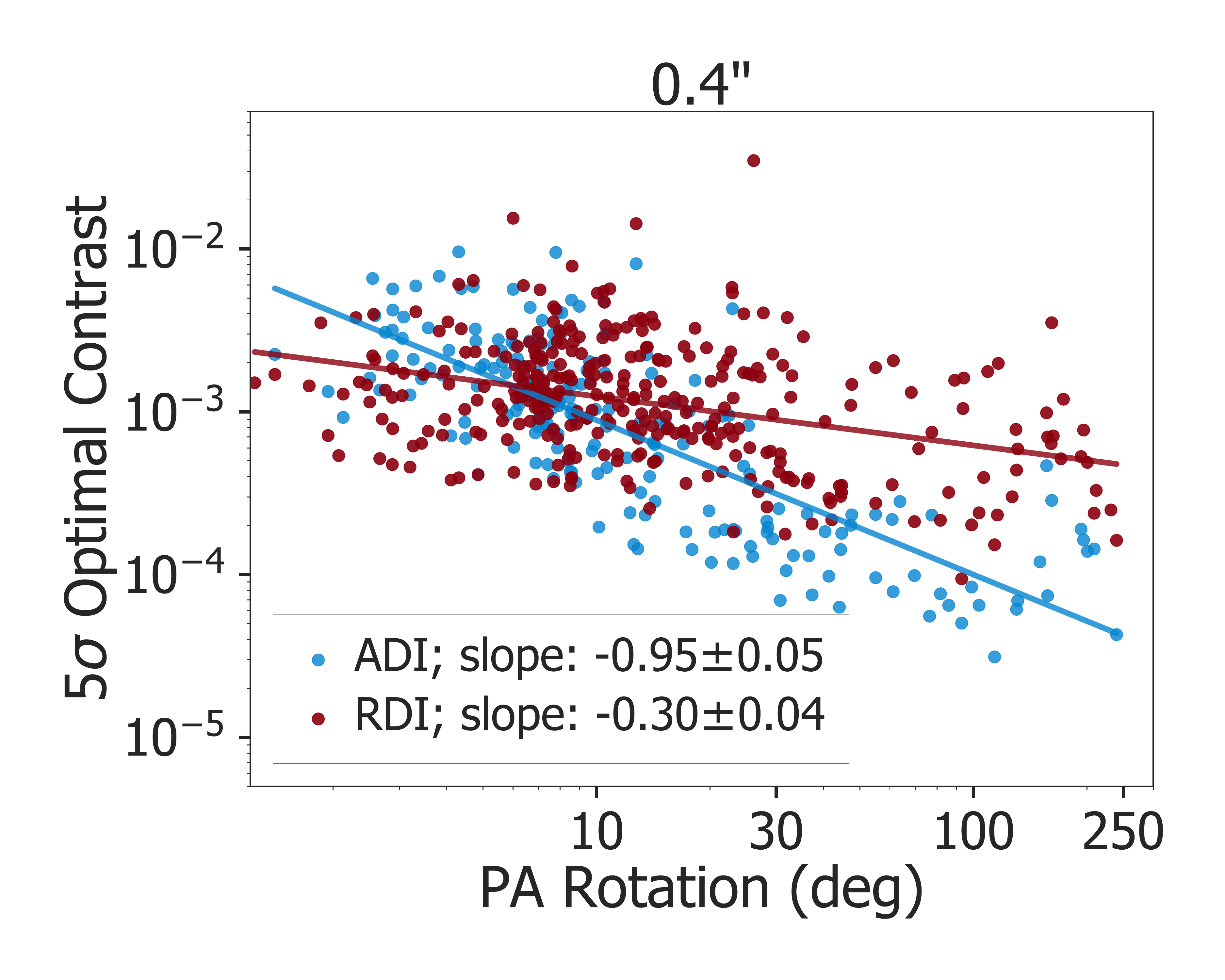}
    \caption{The dependence of the optimal 5$\sigma$ contrast on the PA rotation accumulated during the observation, shown in log-log scale. The linear trends suggest an underlying power-law relation, where the slope is listed in the legends. Blue points represent ADI contrasts, and red points represent RDI contrasts. Slopes are consistently larger in magnitude for ADI than for RDI, and therefore intersect at a point we call the critical PA rotation.}
    \label{fig:pa-contrast-log}
\end{figure*}

\subsubsection{PA rotation: case study of ADI vs. RDI}\label{sec:PA}
We expect that PA rotation is one of the most important variables in determining ADI optimal contrast (confirmed by the statistical models in Section~\ref{sec:stat-models}). We expect its effect on RDI contrast to be less prominent. Importantly, observers can compute a priori how much PA rotation they will acquire on a given target from the target declination, hour angle, and total integration time. In this section, we use data from the same frame sizes and numerical mask sizes for both ADI and RDI (specifically the 1.0\arcsec, 1.5\arcsec frame sizes with the 0.08\arcsec inner radius) to compute optimal contrast. This is done to avoid bias effects from using different zonal geometries on PCA-based PSF subtraction, and therefore to fairly compare ADI and RDI. Because we are interested in comparing ADI and RDI at separations between 0.2\arcsec and 0.4\arcsec, it is natural to use small frame sizes.

We discover that the relationship between final contrast and PA rotation is best described by a power law, with a steeper exponent for ADI than for RDI (see Fig.~\ref{fig:pa-contrast-log}). We calculate the slopes for 21 separations between 0.2\arcsec and 0.4\arcsec, in intervals of 0.01\arcsec. Among these separations, we find slopes for ADI ranging from -0.86 to -0.95, and slopes for RDI ranging from -0.24 to -0.30. We hypothesize that the slopes for RDI are caused solely by the whitening of speckle noise due to increasing angular diversity: a larger PA rotation means that more frames at different angles are combined, so the speckle field becomes more mixed and therefore easier to model with the reference PSF library. On the other hand, we hypothesize that the slopes for ADI come from both speckle noise whitening and the minimizing of self-subtraction with larger PA rotation. However, further study is encouraged to develop precise physical explanations for the exact power laws obtained. 

Due to the differing exponents, the best linear fits for ADI- and RDI-processed data intersect at an angle, which we dub the ``critical PA rotation" ($\theta_{crit}$). Specifically, we see that above the critical PA rotation, ADI results in deeper contrast than RDI, and below that threshold, RDI yields deeper contrast. Furthermore, we see that the intersection moves to larger PA rotations at smaller angular separations. This also confirms our expectations: ADI should perform progressively worse at closer separations due to enhanced self-subtraction. 

Consider a simple geometric expectation for $\theta_{crit}$ as a function of angular separation between the companion and the host star ($\rho$). When constructing a PSF for a target image from a set of reference images, if the companion is present at the same location in the target and reference images, self-subtraction will occur to negatively impact contrast. In ADI, avoiding self-subtraction requires a minimum amount of sky rotation, $\theta_{crit}$ $\approx$ $d_{crit}$ / $\rho$, where $d_{crit}$ is the amount of companion movement on the image necessary to minimize self-subtraction effects (a similar argument is presented in \citet{Marois2005}). We expect that below $\theta_{crit}$, self-subtraction will severely limit ADI performance, so RDI will outperform ADI.

\begin{figure}[t]
    \centering
    \includegraphics[width=8cm]{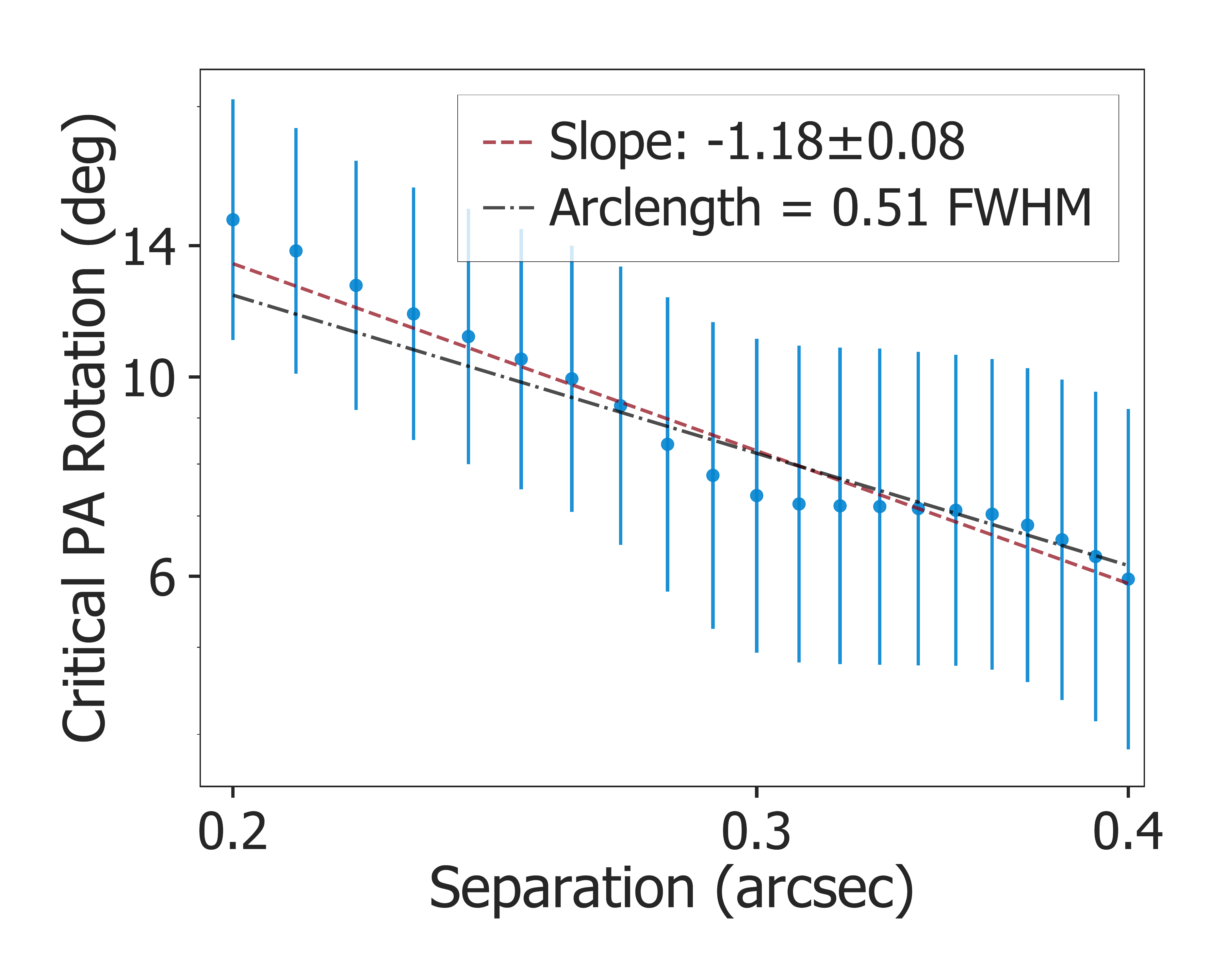}
    \caption{Critical PA rotation, with 1$\sigma$ error bars, as a function of angular separation, shown in log-log scale. PA rotation greater than $\theta_{crit}$ is required for ADI to outperform RDI on average. A two-parameter linear fit (red line) gives a slope of -1.18$\pm$0.08. We overplot a second fit (black line) that assumes a slope of -1, and find that in this case $d_{crit}$ = 0.51$\pm$0.01 FWHM, or about half the FWHM.}
    \label{fig:critical-pa}
\end{figure}

Fig.~\ref{fig:critical-pa} shows $\theta_{crit}$ for separations between 0.2\arcsec and 0.4\arcsec. We compute error bars on $\theta_{crit}$ by propagating the errors of the slopes and intercepts of the two linear fits (in e.g. Fig.~\ref{fig:pa-contrast-log}) to the intersection angle. This is repeated at every 0.01\arcsec in our range of separations to generate error bars on each $\theta_{crit}$. We fit the data points in log-log scale with a two-parameter function for slope and intercept, and discover a power-law relation with a slope of -1.18$\pm$0.08 between $\theta_{crit}$ and $\rho$. The uncertainties for this slope are computed by applying the error bars on $\theta_{crit}$ as Gaussian uncertainties in the fit in Fig.~\ref{fig:critical-pa}. When compared with the geometric expectation of $\theta_{crit}$ $\approx$ $d_{crit}$ / $\rho$, which corresponds to a power-law index of -1, our empirical slope is consistent within 2.25$\sigma$.

We next consider the case of a -1 slope, corresponding to the simple geometric scenario, and fit for the intercept $b$. This intercept can be transformed to a measure of $d_{crit}$ by transforming the fitted power-law expression of log($\theta_{crit}$) $\approx$ $k$ log($\rho$) $+$ $b$ back to linear scale, where $k$ = -1 in this case. This gives $\theta_{crit}$ $\approx$ $10^{b}$ / $\rho$, where we convert the units of $10^{b}$ (degs $\cdot$ \arcsec) to the unitless $d_{crit}$ via the pixel scale and the typical size of the FWHM, determined as the average of the median x FWHM and median y FWHM of our targets. This yields $d_{crit}$ = 0.51$\pm$0.01 FWHM, and we overplot this fit in Fig.~\ref{fig:critical-pa}. We note that this result can be compared to \citet{Marois2005}, who study ADI noise reduction as a function of sky rotation with median-based ADI. Adopting the same assumption of a -1 power law, they find that self-subtraction effects are largely avoided at $d_{crit}$ $\approx$ 1.5 FWHM \citep{Marois2005}, which is a threshold about 3 times more stringent than what we find with PCA-based ADI.

To ensure that we do not bias the results above by defining contrasts within a factor of 3 of the contrast at $\approx$2.6\arcsec to be background-limited (see Section~\ref{sec:noise-reg}), we repeat the fits in Fig.~\ref{fig:critical-pa} twice, changing the threshold to a factor of 1 and 2 respectively. We find that for the two-parameter fit (red line), the new slopes are slightly steeper but all slopes are consistent within 1$\sigma$ with each other. We choose the factor of 3 by balancing the requirement of stringency (we risk including background-limited contrasts with smaller factors) and number of data points (we exclude too many data points with larger factors). For the fit of $d_{crit}$ that assumes a slope of -1 (black line), the derived value of $d_{crit}$ = 0.51$\pm$0.01 is also consistent within 1$\sigma$ with the new values.

In practice, these results allow an observer to decide whether they would achieve better contrast between 0.2\arcsec and 0.4\arcsec using only ADI and all of their time on their target or if they should instead allocate some time for a reference star to enable effective RDI, given that they are operating in the speckle noise regime.

\subsubsection{Stellar Flux}\label{sec:stellar-flux}

\begin{figure*}[t!]
    \centering
    \includegraphics[width=0.4\linewidth]{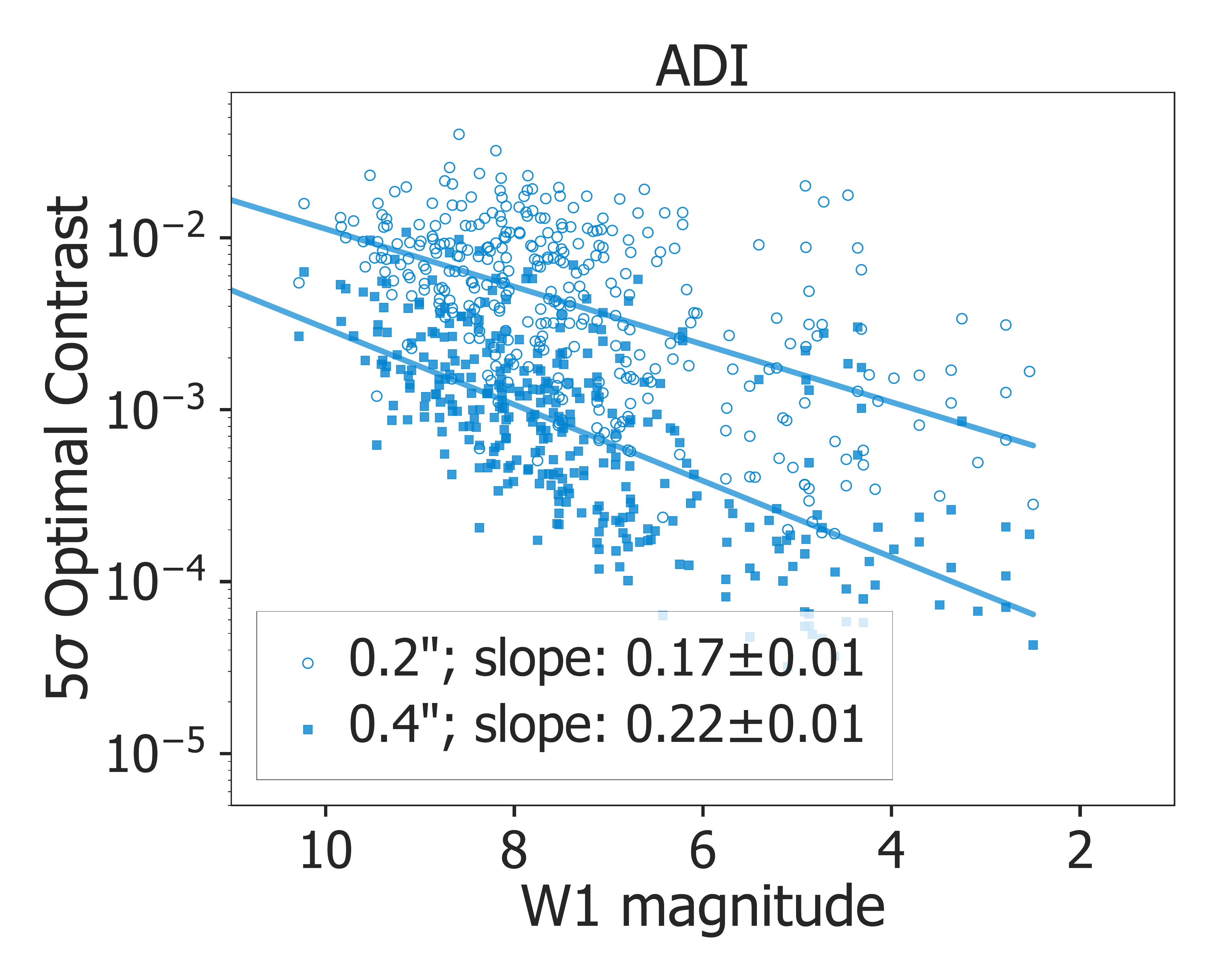}
    \includegraphics[width=0.4\linewidth]{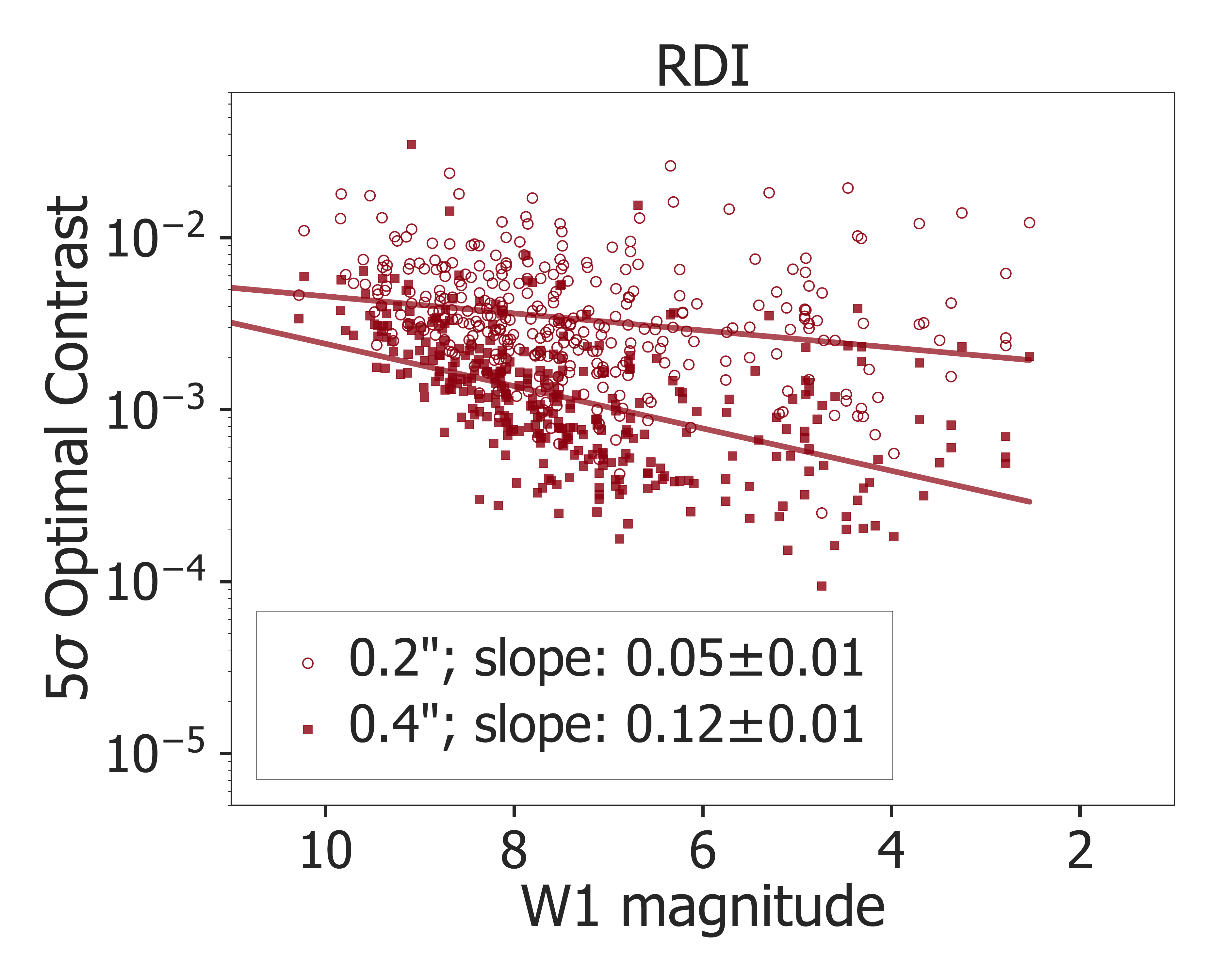}\\
    \includegraphics[width=0.4\linewidth]{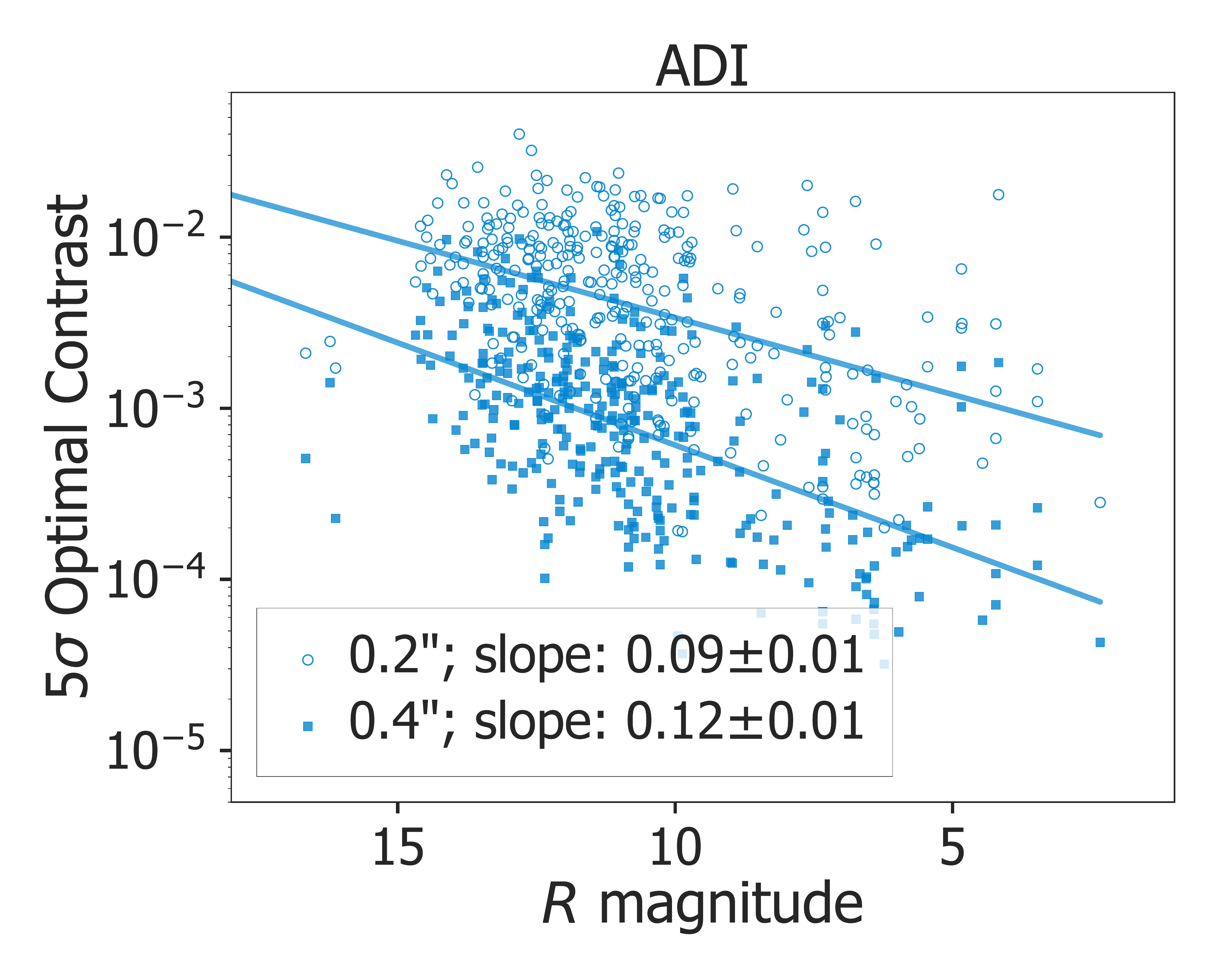}
    \includegraphics[width=0.4\linewidth]{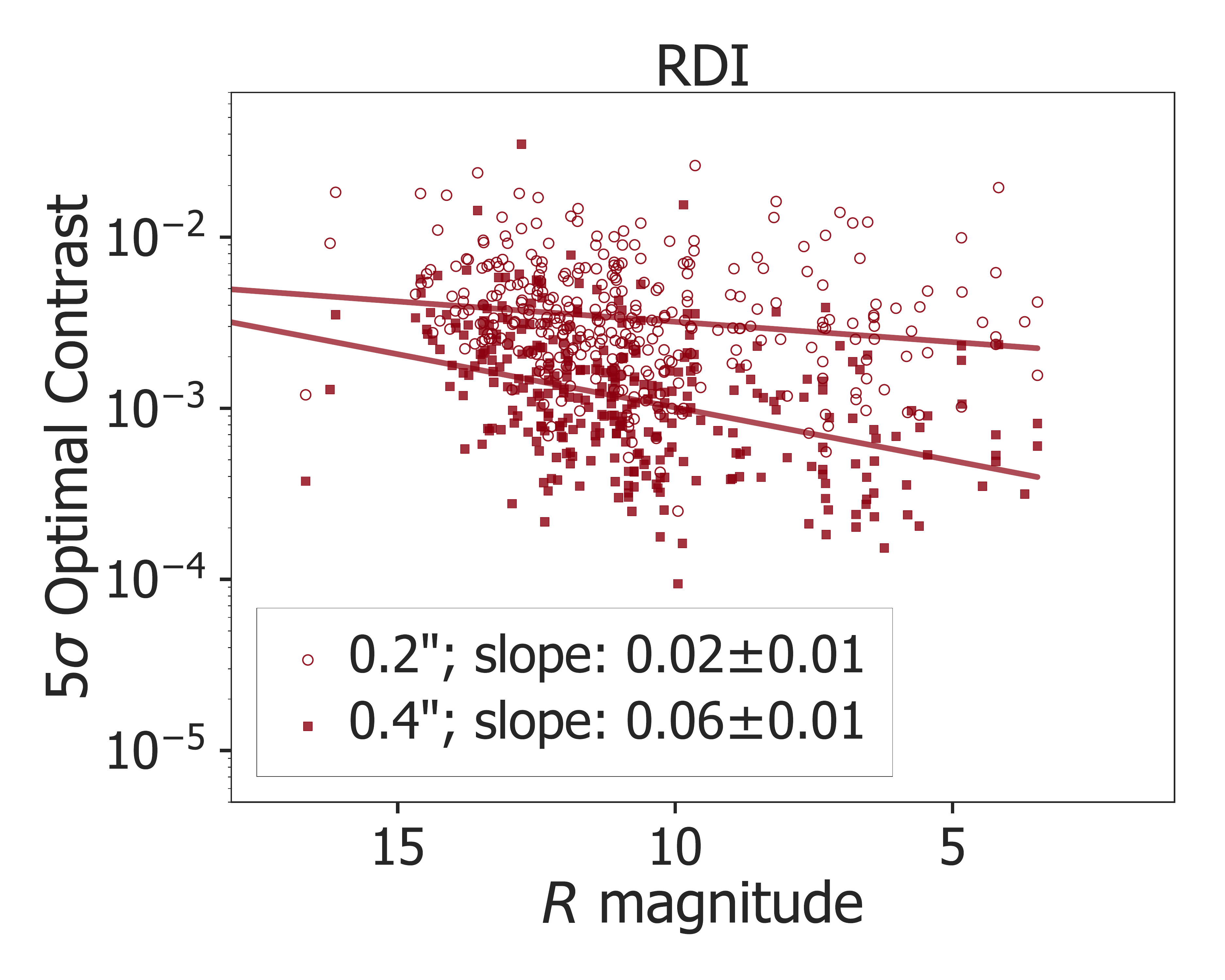}
    \caption{The dependence of contrast limits on the stellar magnitude. WISE W1 (3.4~$\mu$m) and $R$ (0.6-0.7~$\mu$m) band magnitudes are shown.}
    \label{fig:contrast_vs_mags}
\end{figure*}

We search for a relationship between contrast and stellar magnitude in the bandpasses listed in Table~\ref{tab:exp_var} using all data (speckle-limited and background-limited). We find linear relationships between log contrast and magnitude, indicating an underlying power-law relation between contrast and incident flux (see Fig.~\ref{fig:contrast_vs_mags} for the best fit values and uncertainties). We use the WISE W1 (3.4~$\mu$m) magnitude as a proxy for $L^\prime$ magnitude (3.7~$\mu$m). We also analyze the dependence on $R$ band (0.6-0.7~$\mu$m) magnitude because the WFS in the Keck II AO system is sensitive at those wavelengths.

\subsubsection{Seeing}\label{sec:seeing}
The seeing estimation tool, routinely in use at the W.M. Keck Observatory, is used to estimate the seeing from AO system data. The current version uses the closed-loop DM commands. The approach was originally introduced by \citet{Rigaut1991}. As the AO system uses the DM to compensate for wavefront errors induced by atmospheric turbulence, the statistics of the DM shape contains the necessary information to estimate seeing.

We plot ADI and RDI optimal contrast as a function of seeing (see Fig.~\ref{fig:contrast_vs_obs_con}, middle row), and find that seeing is consistent with having no direct correlation with contrast, with slopes and fit uncertainties on the order of $10^{-5}$.

\subsubsection{Atmosphere coherence time $\tau_{0}$}\label{sec:tau0}

Atmospheric turbulence evolves with a characteristic timescale, $\tau_{0}$, called the coherence time. This timescale is defined as the time for the wavefront phase error to change by 1 radian; it is inversely proportional to the velocity of the turbulent wind layer and also inversely proportional to the amplitude of the turbulence (seeing). We estimate $\tau_0$ from a temporal analysis of the DM commands. First, the temporal structure function is estimated from the DM commands, and then a power-law model fit to the temporal structure function is carried out to estimate the coherence time, according to \citet{Davis&Tango1996}. 

Classical AO “lag error” predicts that the mean-square wavefront error $\sigma^{2}$ caused by correction time lag is proportional to $(\tau_0/t)^{-5/3}$ \citep{Fried1990}, where $t$ is the time lag of the AO system. In addition, the Strehl ratio, which is the ratio of the central intensity of the observed PSF to that of a diffraction-limited PSF, is expected to scale as $e^{-\sigma^{2}}$ \citep{Schroeder1987}. Given also the strong dependence found between contrast and Strehl ratio \citep{Milli2017}, we expect that contrast should show a strong power-law relation with $\tau_0/t$. The Keck II AO system has variable correction speed, and is run at a slower speed on fainter stars; therefore, we use the WFS integration time (i.e., inverse of the WFS frame rate) as a proxy for the time lag $t$.

\begin{figure*}[t!]
    \centering
    \includegraphics[width=0.4\linewidth]{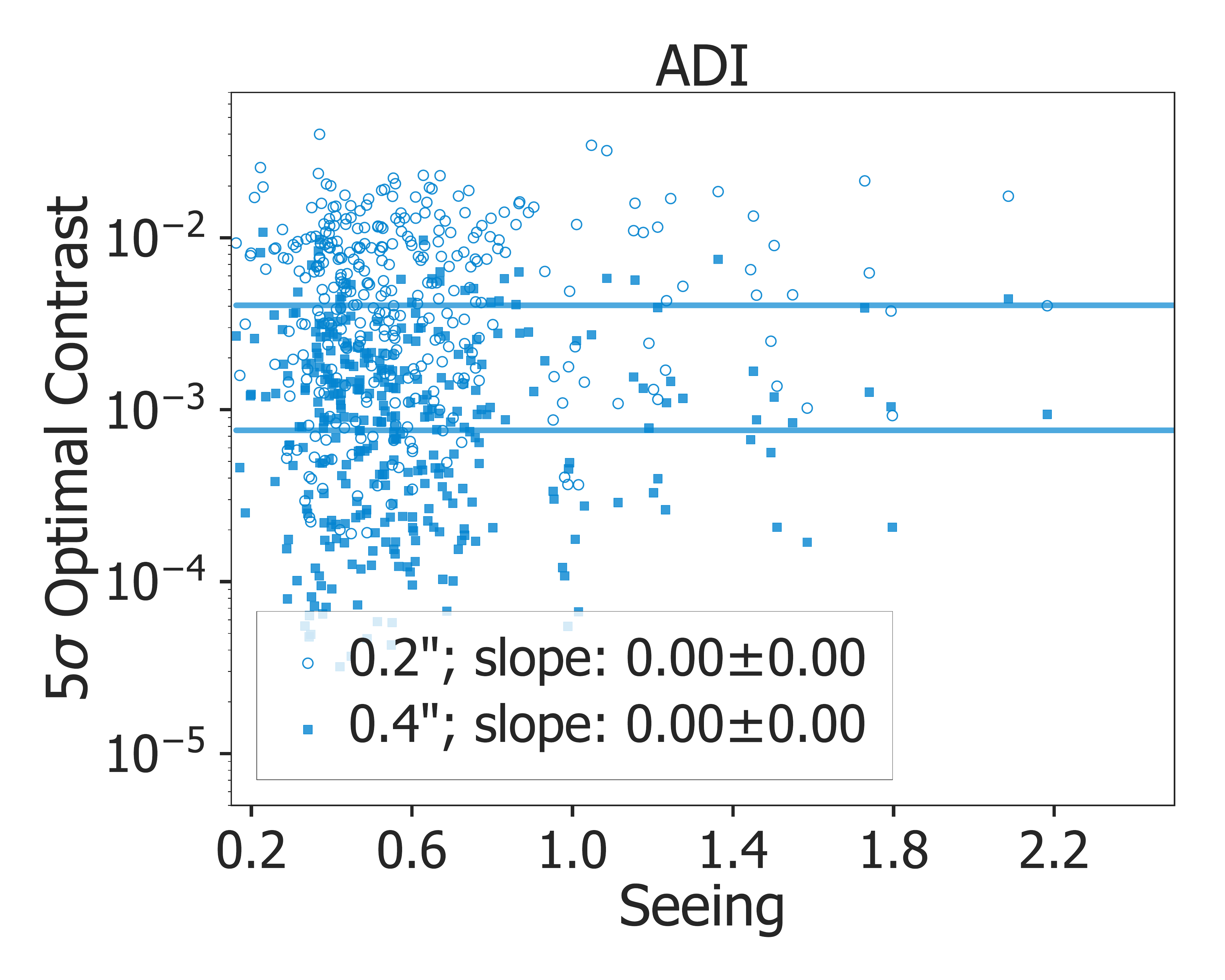}
    \includegraphics[width=0.4\linewidth]{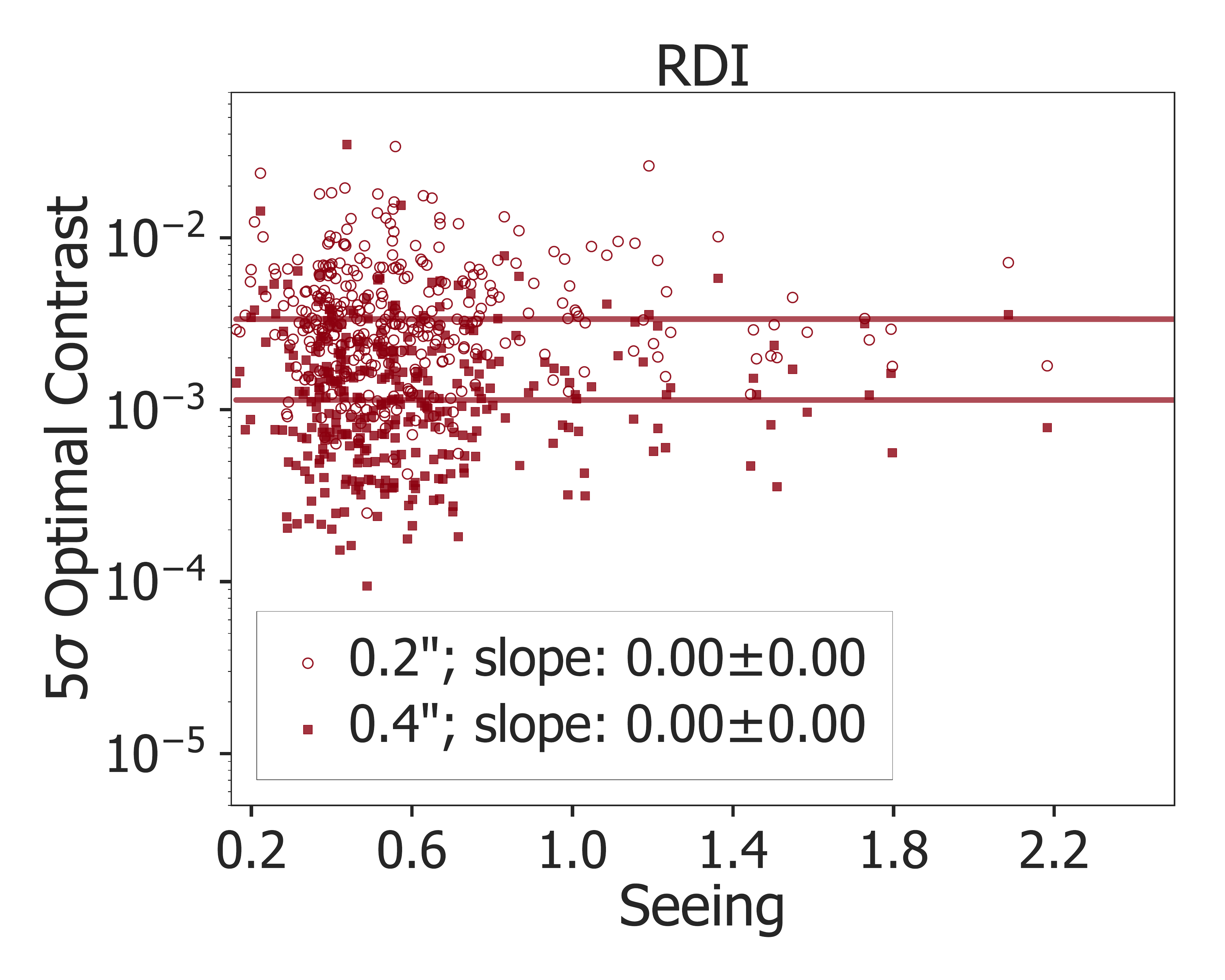}\\
    \includegraphics[width=0.4\linewidth]{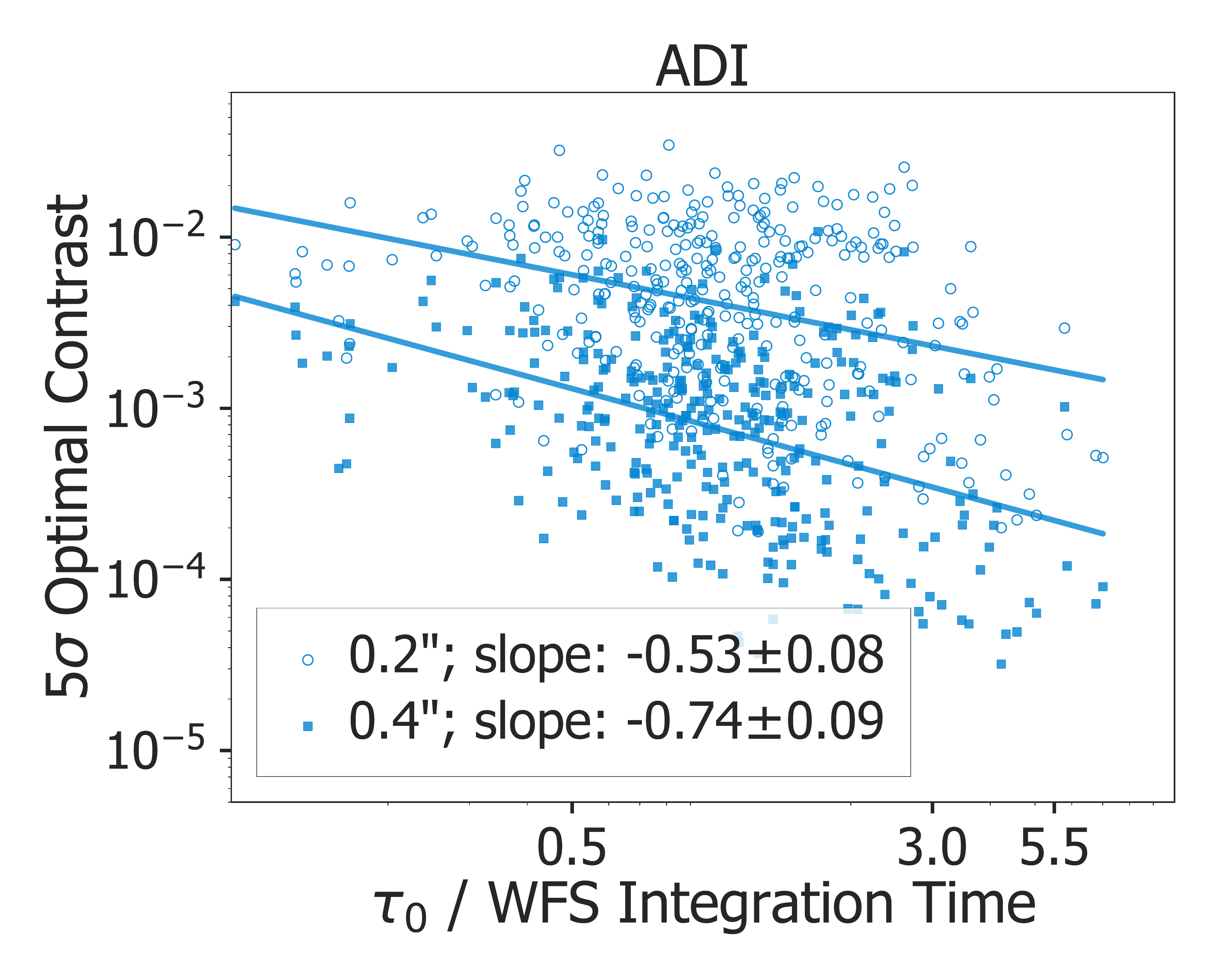}
    \includegraphics[width=0.4\linewidth]{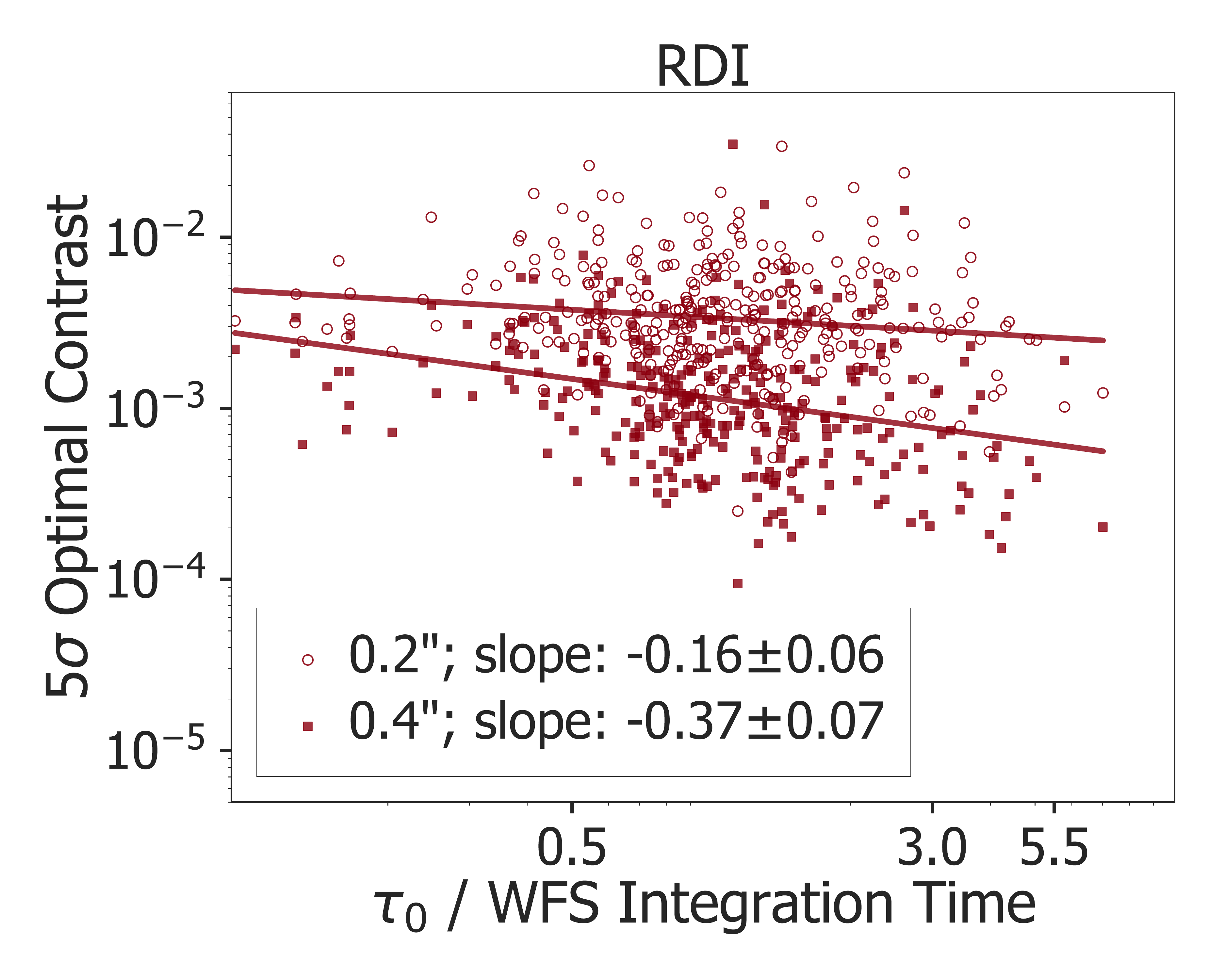}\\
    \includegraphics[width=0.4\linewidth]{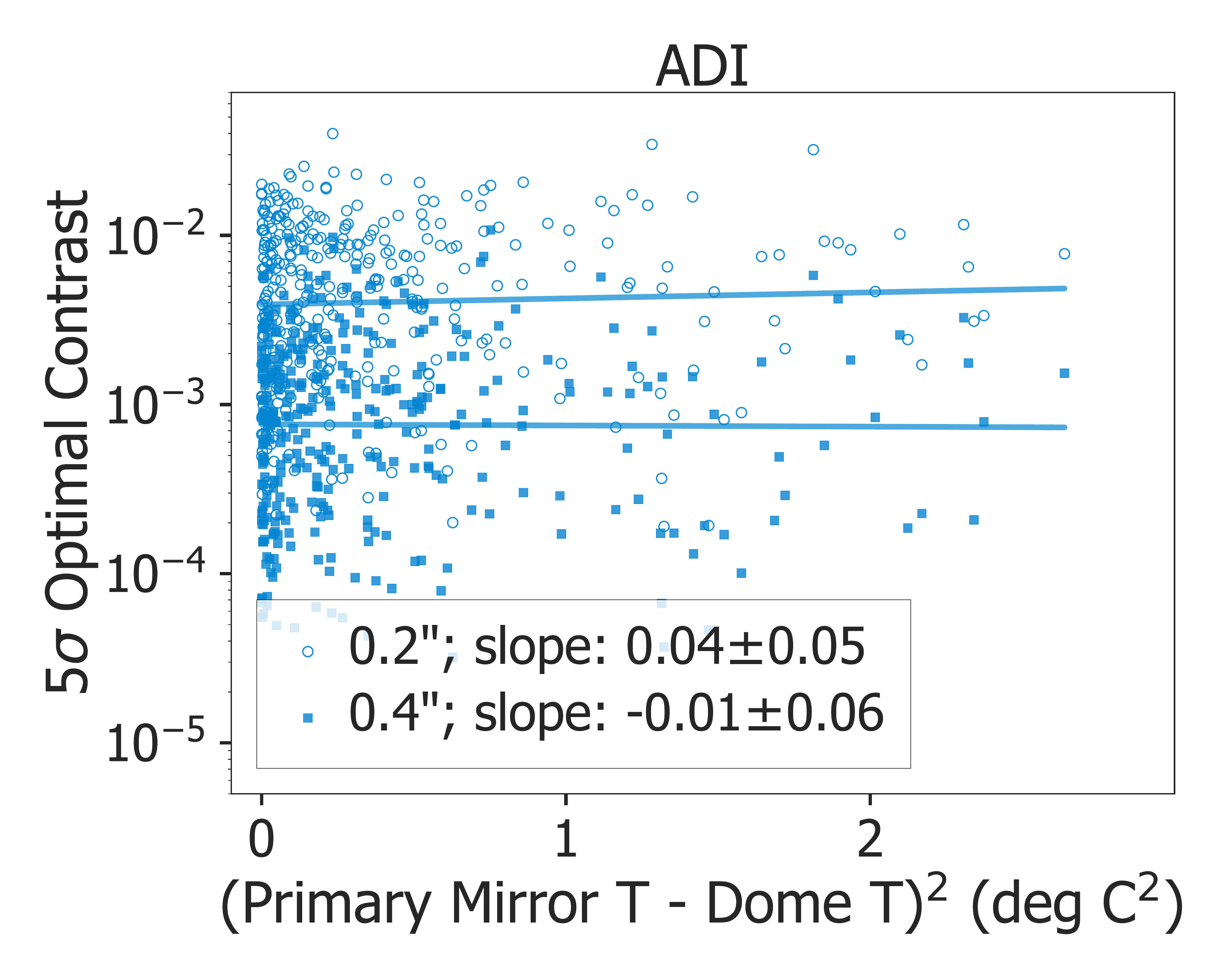}
    \includegraphics[width=0.4\linewidth]{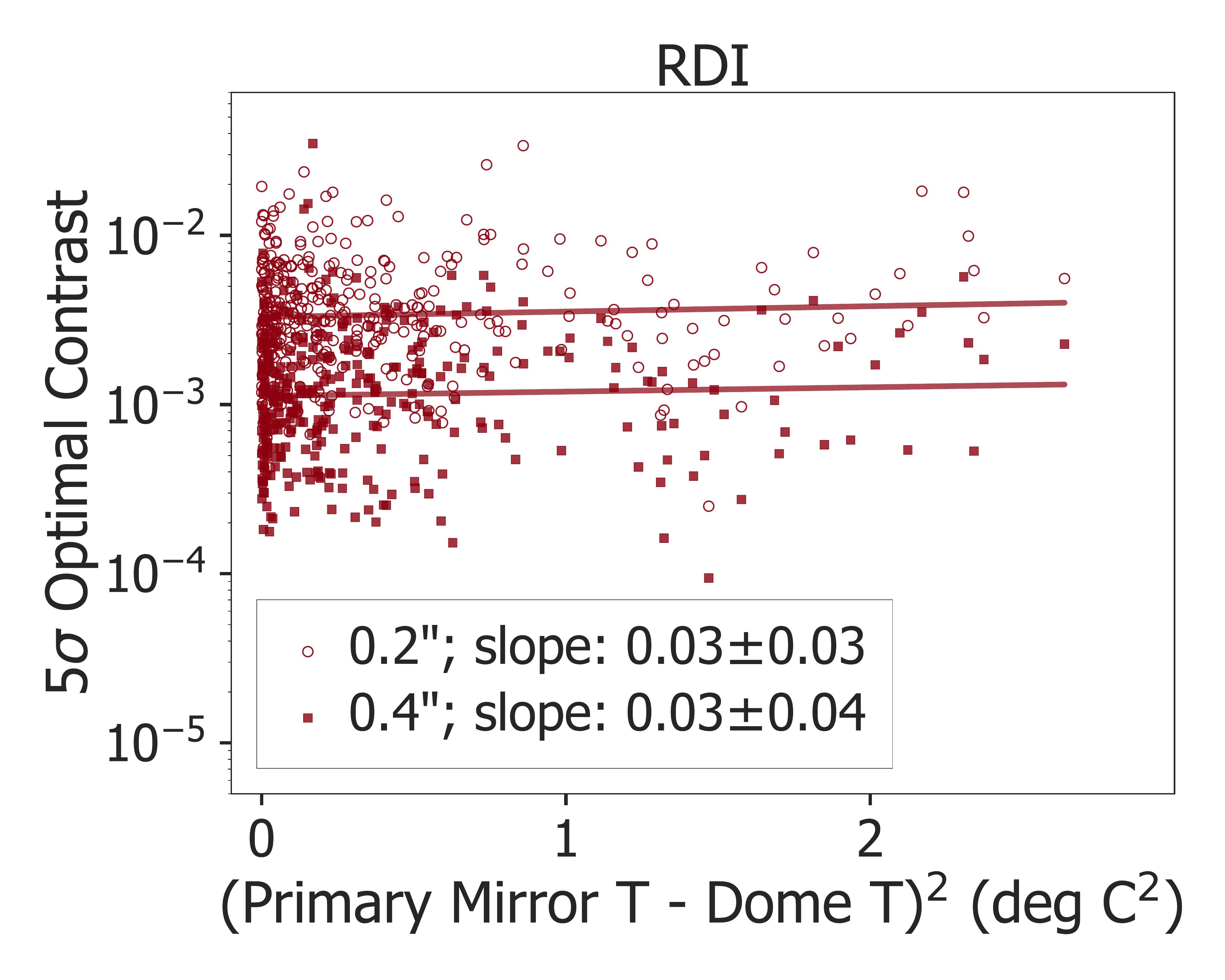}
    \caption{The effects of $\tau_0/t$, seeing, and the temperature difference between the primary mirror and the dome. Whereas $\tau_0$/t is correlated with the achieved contrast limit, both seeing and the temperature difference are consistent with having no relationship or weak relationships in these univariate fits.}
    \label{fig:contrast_vs_obs_con}
\end{figure*}

As with previous studies \citep{Milli2017, Bailey2016}, we see a strong correlation between ($\tau_{0}/t$) and the instrument performance. The performance metric studied is raw contrast in \citet{Bailey2016}, and Strehl ratio in \citet{Milli2017}. Both studies find that the instrument performance is primarily limited by temporal wavefront errors at low coherence times, and that there exists a strong relationship between higher coherence times and better performances. We expand on these studies by quantifying that relationship for post-processed contrasts. Because we expect a power-law relation, we plot $\tau_{0}/t$ in log scale with log contrast. We find slopes of -0.53$\pm$0.08 and -0.74$\pm$0.09 for ADI optimal contrast, and -0.16$\pm$0.06 and -0.37$\pm$0.07 for RDI optimal contrast, at 0.2\arcsec and 0.4\arcsec respectively (Fig.~\ref{fig:contrast_vs_obs_con}, top row). As described in the previous section, NIRC2 vortex contrast is not directly correlated with seeing, so this relationship can be attributed to the velocity of the turbulent wind layer.

\begin{figure*}[t!]
    \centering
    \includegraphics[width=0.4\linewidth]{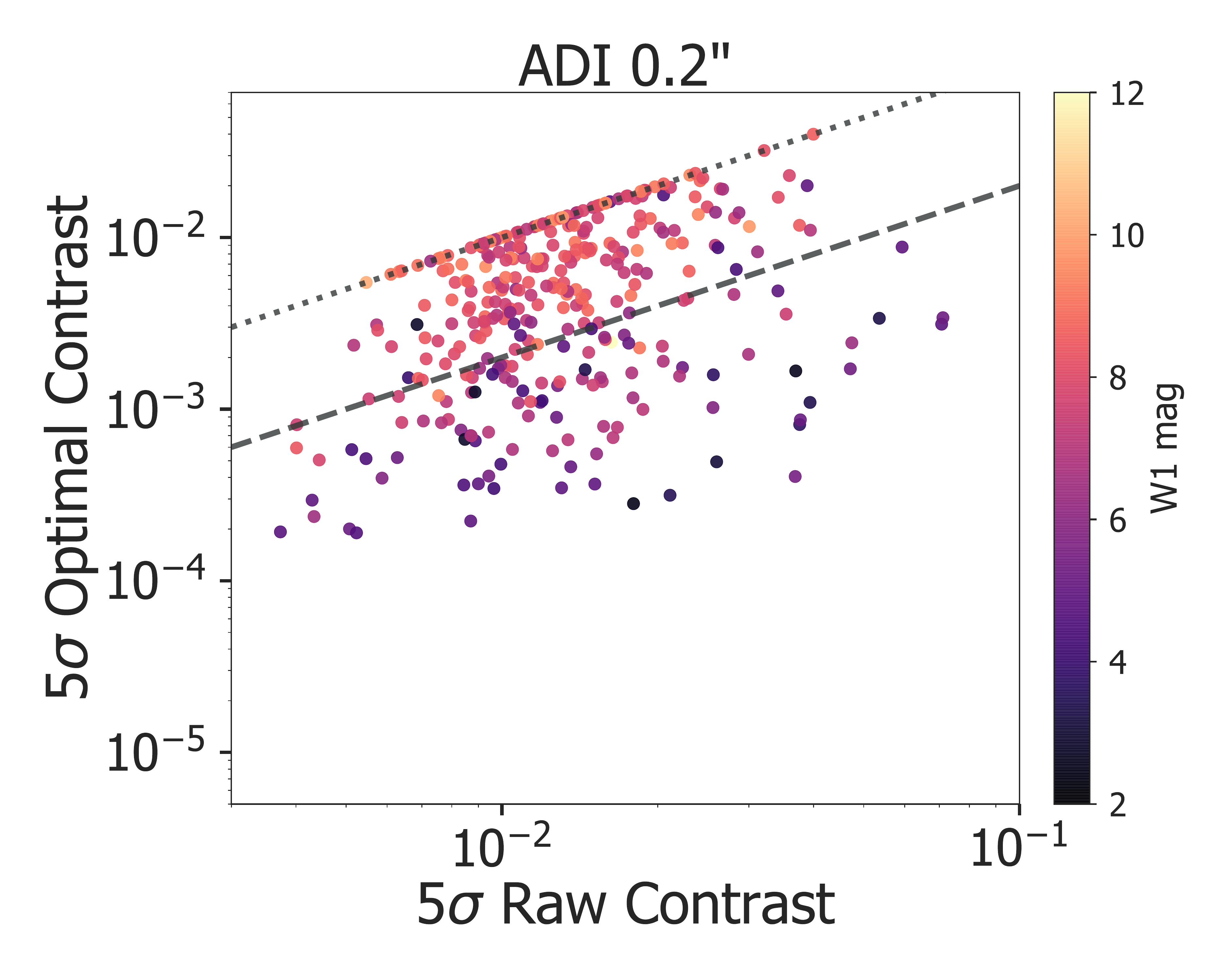}
    \includegraphics[width=0.4\linewidth]{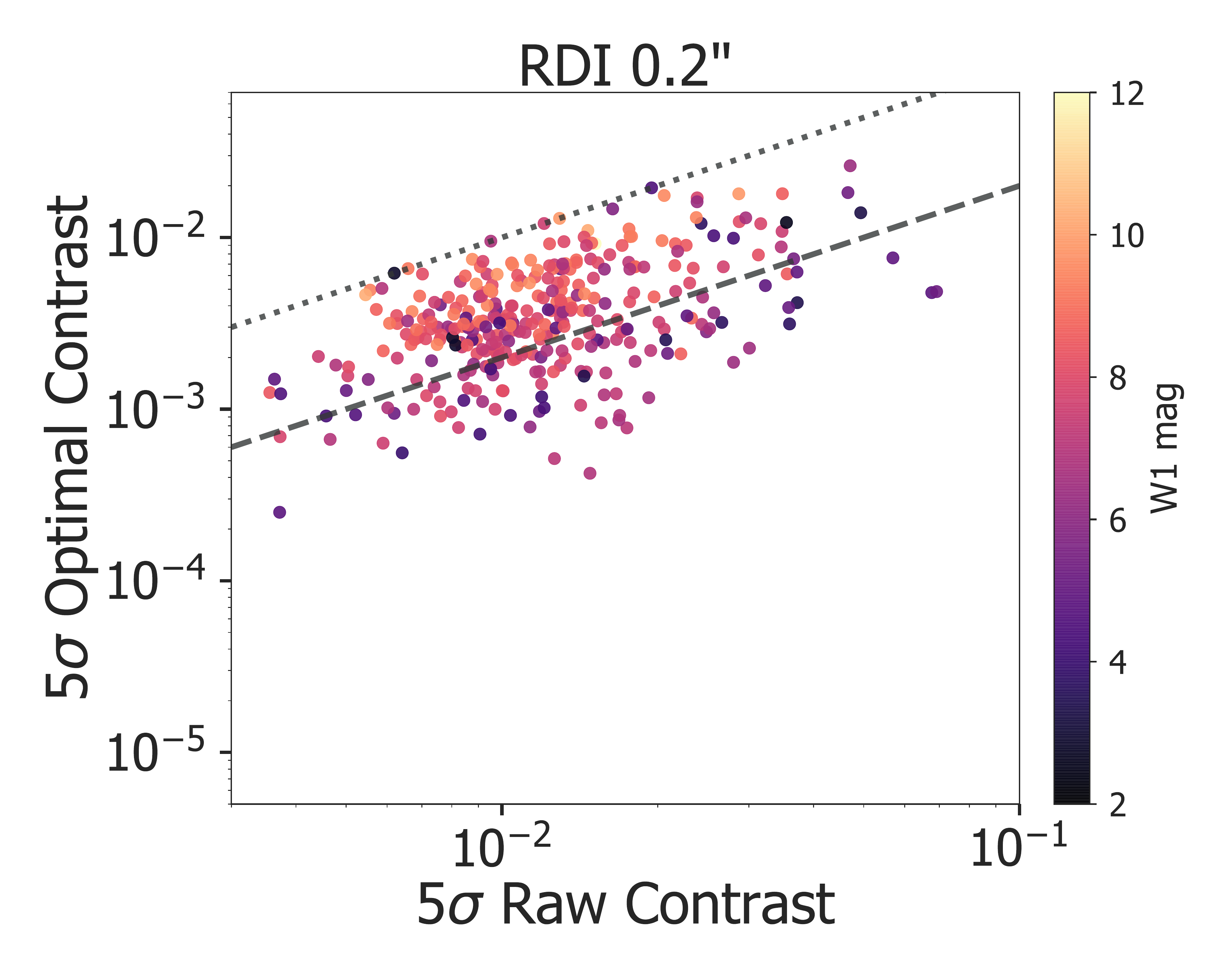}\\
    \includegraphics[width=0.4\linewidth]{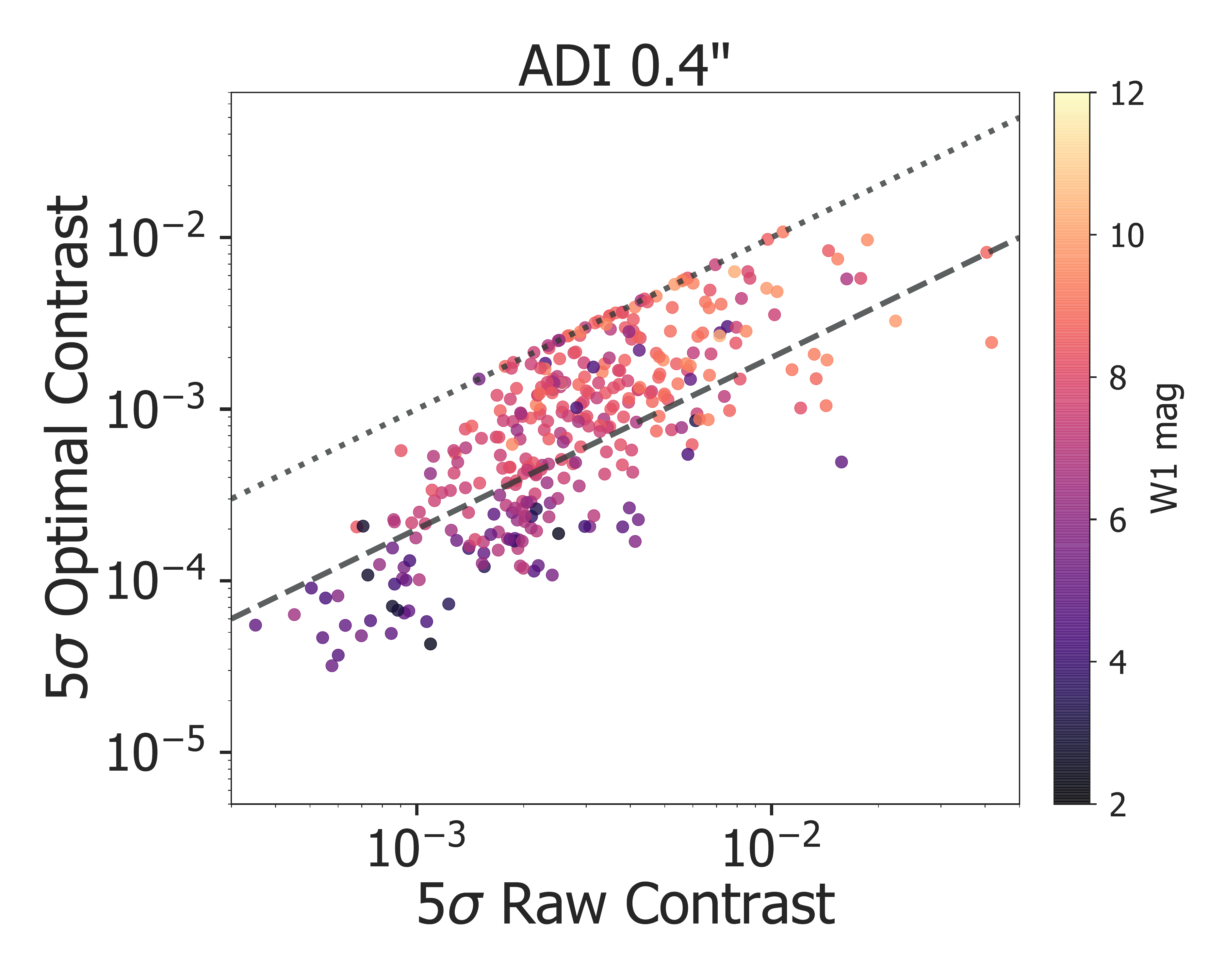}
    \includegraphics[width=0.4\linewidth]{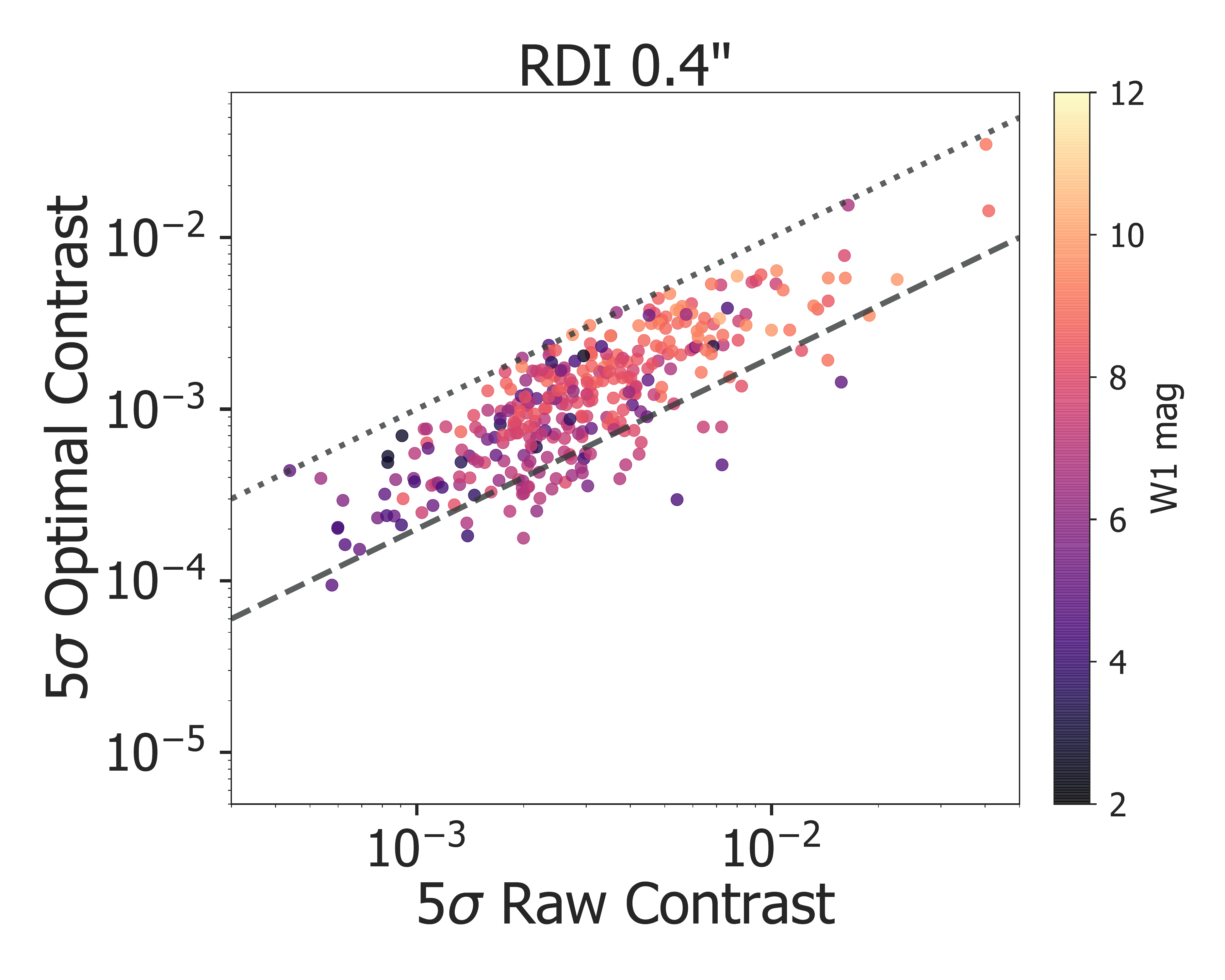}
    \caption{Contrast improvement through ADI and RDI post-processing. The raw contrast is the contrast limit using the median of the de-rotated frames. Points are color-coded by their W1 magnitude. Top dotted line represents a 1:1 ratio, indicating zero gain from post-processing. Bottom dashed line represents a 1:5 ratio, indicating a factor of 5 improvement in contrast from post-processing.}
    \label{fig:contrast_vs_rawcontrast}
\end{figure*}

\subsubsection{Temperature Differentials}\label{sec:temp}
We obtain temperature measurements from YSI Precision Thermistors located at various places on the telescope, dome, and the AO enclosure. The primary mirror temperature is the average reading from 9 thermistors located on the back side of mirror segments measuring the glass temperature. The dome temperature sensor is mounted to the upper rail atop the AO enclosure to measure the dome ambient air temperature. The optical bench temperature measures the ambient air temperature of the AO bench area located close to the WFS. Lastly, the AO acquisition camera (ACAM) enclosure temperature sensor is located inside the ACAM electronics enclosure, on top of the AO bench enclosure and fairly far away from the ACAM camera itself.

We study each possible pair of temperature differential ($\Delta T$), and find a lack of correlation for all pairs. Following \citet{Tallis2018}, who show that the mean-square wavefront error $\sigma^{2}$ should be proportional to $\Delta T^2$, we plot contrast as a function of $\Delta T^2$. As an example, we show optimal contrast against the squared temperature difference between the primary mirror and the dome air (see Fig.~\ref{fig:contrast_vs_obs_con}, bottom row). We find uncertainties in the slopes that are larger than the magnitudes of the slopes for all cases, indicating that temperature differentials are not directly correlated with contrast for the Keck/NIRC2 vortex coronagraph, in contrast to what is found with GPI data by \citet{Tallis2018}. However, we note that the range of $\Delta T$ spanned by our data is small compared to that spanned by the GPI data (about an order of magnitude smaller in terms of $\Delta T^2$), likely due to Keck's superior temperature control.

\subsection{Measuring contrast improvement from PCA-based PSF subtraction}

To measure the performance of our PCA-based PSF subtraction process, we investigate the relationship between optimal contrast and the raw contrast prior to PCA post-processing. Raw contrast is computed following the same procedure as that described in Section~\ref{sec:re-reduction} and is defined as the contrast achieved using the median of the de-rotated pre-processed images, before PSF subtraction. The difference between optimal contrast and raw contrast shows the gain due to post-processing. The typical raw contrast achieved on-sky is on the order of $10^{-2}$ at a separation of 0.2\arcsec (see Fig.~\ref{fig:contrast_vs_rawcontrast}), which is limited by the central obscuration of the Keck telescope \citep{Mawet2011}. The starlight rejection ratio for the same vortex mask was about 10$\times$ better when measured in the lab using a circular entrance aperture \citep{Catalan2016}.

Fig.~\ref{fig:contrast_vs_rawcontrast} shows the optimal contrast limits as a function of the raw contrast, along with lines corresponding to 1:1 (dotted) and 1:5 (dashed) ratios of optimal contrast to raw contrast. The median gain factors from post-processing at 0.2\arcsec are 2.5 for ADI and 3.3 for RDI. At 0.4\arcsec, the median gains are 3.4 and 2.5 for ADI and RDI respectively. Since RDI does not depend directly on PA rotation, it is limited by the quality of raw frames for the target star and the reference stars and how well the references match the target. On the other hand, ADI performance is strongly dependent on PA rotation, which is only indirectly correlated with raw contrast through total integration time and airmass. However, the power of RDI at small separations is evident. For ADI at 0.2\arcsec, a group of points form a straight line with a slope of 1, indicating targets that have raw contrast equal to the optimal contrast. These targets have too little PA rotation ($<$ 5$^{\circ}$) for ADI post-processing to improve the contrast limits beyond the raw contrast and therefore suffer from severe self-subtraction. In these cases, potential companions would move by less than 0.2 of the FWHM size, which could be compared to the $\approx$0.5 FWHM necessary for ADI to outperform RDI as found in Section~\ref{sec:PA}. 

\section{Contrast prediction with random forests}\label{sec:stat-models}

\subsection{Overview}
We analyze the NIRC2 vortex performance to create random forest models \citep{Breiman2001} that predict contrast limits expected in future observations. Our random forests are based on regression trees, since our response variable is continuous. Regression trees are an algorithm to divide the set of observations into many regions through a series of splits. Each split divides a region into two smaller regions at a specific cutoff value of a specific explanatory variable. The cutoff value is chosen to minimize the residual sum of squares
\begin{equation}
RSS = \sum_{j=1}^{J} \sum_{i\in R_j} (y_i - \Test{\ymean})^2,
\label{eq:RSS}
\end{equation}
where $J$ is the total number of regions (one plus the number of splits) in the tree, $R_j$ is the $j$th region, $y_i$ are the individual observations, and $\Test{\ymean}$ is the mean of the response variable for the observations within the $j$th region ($R_j$). The first sum is carried over all regions in the tree, and the second sum is carried over all observations within a region, so the two sums incorporate all observations used to create a given tree. Random forests are a conglomerate of many trees, where each individual tree is made from a bootstrapped sample of the data, created by randomly sampling the original sample with replacement. The predictions from each tree are averaged to yield the final result of the random forest. A detailed description of the random forest algorithm is beyond the scope of this paper. For details, we refer the reader to \citet{Louppe2014understanding}.

\subsection{Model construction}
\begin{figure}[t]
    \centering
    \includegraphics[width=\linewidth]{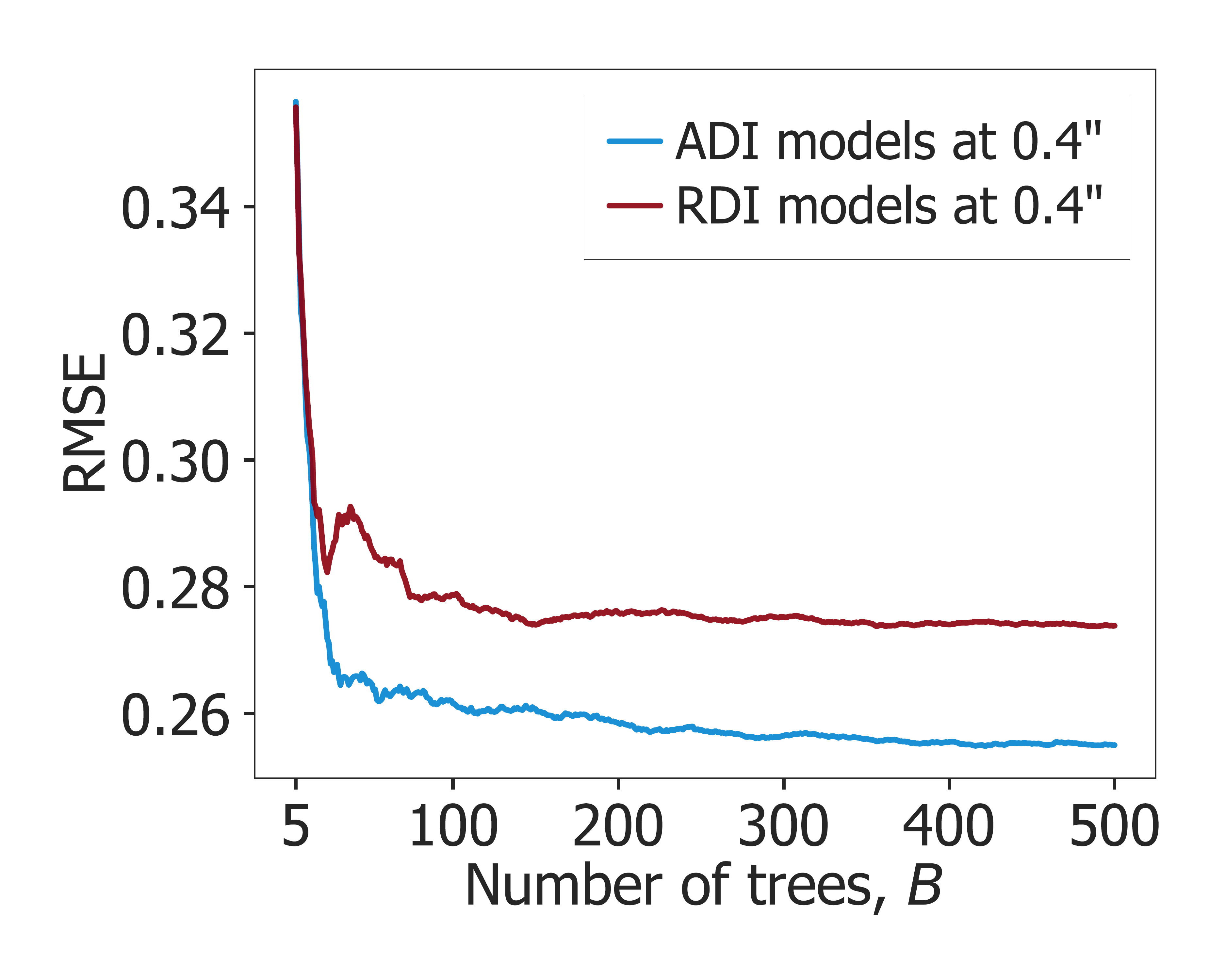}
    \caption{The RMSE of random forest models as a function of $B$, the number of trees used in the model, for ADI and RDI contrasts at 0.4\arcsec. Models are shown starting from $B$ = 5. The error stabilizes around 500 trees.
    \label{fig:num_trees}}
\end{figure}

In the R programming language, we build random forests for our data using the \texttt{caret} package~\citep{Kuhn2008} for predictive models. This implementation has two tunable parameters, $B$, the number of trees, and $m$, the number of explanatory variables used for splitting. Each time a split in a tree is considered, a random subset of $m$ variables is chosen from the full set of explanatory variables to de-correlate individual trees in a given model and thus reduce the variance of the model. We select the optimal $m$ for each model as the one that gives the lowest prediction error (defined below in Section~\ref{sec:metrics}). We keep $B$ constant at 500 trees in each model, since the error plateaus around that number, as shown in Fig.~\ref{fig:num_trees}; the error only decreases by less than 0.4\% when increasing $B$ to 1000.

We build models to predict three different response variables: ADI optimal contrast, RDI optimal contrast, and raw contrast. For ADI optimal contrast, we construct models from 0.2\arcsec to 1.0\arcsec, in intervals of 0.01\arcsec. For RDI optimal contrast and raw contrast, we construct models from 0.2\arcsec to 0.4\arcsec. For each model, we transform the response variable into log scale, so the models predict log contrast. Because raw contrast and optimal contrast are not independent, we exclude raw contrasts from our model when predicting optimal contrast, and vice versa. In addition, we keep only one variable when a pair of variables has a Pearson's correlation coefficient of \textgreater 0.9. The final subset of explanatory variables used in our statistical models is listed in Table~\ref{tab:exp_var}. Note that we include variables even if they show no correlation in a linear fit because random forests work in a non-linear manner by splitting the parameter space. Lastly, for each variable, we remove values that are unquestionably erroneous. Specifically, we remove a few faulty seeing values that conglomerate around 10000\arcsec and a few $\tau_0$ values around 10000ms, which are likely caused by cloud cover or AO anomalies. We also remove a few faulty PSF FWHM values that center around 100 pixels and around 2 pixels, which are caused by a failure of the PSF fitting routine due to extremely poor image quality, and PA rotation values smaller than 0, caused by faulty metadata in the fits headers. In total, we remove less than 4\% of the data in this way.

\subsection{Performance metrics}\label{sec:metrics}
We use two performance metrics to characterize our models: $R^{2}$ is the amount of variance explained and the root-mean-square error (RMSE) is calculated by predicting independent observations not used in creating a given model. In this way, the RMSE represents an unbiased estimate of the model's performance and is applicable for future observations. As mentioned, random forests use a bootstrapped sample of the data to build each tree. On average, 1/3 of data would be omitted in a given bootstrapped sample, and that portion forms the so-called out-of-bag (OOB) observations. To compute the RMSE, each regression tree is applied onto its associated OOB observations and the mean error of predictions on the OOB data is computed. These errors are averaged over all trees to yield the final RMSE of the random forest. Since we transform contrast into log scale, the RMSE is equivalent to a dex error (i.e., predicting $10^{-2.8}$ instead of the actual $10^{-3.0}$ is an RMSE of 0.2dex).

\subsection{Prediction accuracy}
\begin{figure*}[t]
    \centering
    \includegraphics[width=0.4\linewidth]{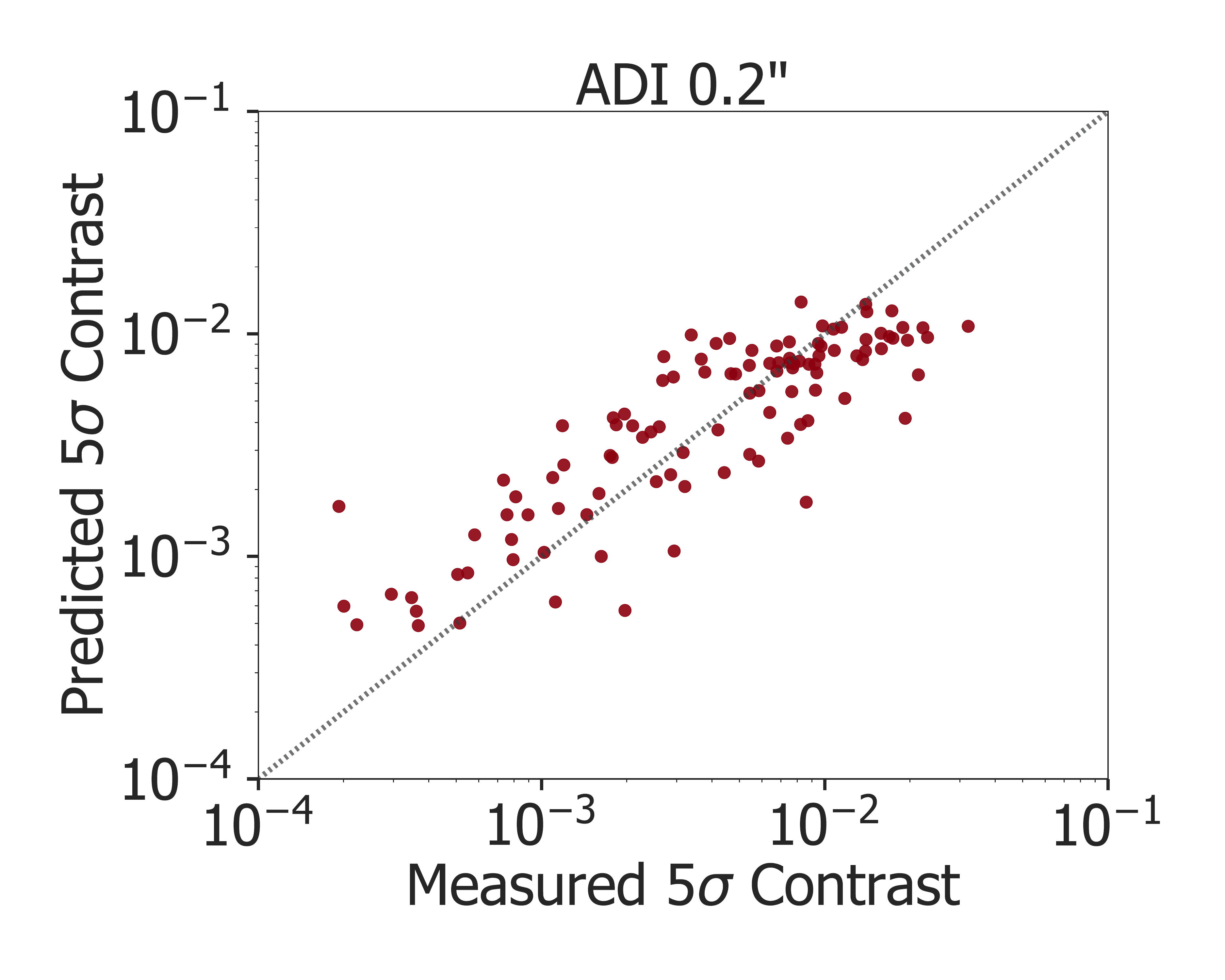}
    \includegraphics[width=0.4\linewidth]{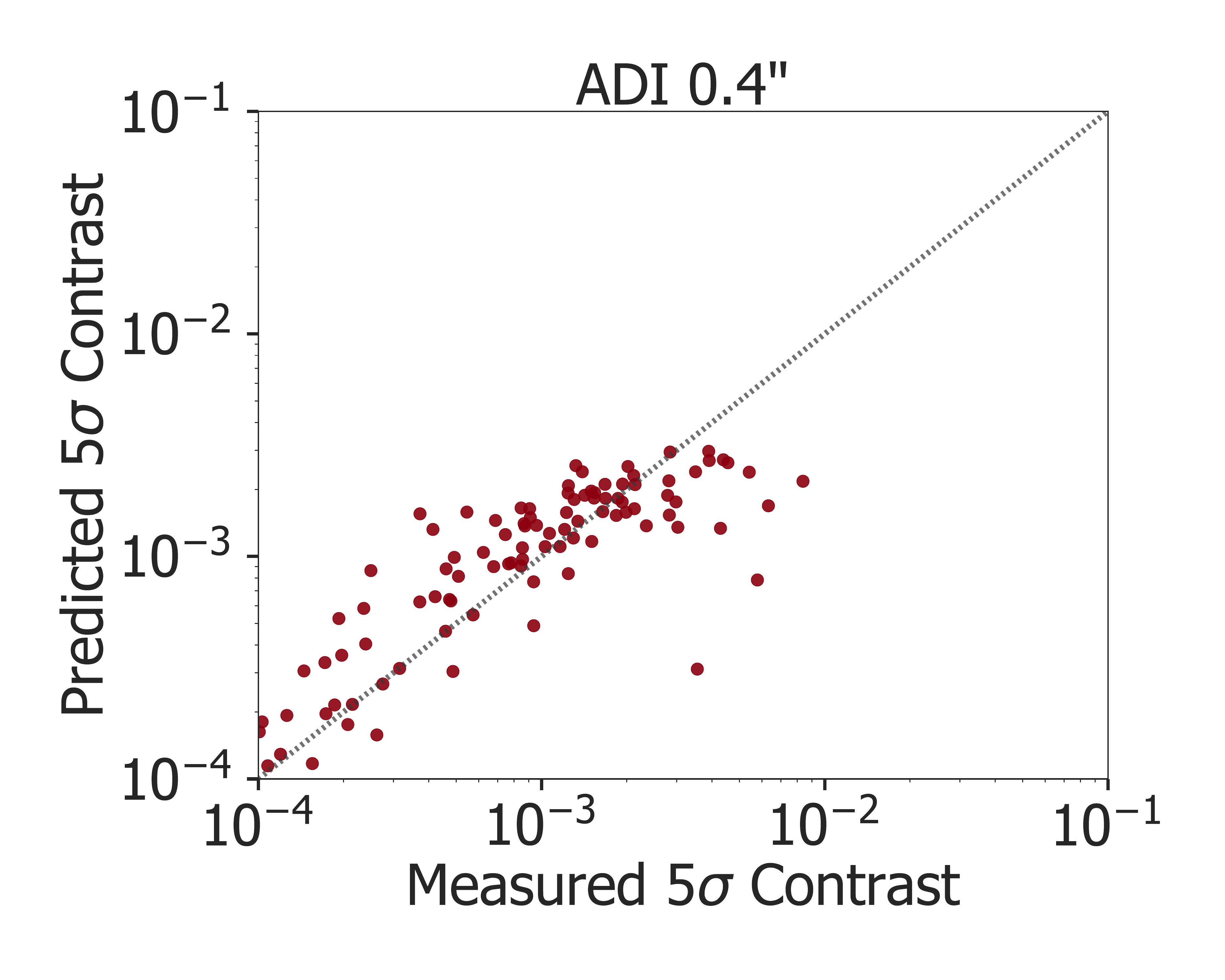}\\
    \includegraphics[width=0.4\linewidth]{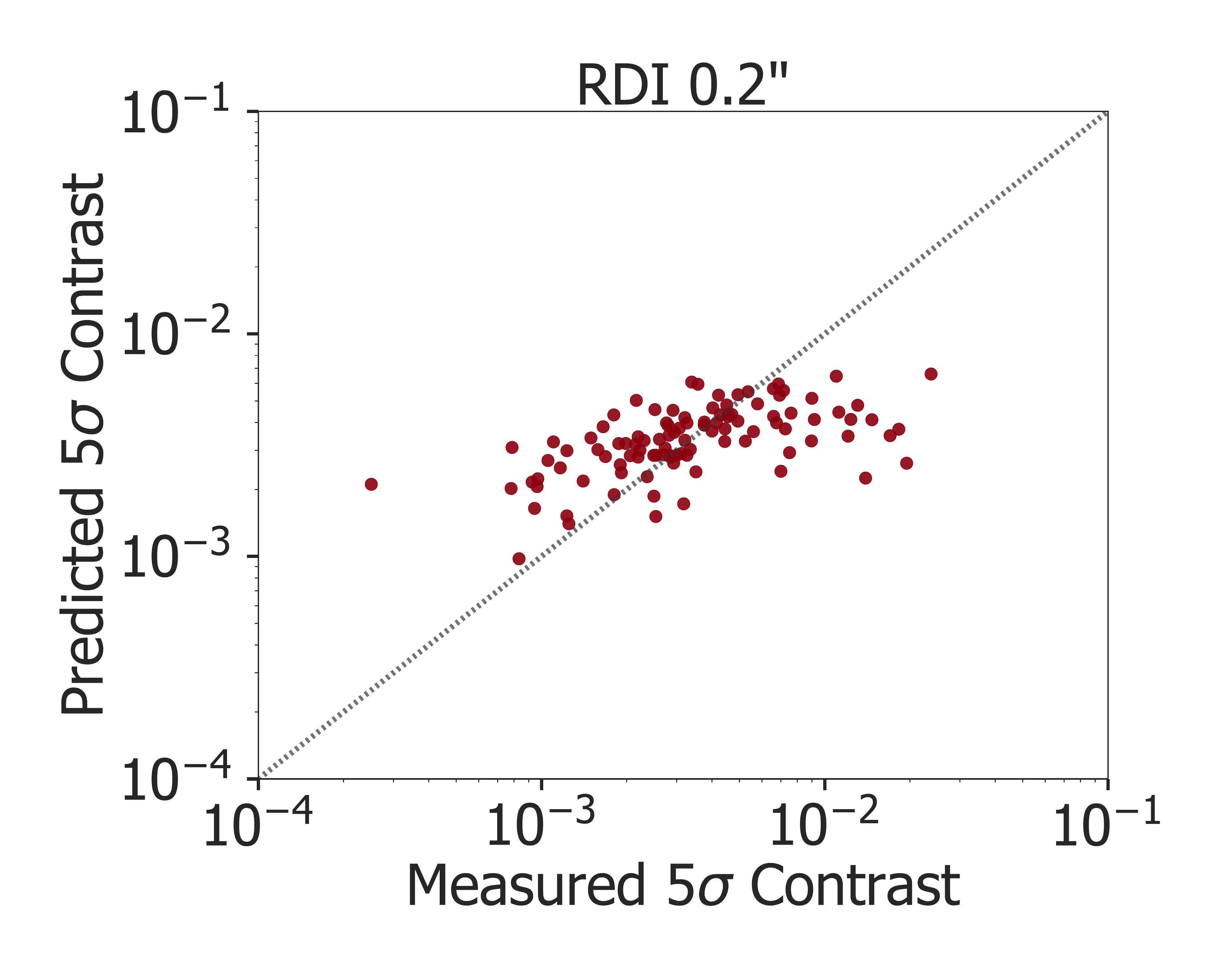}
    \includegraphics[width=0.4\linewidth]{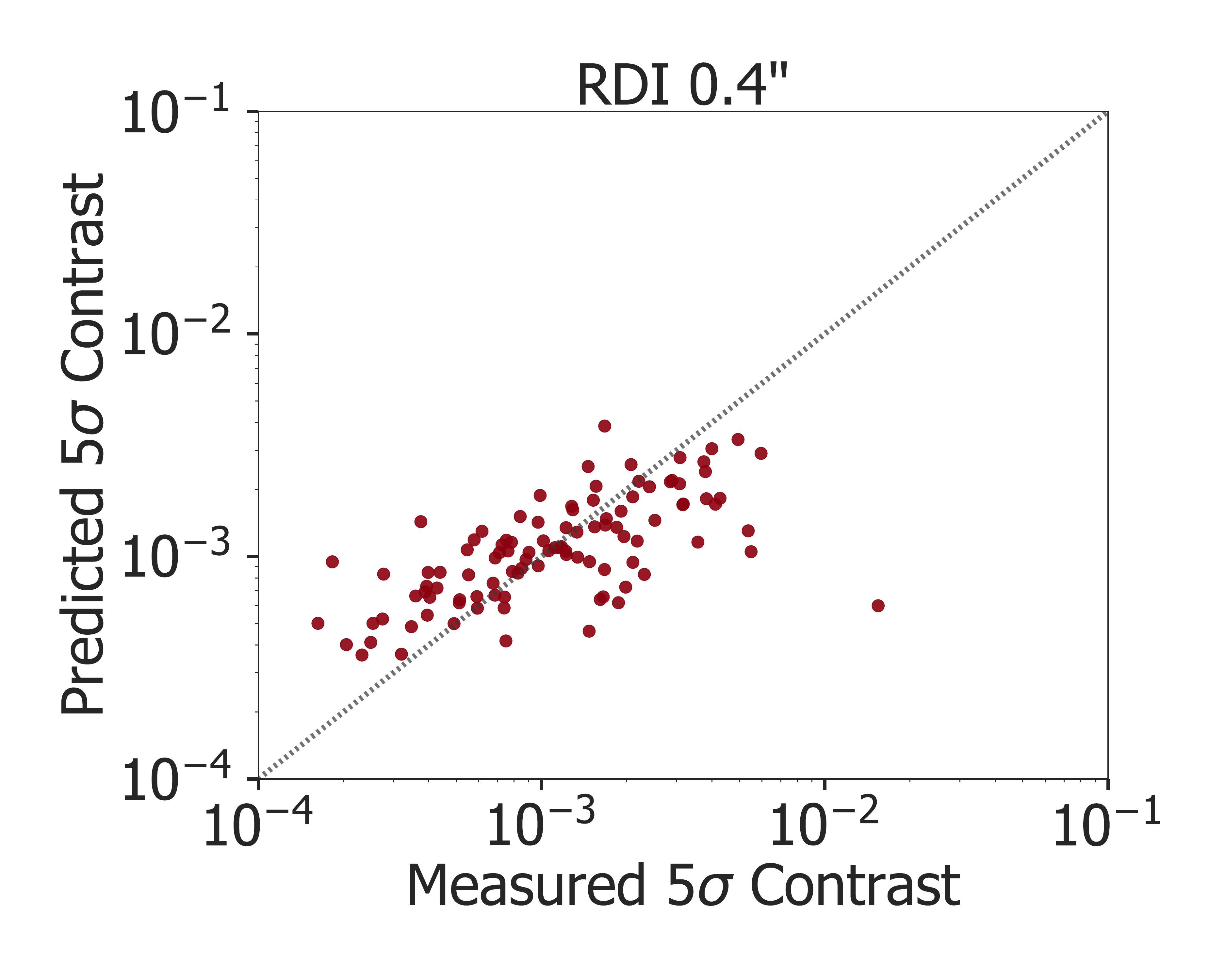}\\
    \includegraphics[width=0.4\linewidth]{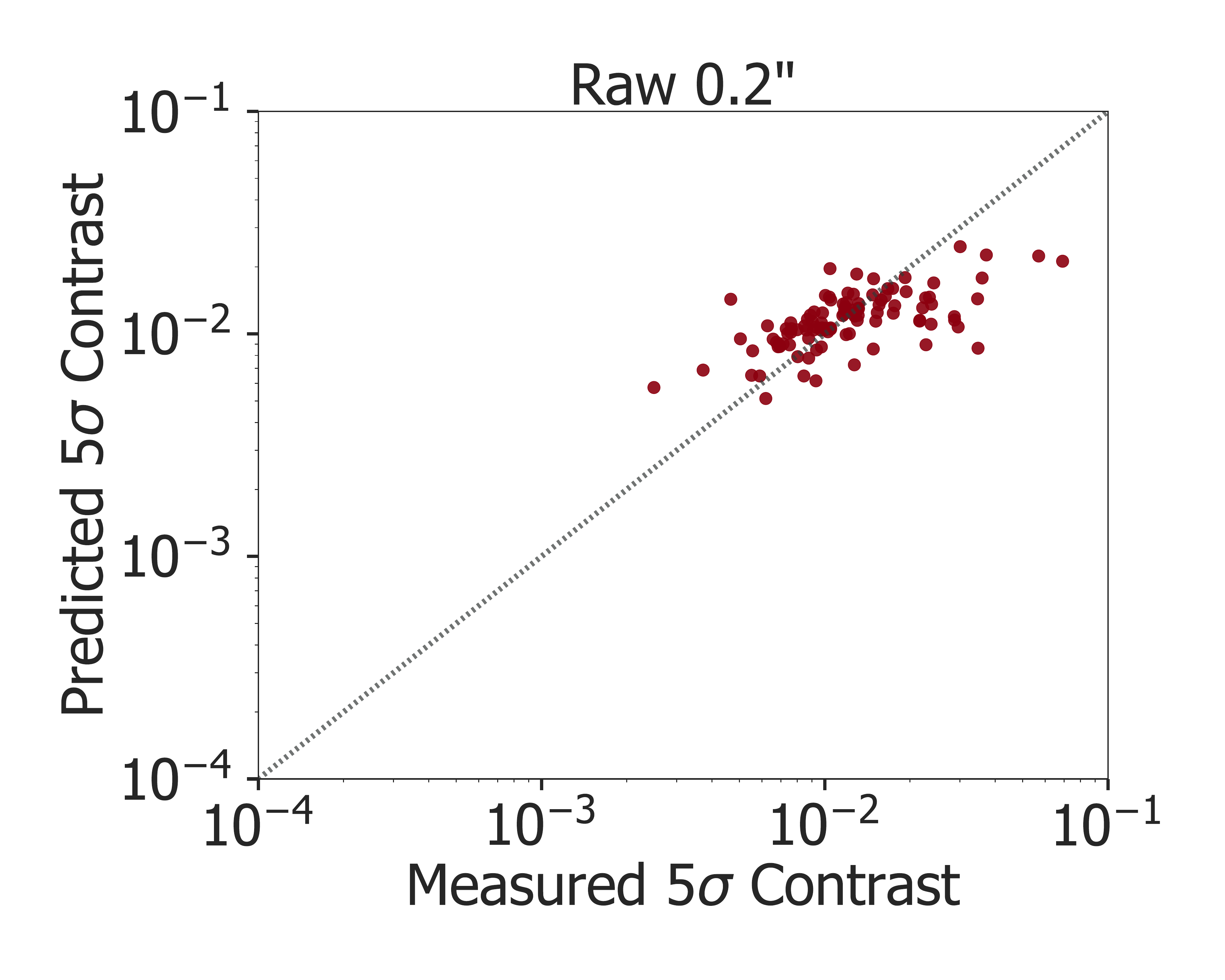}
    \includegraphics[width=0.4\linewidth]{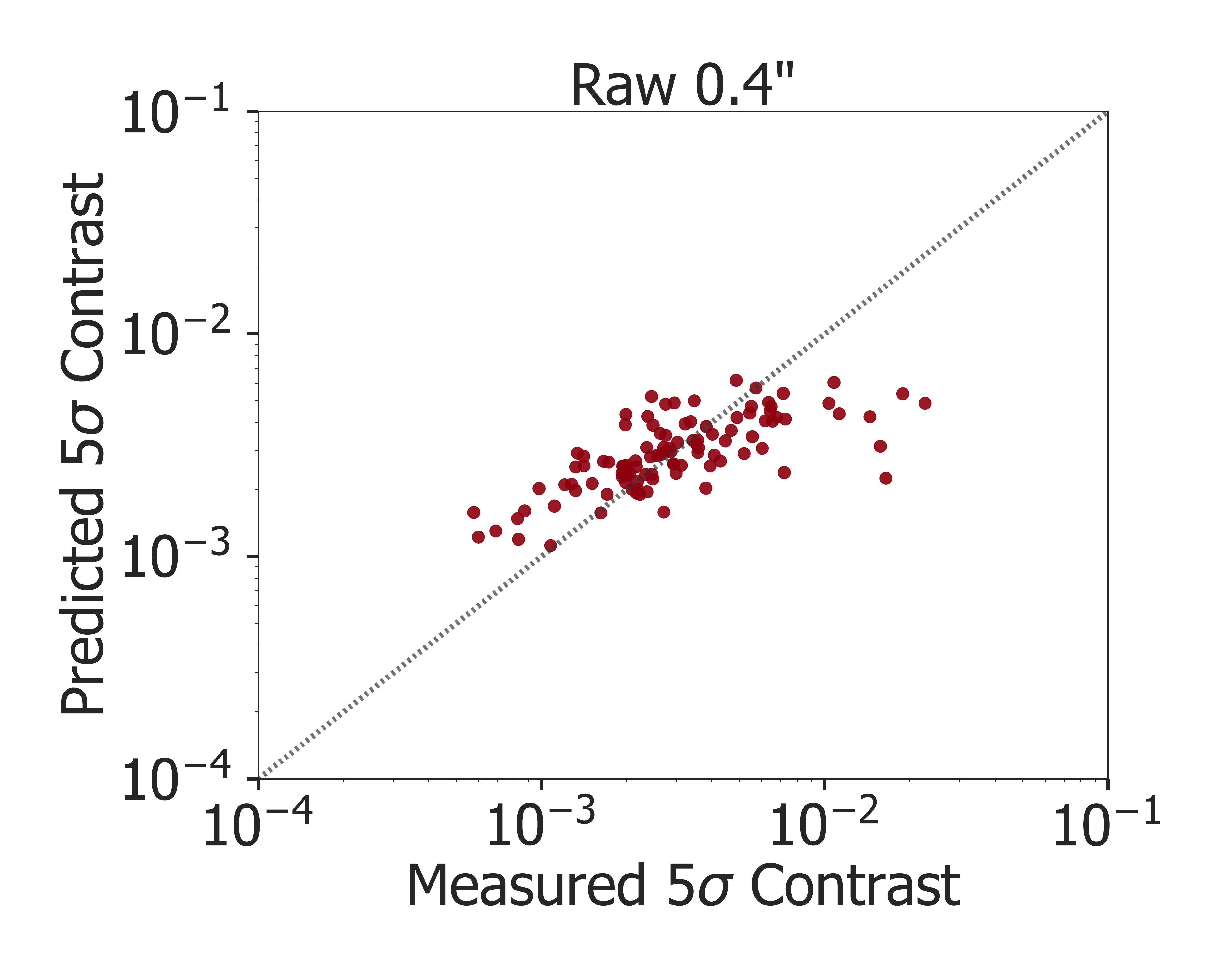}
    \caption{Predicted 5$\sigma$ contrasts for an independent 35\% of the data, using random forest models built with the other 65\% of data. 1:1 ratio lines are overplotted.
    \label{fig:rf-predict-full}}
\end{figure*}

For an initial assessment of the prediction accuracy, we first construct models with a random subset of 65\% of the observations and use them to predict the remaining 35\%. The predicted contrast values are plotted against the measured contrast values in Fig.~\ref{fig:rf-predict-full}. The predicted values align closely with the measured values for the most part, with the exception of RDI optimal contrast at 0.2\arcsec, which is evidently harder to predict than the other contrasts.

\begin{deluxetable*}{llllllll}
\tablecolumns{8}
\tabletypesize{\scriptsize}
\tablewidth{\textwidth}
\tablecaption{Relative variable importance and performance metrics of random forest models \label{tab:var-imp-rf}}
\tablehead{\colhead{Variable} & \colhead{ADI 0.2\arcsec} & \colhead{ADI 0.4\arcsec} & \colhead{ADI 1.0\arcsec} & \colhead{RDI 0.2\arcsec} & \colhead{RDI 0.4\arcsec} & \colhead{Raw 0.2\arcsec} & \colhead{Raw 0.4\arcsec}}
\startdata
\tableline
\sidehead{\textbf{Explanatory variable}}
PA Rotation & \textbf{100.0} & \textbf{100.0} & \textbf{100.0} & \textbf{71.1} & \textbf{36.7} & 44.8 & \textbf{83.1}\\
Total Integration Time & \textbf{51.1} & \textbf{30.5} & \textbf{38.9} & \textbf{81.9} & \textbf{33.4} & 37.3 & \textbf{43.9} \\
W1 magnitude & \textbf{43.2} & \textbf{76.0} & \textbf{84.1} & \textbf{100.0} & \textbf{100.0} & \textbf{46.7} & \textbf{100.0} \\
Airmass & \textbf{28.6} & 14.1 & 18.1 & 21.6 & 9.2 & \textbf{100.0} & 39.8 \\
$R$ magnitude & \textbf{26.2} & \textbf{22.0} & \textbf{36.4} & 54.5 & 28.6 & \textbf{59.0} & 32.7 \\
PSF x FWHM & 22.8 & 20.3 & 15.9 & \textbf{77.5} &
\textbf{34.7} & \textbf{88.3} & \textbf{49.4}\\
PSF y FWHM & 20.8 & \textbf{23.2} & \textbf{28.7} & \textbf{80.0} & \textbf{35.5} & \textbf{85.6} & \textbf{68.6} \\
Seeing & 15.2 & 8.5 & 0.0 & 45.6 & 20.4 & 42.3 & 40.2 \\
Optical Bench Temperature & 12.7 & 2.6 & 11.0 & 36.7 & 8.2 & 19.6 & 20.5\\
$\tau_{0}$ / WFS Integration Time & 12.6 & 11.2 & 20.7 & 59.3 & 22.3 & 12.5 & 29.2\\
$\lvert \textrm{ACAM - Dome} \rvert$ Temperature & 10.7 & 2.4 & 12.4 & 27.3 & 13.6 & 2.2 & 11.8\\
ACAM Temperature & 9.0 & 3.5 & 12.9 & 44.3 & 14.1 & 17.4 & 29.6 \\
Dome Pressure & 8.9 & 4.2 & 9.0 & 16.5 & 1.8 & 41.5 & 4.0\\
Ground Wind Speed & 6.2 & 0.0 & 8.7 & 0.0 & 0.0 & 0.0 & 0.0 \\
Primary Mirror Temperature & 5.0 & 7.7 & 13.6 & 44.4 & 10.1 & 35.7 & 23.1\\
Dome Humidity & 1.8 & 1.2 & 6.0 & 34.1 & 0.5 & 17.8 & 12.4 \\
$\lvert \textrm{Primary Mirror - Dome} \rvert$ Temperature & 0.0 & 2.8 & 15.5 & 4.6 & 5.5 & 4.6 & 7.5\\
RDI Reference Library Size & n/a & n/a & n/a & 57.4 & 21.6 & n/a & n/a \\
\sidehead{\textbf{Performance metric}}
$R^{2}$ & 0.733 & 0.776 & 0.707 & 0.329 & 0.506 & 0.463 & 0.487 \\
RMSE (dex) & 0.255 & 0.252 & 0.368 & 0.274 & 0.271 & 0.176 & 0.222 \\
\enddata
\tablecomments{Normalized relative variable importance, $R^{2}$, and RMSE values for random forest models predicting PCA-based optimal contrast with the given observing strategy and separation (columns 2 through 6) and predicting raw contrast at the given separation (columns 7 and 8). Variables are arranged in descending importance for ADI contrast at 0.2\arcsec. For each model, the top five most significant variables appear in bold. We do not construct models for optimal RDI contrast at 1.0\arcsec because RDI contrast is poorly sampled at this separation in our database. Each column is normalized independently. However, within each column, values indicate relative significance of variables. For instance, PA rotation is roughly twice as important as total integration time for predicting ADI contrast at 0.2\arcsec, but the two variables are similar in importance for predicting RDI contrast at the same separation. Note that the smaller RMSE values for raw contrast models compared to that of the ADI models, despite their lower $R^{2}$ than the ADI models, result from the fact that raw contrasts span a smaller range of contrast values.}
\end{deluxetable*}

Next, we construct models with 100\% of data and record the $R^{2}$ and RMSE of each model. These are reported in the bottom two rows of Table~\ref{tab:var-imp-rf} for a set of selected separations. We find that the ADI models (0.2\arcsec to 1.0\arcsec) reach $R^{2}$ values between 69.9\%-82.3\%, with RMSE values between 0.25-0.37dex. The RDI models (0.2\arcsec to 0.4\arcsec) have $R^{2}$ values between 31.5\% and 50.1\%, significantly less than their ADI counterparts, and RMSE values between 0.27-0.28dex. On average, both the ADI and RDI models are able to predict contrast within a factor of two. The raw contrast models have slightly higher $R^{2}$ values than the RDI models at small separations, but similar $R^{2}$ values compared to the RDI models at larger separations. Note that the similarity in prediction errors between the ADI and RDI models, notwithstanding the big difference in their $R^{2}$ values, is due to the fact that RDI contrasts span a smaller range than ADI contrasts. Therefore, $R^{2}$ is the suitable metric to use in comparing the different models, while RMSE is a valuable measure only within a given model.

We also find that the RMSE is smaller at larger separations for the RDI models, indicating a growing difficulty of predicting RDI contrast at smaller separations. The ADI models show a similar increase in $R^{2}$ from 0.2\arcsec up to 0.75\arcsec, after which predictions become more difficult. We attribute the increasing difficulty of predicting ADI contrasts at larger separations to the background limit. At 1.0\arcsec, 93\% of ADI contrasts are background-limited, following our definition in Section~\ref{sec:noise-reg}. We expect that background-limited contrasts hinge on the dynamic extended structures in the thermal background, a feature not measured by our explanatory variables.

We expect RDI performance to be highly dependent on factors such as the correlation between the reference PSF and the target PSF, which is determined by the homogeneity of the stellar properties, the similarity in integration time and the observing conditions for the reference star and target star, as well as the centering accuracy of the targets onto the vortex core, and these factors are not yet quantified in our workflow. Future studies are encouraged to take into account these additional variables to systematically understand RDI performance and increase $R^{2}$ for the RDI models.

\subsection{Variable importance}
For each explanatory variable, we quantify its relative importance by building a model where this variable is randomly permuted so that it has no influence on the predictions (but is kept so that the number of variables remains the same). We compare the RMSE from predicting on this model with the RMSE from predicting on the model including all explanatory variables. A larger decrease of the error means that the variable at hand is relatively more important. Table~\ref{tab:var-imp-rf} reports relative variable importance on a linear scale from 0 to 100, where the least important variable is assigned 0 and the most important variable is assigned 100. In the R \texttt{caret} package, these calculations are part of the ``varImp'' function.

We find that for ADI optimal contrast, the top three most important variables are PA rotation (most important across all three separations), W1 magnitude, and total integration time. For RDI optimal contrast, the x and y sizes of the PSF FWHM become more important with respect to other variables, especially for 0.2\arcsec, and W1 magnitude becomes the most important variable. We find that airmass is the most significant factor for raw contrast at 0.2\arcsec, while it is much less important for RDI or ADI optimal contrasts at the same separation. For the raw contrasts, seeing also becomes an important variable, about 40\% as important as the most important variable at the same separation. In comparison, seeing is completely inconsequential for ADI optimal contrast at 1.0\arcsec. Most of these relationships indicate that ADI is generally more robust to poor conditions and poor AO performance.

\subsection{Implementation of contrast predictor for future observations}

\begin{figure*}[t!]
    \centering
    \includegraphics[width=0.4\linewidth]{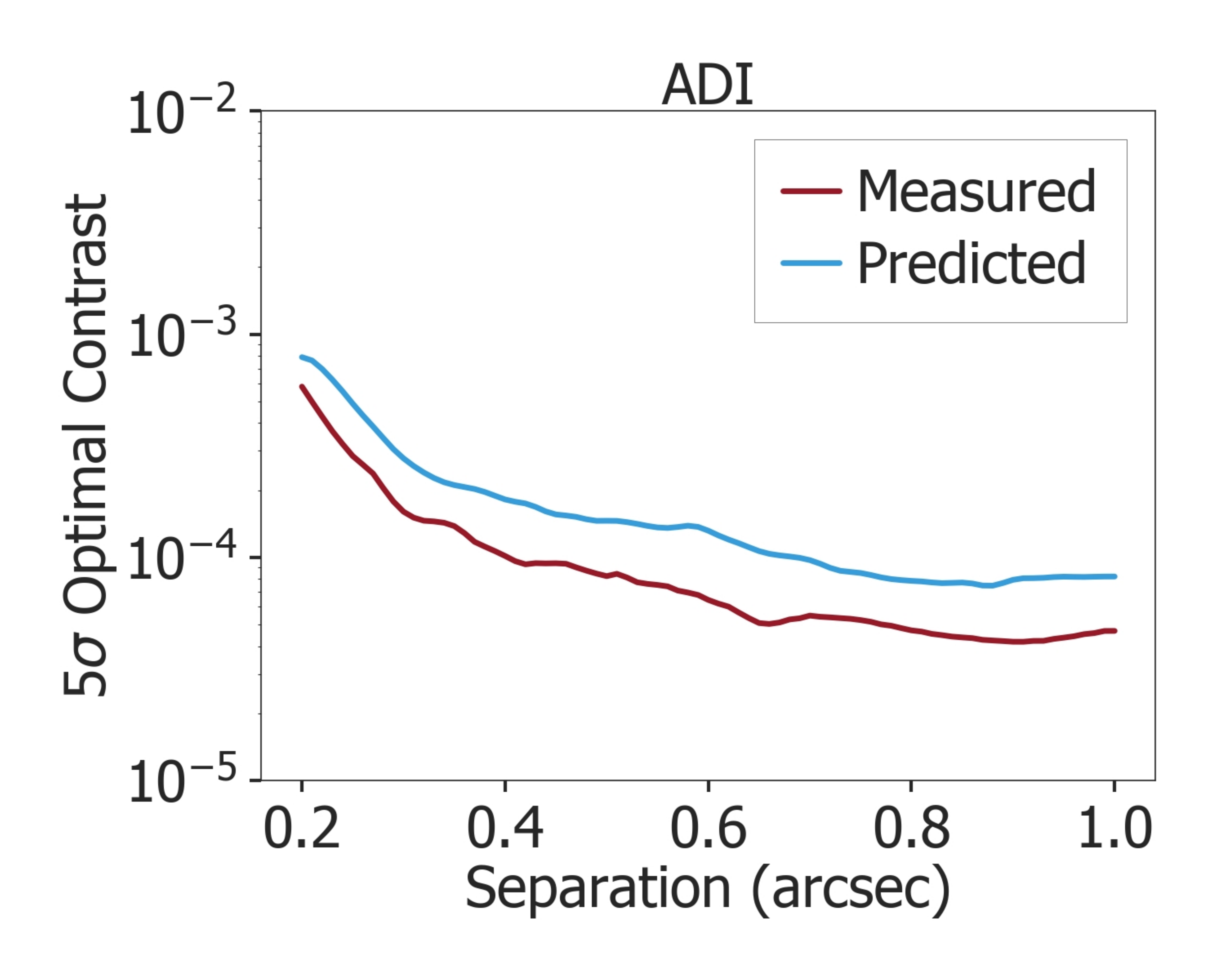}
    \includegraphics[width=0.4\linewidth]{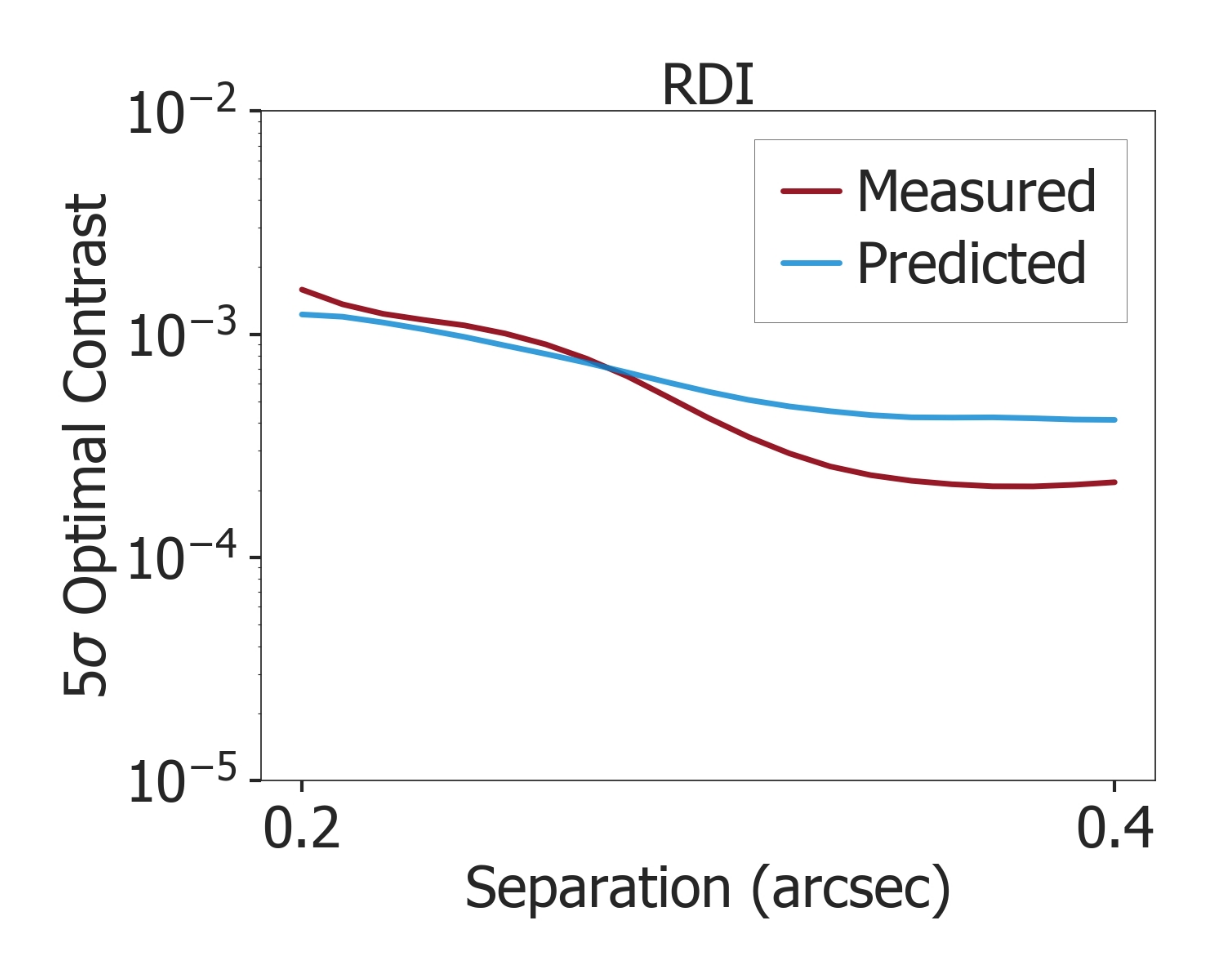}\\
    \includegraphics[width=0.4\linewidth]{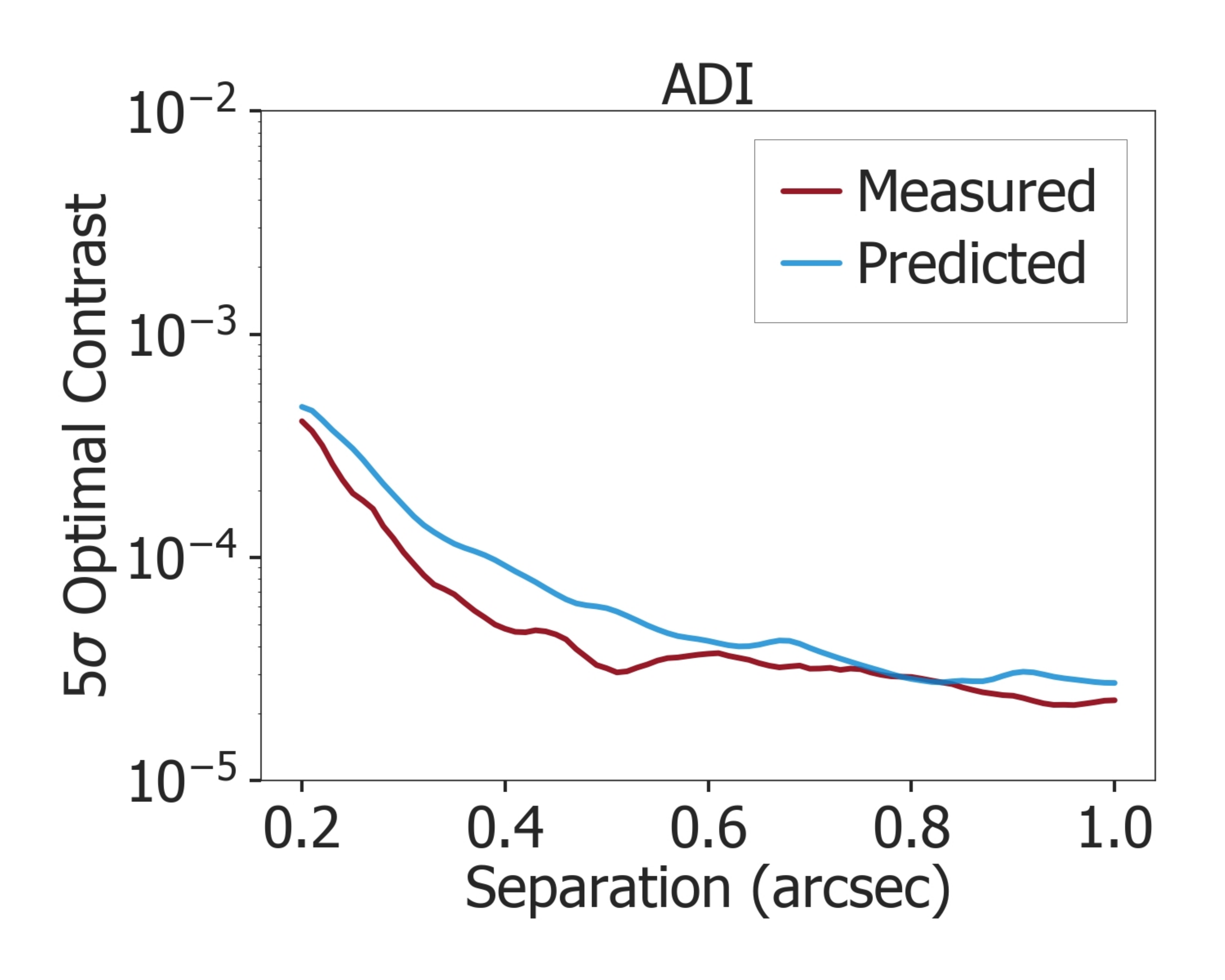}
    \includegraphics[width=0.4\linewidth]{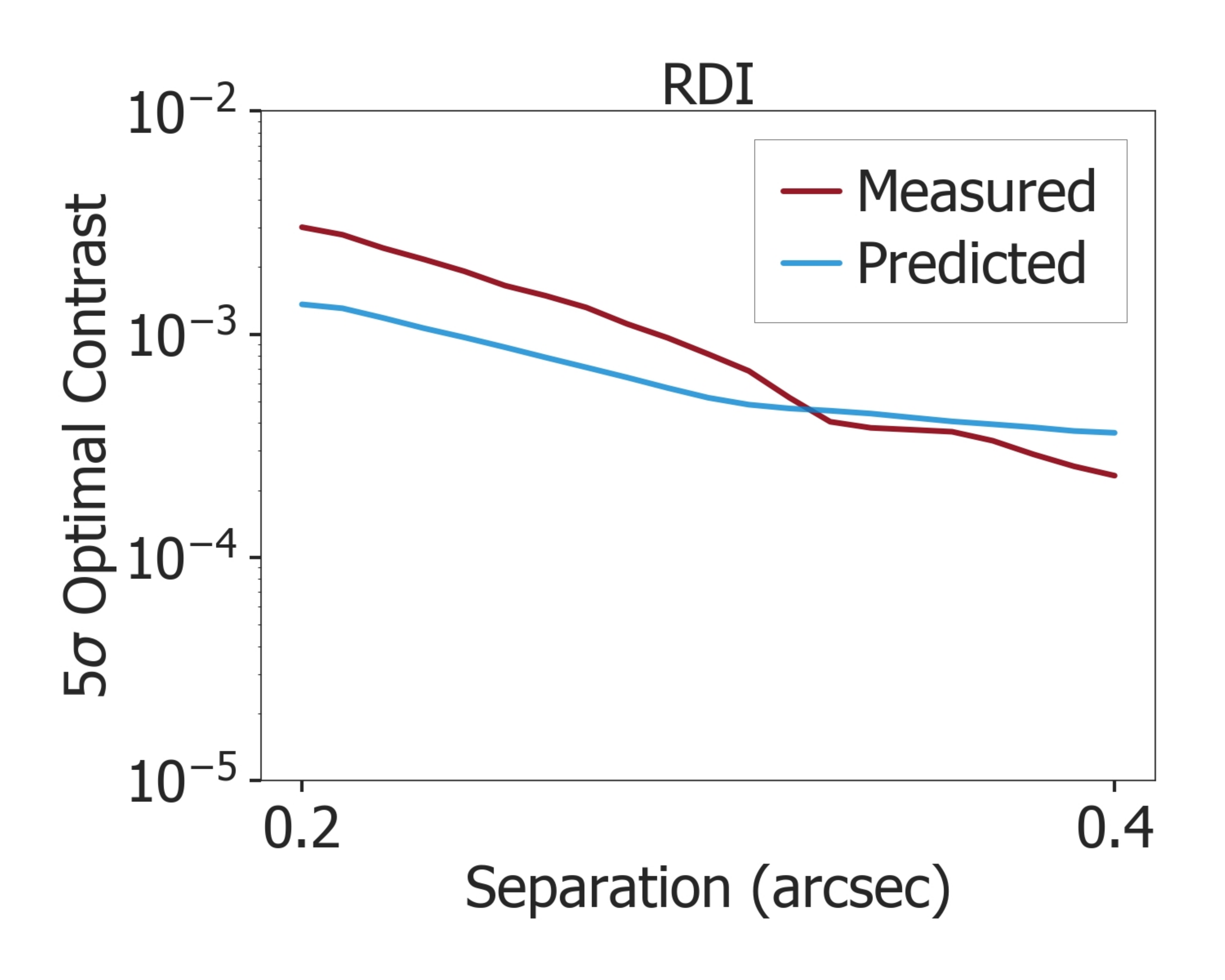}\\
    \includegraphics[width=0.4\linewidth]{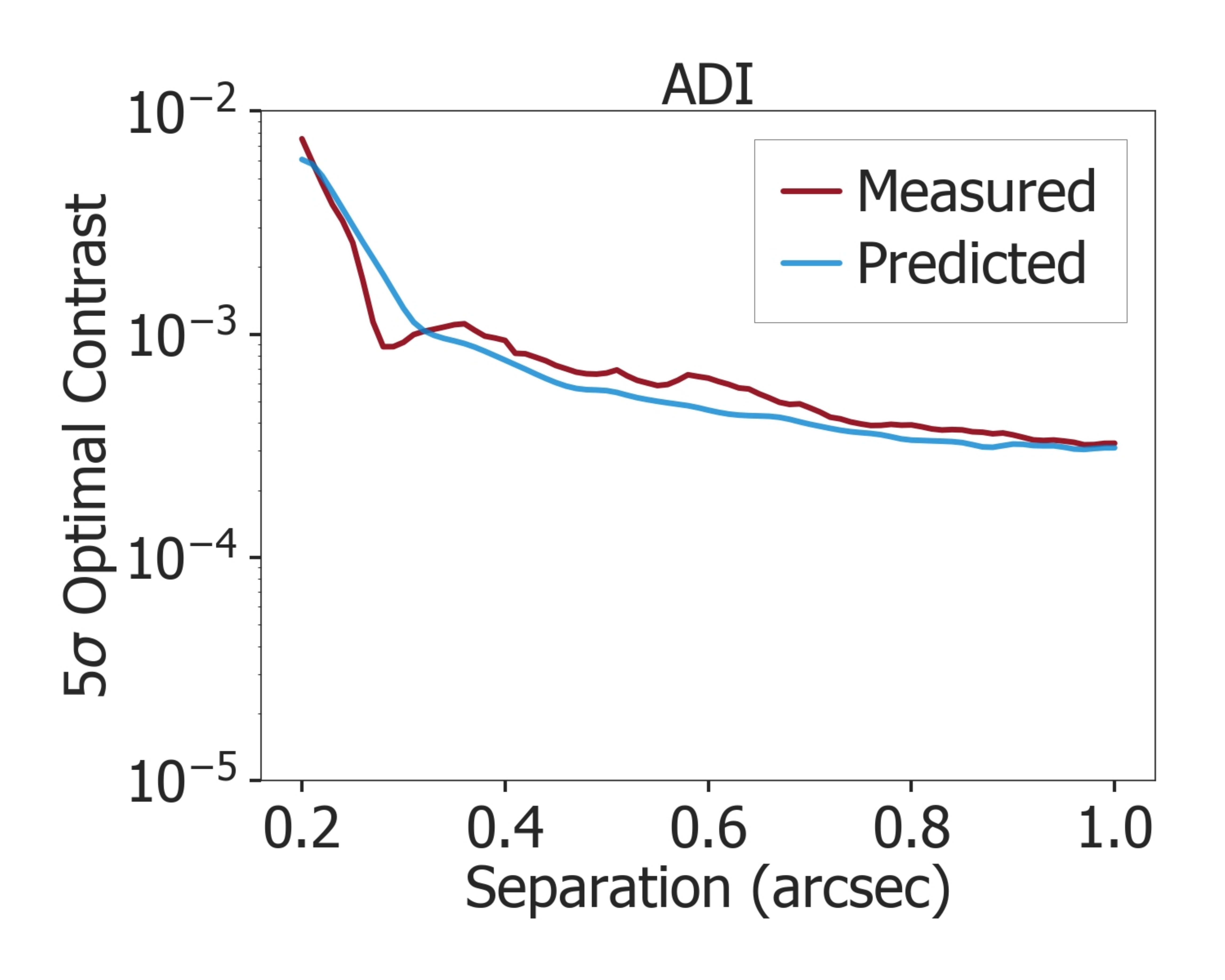}
    \includegraphics[width=0.4\linewidth]{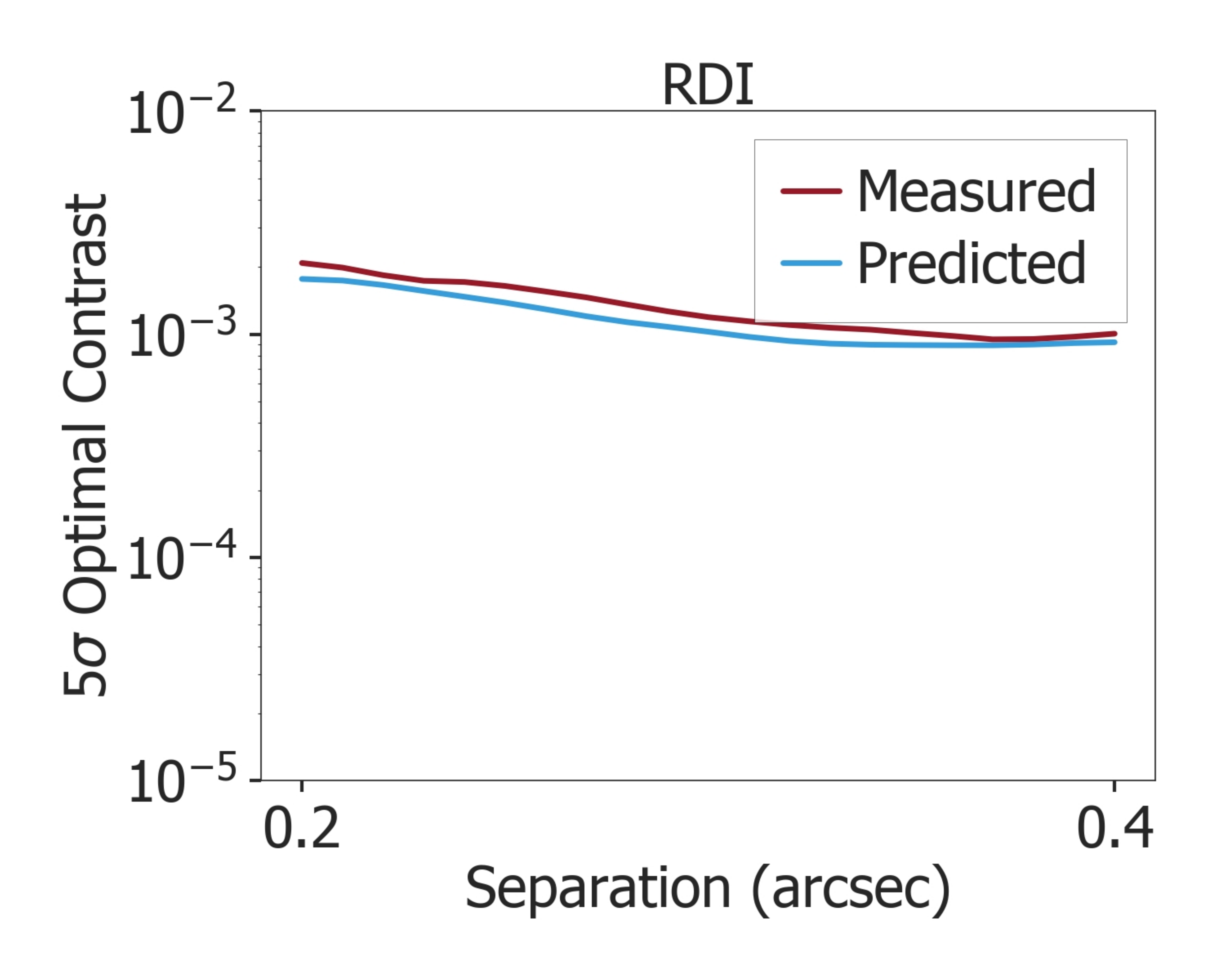}
    \caption{Predicted contrast curves and measured contrast curves for three targets in our sample set. For reference, the target in the top row has W1 magnitude: 6.8, PA rotation: 42.1$^{\circ}$, total integration time: 62.5 mins, the middle target has W1 magnitude: 5.5, PA rotation: 115.7$^{\circ}$, total integration time: 32.5 mins, and the bottom target has W1 magnitude: 7.8, PA rotation: 8.9$^{\circ}$, total integration time: 12.5 mins. For each target, we create a set of models (in intervals of 0.01\arcsec) for ADI and RDI, and feed the target's explanatory variables into the models in order to predict its contrast at different separations. We have applied a smoothing algorithm with a window size equal to the average FWHM of the target to the predicted contrast curves: in raw form, they show minor sawtooth-like structures that arise because there is different random forest model for every 0.01\arcsec, creating artificial noise from model-to-model variations.}
    \label{fig:sample-predicted-cc}
\end{figure*}

Using our random forest models, we implement a contrast prediction tool, the Vortex Imaging Contrast Oracle (VICO)\footnote{\href{ http://vortex.astro.caltech.edu/predict}{http://vortex.astro.caltech.edu/predict}}, for the Keck/NIRC2 vortex coronagraph. VICO predicts contrast by running pre-made random forest models that are built on data reduced by our standardized pipeline. Rather than limiting the models to use the same dataset as presented in this paper, the models in VICO are updated periodically to incorporate data from new NIRC2 vortex observations. Although the contrast predictor could take all the variables listed in Table~\ref{tab:var-imp-rf} as input, in practice, many variables depend on the conditions of the observing time period, and cannot be known precisely before the observation. Therefore, to make VICO more user-friendly, we remove a few environmental variables such as the instrument temperatures, which are both difficult to estimate and relatively insignificant in terms of predicting contrast.

To demonstrate contrast prediction, we show ADI and RDI predicted contrast curves for three randomly selected targets in Fig.~\ref{fig:sample-predicted-cc}. The curves are made from aggregates of 81 ADI models and 21 RDI models (one for every 0.01\arcsec). We remove the selected target from the sample set and use the remaining dataset to make these models, so these predicted contrast curves represent an unbiased and realistic example of the prediction powers of our random forest models.

\subsection{Linear regression model}
We also considered linear regression models for predicting contrast. Since we expect power-law relations with contrast for many of our variables except for stellar magnitude, we take the log of all explanatory variables except the magnitudes for the linear regression models (this is unnecessary for random forests, which are invariant to monotonic transformations of explanatory variables). To compute a measure of error for linear regression models that is equivalent to the random forest RMSE (see Section~\ref{sec:metrics}), we bootstrap the data 500 times (equal to $B$ for the random forests) to create 500 sub-models, record the OOB observations for each sub-model, predict on them, and average the errors from all sub-models to get the RMSE of the final linear regression model. We find that the random forest models yield lower errors than their linear regression counterparts in every scenario. On average, the linear regression RMSE values are larger than the random forest RMSE values by more than 25\%. This indicates that the random forest approach better suits our needs.

\section{Conclusions}\label{sec:conclusion}
We characterize the performance of the Keck/NIRC2 vector vortex coronagraph with a sample set of 359 targets observed from December 2015 to January 2018. Using a streamlined workflow, we uniformly re-process the data in our sample set. Using a full-frame PCA-based post-processing technique, we compare the performance from the two observing strategies ADI and RDI, and find an empirical power-law index of -1.18 between angular separation and the minimum amount of PA rotation required for ADI to yield deeper contrasts than RDI. In addition, we find strong negative correlations between contrast and WISE W1 and $R$ stellar magnitudes. The ratio of atmospheric coherence time to WFS integration time ($\tau_{0}/t$) also shows a strong negative power-law relation with contrast, with slopes of -0.53$\pm$0.08 and -0.74$\pm$0.09 for ADI contrast at 0.2\arcsec and 0.4\arcsec respectively. For both stellar magnitudes and $\tau_{0}/t$, the slopes are, on average, larger by factors of two for ADI contrast than for RDI contrast. On the other hand, we find that seeing and temperature differentials show no significant direct correlations with contrast.

Furthermore, we create random forest models in order to predict contrast as a function of separation using a range of explanatory variables that describe the observing conditions, stellar magnitudes, and observation parameters. Using these models, we implement a website (\url{http://vortex.astro.caltech.edu/predict}) where observers can predict ADI and RDI contrast curves for future observations with the NIRC2 vortex. Currently, our random forest models can predict both ADI and RDI contrast to within a factor of two, and they will continue to improve as new observations are added to the sample set. In general, ADI contrast is better described by the random forests, with the ADI models showing $R^2$ values roughly twice as large as their RDI counterparts. The correlation between the reference PSF library and the target PSF, a factor not included in this study, is expected to strongly correlate with RDI performance. We also determine variable importance from the random forests, and find that ADI contrast is dominated by PA rotation, total integration time, and stellar magnitude (in the $L^\prime$ bandpass of the instrument), while RDI contrast is also strongly limited by the FWHM of the target PSF. The seeing and airmass play important roles in determining raw contrast, but weakly influence the ADI and RDI post-processed contrast.

In the long term, our study informs what updates to the current instrument we could make to enhance it the most. The strong dependence on $\tau_{0}/t$ suggests that the NIRC2 vortex would benefit greatly from future improvements in AO loop speed and the implementation of predictive wavefront control. More importantly, our study systematically compares the ADI and RDI techniques and provides a suite of accurate predictive models, thereby enhancing observing strategies for future high-contrast imaging campaigns.

\acknowledgments
The data presented herein were obtained at the W. M. Keck Observatory, which is operated as a scientific partnership among the California Institute of Technology, the University of California, and the National Aeronautics and Space Administration (NASA). The Observatory was made possible by the generous financial support of the W. M. Keck Foundation. The authors wish to recognize and acknowledge the very significant cultural role and reverence that the summit of Maunakea has always had within the indigenous Hawaiian community. We are most fortunate to have the opportunity to conduct observations from this mountain. This work was funded, in part, by a Summer Undergraduate Research Fellowship (SURF) from California Institute of Technology. W.J.X. would like to thank Johanna Hardin at Pomona College for her advice on implementing the statistical models. G.R. is supported by an NSF Astronomy and Astrophysics Postdoctoral Fellowship under award AST-1602444. E.C. acknowledges support from NASA through Hubble Fellowship grant HF2-51355 awarded by STScI, which is operated by AURA, Inc. for NASA under contract NAS5-26555. O.A. is an F.R.S-FNRS research associate. V.B. acknowledges government sponsorship; this research was carried out in part at the Jet Propulsion Laboratory, California Institute of Technology, under a contract with the National Aeronautics and Space Administration. 

\facilities{Keck II (NIRC2)}
\software{\texttt{VIP}~\citep{GomezGonzalez2017}, \texttt{QACITS}~\citep{Huby2015,Huby2017}, \texttt{Astropy}~\citep{Astropy2018}, \texttt{Matplotlib}~\citep{Matplotlib2007}, \texttt{caret}~\citep{Kuhn2008}, \texttt{scikit-image}~\citep{scikitimage}, \texttt{Mongo}~(\url{https://docs.mongodb.com/})}

\bibliography{XuanLibrary}

\begin{thebibliography}{}
\expandafter\ifx\csname natexlab\endcsname\relax\def\natexlab#1{#1}\fi
\providecommand{\url}[1]{\href{#1}{#1}}
\providecommand{\dodoi}[1]{doi:~\href{http://doi.org/#1}{\nolinkurl{#1}}}
\providecommand{\doeprint}[1]{\href{http://ascl.net/#1}{\nolinkurl{http://ascl.net/#1}}}
\providecommand{\doarXiv}[1]{\href{https://arxiv.org/abs/#1}{\nolinkurl{https://arxiv.org/abs/#1}}}

\bibitem[{{Bailey} {et~al.}(2016){Bailey}, {Poyneer}, {Macintosh}, {Savransky},
  {Wang}, {De Rosa}, {Follette}, {Ammons}, {Hayward}, {Ingraham}, {Maire},
  {Palmer}, {Perrin}, {Rajan}, {Rantakyr{\"o}}, {Thomas}, \&
  {V{\'e}ran}}]{Bailey2016}
{Bailey}, V.~P., {Poyneer}, L.~A., {Macintosh}, B.~A., {et~al.} 2016,
  \procspie, 9909, 99090V, \dodoi{10.1117/12.2233172}

\bibitem[{{Bowler}(2016)}]{Bowler2016}
{Bowler}, B.~P. 2016, \pasp, 128, 102001,
  \dodoi{10.1088/1538-3873/128/968/102001}

\bibitem[{{Bowler} {et~al.}(2012){Bowler}, {Liu}, {Shkolnik}, {Dupuy}, {Cieza},
  {Kraus}, \& {Tamura}}]{Bowler2012}
{Bowler}, B.~P., {Liu}, M.~C., {Shkolnik}, E.~L., {et~al.} 2012, \apj, 753,
  142, \dodoi{10.1088/0004-637X/753/2/142}

\bibitem[{Breiman(2001)}]{Breiman2001}
Breiman, L. 2001, Machine Learning, 45, 5, \dodoi{10.1023/A:1010933404324}

\bibitem[{{Chauvin} {et~al.}(2017){Chauvin}, {Desidera}, {Lagrange}, {Vigan},
  {Gratton}, {Langlois}, {Bonnefoy}, {Beuzit}, {Feldt}, {Mouillet}, {Meyer},
  {Cheetham}, {Biller}, {Boccaletti}, {D'Orazi}, {Galicher}, {Hagelberg},
  {Maire}, {Mesa}, {Olofsson}, {Samland}, {Schmidt}, {Sissa}, {Bonavita},
  {Charnay}, {Cudel}, {Daemgen}, {Delorme}, {Janin-Potiron}, {Janson},
  {Keppler}, {Le Coroller}, {Ligi}, {Marleau}, {Messina}, {Molli{\`e}re},
  {Mordasini}, {M{\"u}ller}, {Peretti}, {Perrot}, {Rodet}, {Rouan}, {Zurlo},
  {Dominik}, {Henning}, {Menard}, {Schmid}, {Turatto}, {Udry}, {Vakili}, {Abe},
  {Antichi}, {Baruffolo}, {Baudoz}, {Baudrand}, {Blanchard}, {Bazzon}, {Buey},
  {Carbillet}, {Carle}, {Charton}, {Cascone}, {Claudi}, {Costille}, {Deboulbe},
  {De Caprio}, {Dohlen}, {Fantinel}, {Feautrier}, {Fusco}, {Gigan}, {Giro},
  {Gisler}, {Gluck}, {Hubin}, {Hugot}, {Jaquet}, {Kasper}, {Madec}, {Magnard},
  {Martinez}, {Maurel}, {Le Mignant}, {M{\"o}ller-Nilsson}, {Llored}, {Moulin},
  {Orign{\'e}}, {Pavlov}, {Perret}, {Petit}, {Pragt}, {Puget}, {Rabou},
  {Ramos}, {Rigal}, {Rochat}, {Roelfsema}, {Rousset}, {Roux}, {Salasnich},
  {Sauvage}, {Sevin}, {Soenke}, {Stadler}, {Suarez}, {Weber}, {Wildi},
  {Antoniucci}, {Augereau}, {Baudino}, {Brandner}, {Engler}, {Girard}, {Gry},
  {Kral}, {Kopytova}, {Lagadec}, {Milli}, {Moutou}, {Schlieder},
  {Szul{\'a}gyi}, {Thalmann}, \& {Wahhaj}}]{Chauvin2017}
{Chauvin}, G., {Desidera}, S., {Lagrange}, A.-M., {et~al.} 2017, \aap, 605, L9,
  \dodoi{10.1051/0004-6361/201731152}

\bibitem[{{Cutri} \& {et al.}(2012)}]{Cutri2012}
{Cutri}, R.~M., \& {et al.} 2012, VizieR Online Data Catalog, 2311

\bibitem[{{Cutri} \& {et al.}(2014)}]{Cutri2014}
---. 2014, VizieR Online Data Catalog, 2328

\bibitem[{{Davis} \& {Tango}(1996)}]{Davis&Tango1996}
{Davis}, J., \& {Tango}, W. 1996, \pasp, 108, 456, \dodoi{10.1086/133747}

\bibitem[{{Fried}(1990)}]{Fried1990}
{Fried}, D.~L. 1990, J. Opt. Soc. Am. A, 7, 1224,
  \dodoi{10.1364/JOSAA.7.001224}

\bibitem[{{Gerard} \& {Marois}(2016)}]{Gerard2016}
{Gerard}, B.~L., \& {Marois}, C. 2016, \procspie, 9909, 990958,
  \dodoi{10.1117/12.2231905}

\bibitem[{{Gomez Gonzalez} {et~al.}(2017){Gomez Gonzalez}, {Wertz}, {Absil},
  {Christiaens}, {Defr{\`e}re}, {Mawet}, {Milli}, {Absil}, {Van Droogenbroeck},
  {Cantalloube}, {Hinz}, {Skemer}, {Karlsson}, \& {Surdej}}]{GomezGonzalez2017}
{Gomez Gonzalez}, C.~A., {Wertz}, O., {Absil}, O., {et~al.} 2017, \aj, 154, 7,
  \dodoi{10.3847/1538-3881/aa73d7}

\bibitem[{Guizar-Sicairos {et~al.}(2008)Guizar-Sicairos, Thurman, \&
  Fienup}]{Guizar-Sicairos2008}
Guizar-Sicairos, M., Thurman, S.~T., \& Fienup, J.~R. 2008, Opt. Lett., 33,
  156, \dodoi{10.1364/OL.33.000156}

\bibitem[{{Hinkley} {et~al.}(2011){Hinkley}, {Oppenheimer}, {Zimmerman},
  {Brenner}, {Parry}, {Crepp}, {Vasisht}, {Ligon}, {King}, {Soummer},
  {Sivaramakrishnan}, {Beichman}, {Shao}, {Roberts}, {Bouchez}, {Dekany},
  {Pueyo}, {Roberts}, {Lockhart}, {Zhai}, {Shelton}, \&
  {Burruss}}]{Hinkley2011}
{Hinkley}, S., {Oppenheimer}, B.~R., {Zimmerman}, N., {et~al.} 2011, \pasp,
  123, 74, \dodoi{10.1086/658163}

\bibitem[{{Huby} {et~al.}(2015){Huby}, {Baudoz}, {Mawet}, \&
  {Absil}}]{Huby2015}
{Huby}, E., {Baudoz}, P., {Mawet}, D., \& {Absil}, O. 2015, \aap, 584,
  \dodoi{10.1051/0004-6361/201527102}

\bibitem[{{Huby} {et~al.}(2017){Huby}, {Bottom}, {Femenia}, {Ngo}, {Mawet},
  {Serabyn}, \& {Absil}}]{Huby2017}
{Huby}, E., {Bottom}, M., {Femenia}, B., {et~al.} 2017, \aap, 600,
  \dodoi{10.1051/0004-6361/201630232}

\bibitem[{Hunter(2007)}]{Matplotlib2007}
Hunter, J.~D. 2007, Computing In Science \& Engineering, 9, 90,
  \dodoi{10.1109/MCSE.2007.55}

\bibitem[{{Jovanovic} {et~al.}(2015){Jovanovic}, {Martinache}, {Guyon},
  {Clergeon}, {Singh}, {Kudo}, {Garrel}, {Newman}, {Doughty}, {Lozi}, {Males},
  {Minowa}, {Hayano}, {Takato}, {Morino}, {Kuhn}, {Serabyn}, {Norris},
  {Tuthill}, {Schworer}, {Stewart}, {Close}, {Huby}, {Perrin}, {Lacour},
  {Gauchet}, {Vievard}, {Murakami}, {Oshiyama}, {Baba}, {Matsuo}, {Nishikawa},
  {Tamura}, {Lai}, {Marchis}, {Duchene}, {Kotani}, \&
  {Woillez}}]{Jovanovic2015}
{Jovanovic}, N., {Martinache}, F., {Guyon}, O., {et~al.} 2015, \pasp, 127, 890,
  \dodoi{10.1086/682989}

\bibitem[{Kuhn(2008)}]{Kuhn2008}
Kuhn, M. 2008, Journal of Statistical Software, Articles, 28, 1,
  \dodoi{10.18637/jss.v028.i05}

\bibitem[{{Lafreni{\`e}re} {et~al.}(2009){Lafreni{\`e}re}, {Marois}, {Doyon},
  \& {Barman}}]{Lafrenière2009}
{Lafreni{\`e}re}, D., {Marois}, C., {Doyon}, R., \& {Barman}, T. 2009, \apjl,
  694, L148, \dodoi{10.1088/0004-637X/694/2/L148}

\bibitem[{Louppe(2014)}]{Louppe2014understanding}
Louppe, G. 2014, PhD thesis, University of Liege, Belgium.
\newblock \doarXiv{1407.7502}

\bibitem[{{Macintosh} {et~al.}(2014){Macintosh}, {Graham}, {Ingraham},
  {Konopacky}, {Marois}, {Perrin}, {Poyneer}, {Bauman}, {Barman}, {Burrows},
  {Cardwell}, {Chilcote}, {De Rosa}, {Dillon}, {Doyon}, {Dunn}, {Erikson},
  {Fitzgerald}, {Gavel}, {Goodsell}, {Hartung}, {Hibon}, {Kalas}, {Larkin},
  {Maire}, {Marchis}, {Marley}, {McBride}, {Millar-Blanchaer}, {Morzinski},
  {Norton}, {Oppenheimer}, {Palmer}, {Patience}, {Pueyo}, {Rantakyro},
  {Sadakuni}, {Saddlemyer}, {Savransky}, {Serio}, {Soummer},
  {Sivaramakrishnan}, {Song}, {Thomas}, {Wallace}, {Wiktorowicz}, \&
  {Wolff}}]{Macintosh2014}
{Macintosh}, B., {Graham}, J.~R., {Ingraham}, P., {et~al.} 2014, Proceedings of
  the National Academy of Science, 111, 12661, \dodoi{10.1073/pnas.1304215111}

\bibitem[{{Macintosh} {et~al.}(2015){Macintosh}, {Graham}, {Barman}, {De Rosa},
  {Konopacky}, {Marley}, {Marois}, {Nielsen}, {Pueyo}, {Rajan}, {Rameau},
  {Saumon}, {Wang}, {Patience}, {Ammons}, {Arriaga}, {Artigau}, {Beckwith},
  {Brewster}, {Bruzzone}, {Bulger}, {Burningham}, {Burrows}, {Chen}, {Chiang},
  {Chilcote}, {Dawson}, {Dong}, {Doyon}, {Draper}, {Duch{\^e}ne}, {Esposito},
  {Fabrycky}, {Fitzgerald}, {Follette}, {Fortney}, {Gerard}, {Goodsell},
  {Greenbaum}, {Hibon}, {Hinkley}, {Cotten}, {Hung}, {Ingraham},
  {Johnson-Groh}, {Kalas}, {Lafreniere}, {Larkin}, {Lee}, {Line}, {Long},
  {Maire}, {Marchis}, {Matthews}, {Max}, {Metchev}, {Millar-Blanchaer},
  {Mittal}, {Morley}, {Morzinski}, {Murray-Clay}, {Oppenheimer}, {Palmer},
  {Patel}, {Perrin}, {Poyneer}, {Rafikov}, {Rantakyr{\"o}}, {Rice}, {Rojo},
  {Rudy}, {Ruffio}, {Ruiz}, {Sadakuni}, {Saddlemyer}, {Salama}, {Savransky},
  {Schneider}, {Sivaramakrishnan}, {Song}, {Soummer}, {Thomas}, {Vasisht},
  {Wallace}, {Ward-Duong}, {Wiktorowicz}, {Wolff}, \&
  {Zuckerman}}]{Macintosh2015}
{Macintosh}, B., {Graham}, J.~R., {Barman}, T., {et~al.} 2015, Science, 350,
  64, \dodoi{10.1126/science.aac5891}

\bibitem[{{Marois} {et~al.}(2006){Marois}, {Lafreni{\`e}re}, {Doyon},
  {Macintosh}, \& {Nadeau}}]{Marois2005}
{Marois}, C., {Lafreni{\`e}re}, D., {Doyon}, R., {Macintosh}, B., \& {Nadeau},
  D. 2006, \apj, 641, 556, \dodoi{10.1086/500401}

\bibitem[{{Marois} {et~al.}(2008){Marois}, {Macintosh}, {Barman}, {Zuckerman},
  {Song}, {Patience}, {Lafreni{\`e}re}, \& {Doyon}}]{Marois2008}
{Marois}, C., {Macintosh}, B., {Barman}, T., {et~al.} 2008, Science, 322, 1348,
  \dodoi{10.1126/science.1166585}

\bibitem[{{Mawet} {et~al.}(2005){Mawet}, {Riaud}, {Absil}, \&
  {Surdej}}]{Mawet2005}
{Mawet}, D., {Riaud}, P., {Absil}, O., \& {Surdej}, J. 2005, \apj, 633, 1191,
  \dodoi{10.1086/462409}

\bibitem[{Mawet {et~al.}(2011)Mawet, Serabyn, Wallace, \& Pueyo}]{Mawet2011}
Mawet, D., Serabyn, E., Wallace, J.~K., \& Pueyo, L. 2011, Opt. Lett., 36,
  1506, \dodoi{10.1364/OL.36.001506}

\bibitem[{{Mawet} {et~al.}(2014){Mawet}, {Milli}, {Wahhaj}, {Pelat}, {Absil},
  {Delacroix}, {Boccaletti}, {Kasper}, {Kenworthy}, {Marois}, {Mennesson}, \&
  {Pueyo}}]{Mawet2014}
{Mawet}, D., {Milli}, J., {Wahhaj}, Z., {et~al.} 2014, \apj, 792, 97,
  \dodoi{10.1088/0004-637X/792/2/97}

\bibitem[{{Milli} {et~al.}(2017){Milli}, {Mouillet}, {Fusco}, {Girard},
  {Masciadri}, {Pena}, {Sauvage}, {Reyes}, {Dohlen}, {Beuzit}, {Kasper},
  {Sarazin}, \& {Cantalloube}}]{Milli2017}
{Milli}, J., {Mouillet}, D., {Fusco}, T., {et~al.} 2017, AO4ELT5 Proceedings,
  \dodoi{10.26698/AO4ELT5.0034}

\bibitem[{Poyneer {et~al.}(2016)Poyneer, Palmer, Macintosh, Savransky,
  Sadakuni, Thomas, V\'{e}ran, Follette, Greenbaum, Ammons, Bailey, Bauman,
  Cardwell, Dillon, Gavel, Hartung, Hibon, Perrin, Rantakyr\"{o},
  Sivaramakrishnan, \& Wang}]{Poyneer2016}
Poyneer, L.~A., Palmer, D.~W., Macintosh, B., {et~al.} 2016, Appl. Opt., 55,
  323, \dodoi{10.1364/AO.55.000323}

\bibitem[{{Rigaut} {et~al.}(1991){Rigaut}, {Rousset}, {Kern}, {Fontanella},
  {Gaffard}, {Merkle}, \& {L{\'e}na}}]{Rigaut1991}
{Rigaut}, F., {Rousset}, G., {Kern}, P., {et~al.} 1991, \aap, 250, 280

\bibitem[{{Ruane} {et~al.}(2017){Ruane}, {Mawet}, {Kastner}, {Meshkat},
  {Bottom}, {Femen{\'{\i}}a Castell{\'a}}, {Absil}, {Gomez Gonzalez}, {Huby},
  {Zhu}, {Jenson-Clem}, {Choquet}, \& {Serabyn}}]{Ruane2017}
{Ruane}, G., {Mawet}, D., {Kastner}, J., {et~al.} 2017, \aj, 154, 73,
  \dodoi{10.3847/1538-3881/aa7b81}

\bibitem[{{Sauvage} {et~al.}(2016){Sauvage}, {Fusco}, {Lamb}, {Girard},
  {Brinkmann}, {Guesalaga}, {Wizinowich}, {O'Neal}, {N'Diaye}, {Vigan},
  {Mouillet}, {Beuzit}, {Kasper}, {Le Louarn}, {Milli}, {Dohlen}, {Neichel},
  {Bourget}, {Haguenauer}, \& {Mawet}}]{Sauvage2016}
{Sauvage}, J.-F., {Fusco}, T., {Lamb}, M., {et~al.} 2016, \procspie, 9909,
  990916, \dodoi{10.1117/12.2232459}

\bibitem[{{Schroeder}(1987)}]{Schroeder1987}
{Schroeder}, D.~J. 1987, {Astronomical Optics} (San Diego, Academic Press)

\bibitem[{{Serabyn} {et~al.}(2017){Serabyn}, {Huby}, {Matthews}, {Mawet},
  {Absil}, {Femenia}, {Wizinowich}, {Karlsson}, {Bottom}, {Campbell},
  {Carlomagno}, {Defr{\`e}re}, {Delacroix}, {Forsberg}, {Gomez Gonzalez},
  {Habraken}, {Jolivet}, {Liewer}, {Lilley}, {Piron}, {Reggiani}, {Surdej},
  {Tran}, {Vargas Catal{\'a}n}, \& {Wertz}}]{Serabyn2017}
{Serabyn}, E., {Huby}, E., {Matthews}, K., {et~al.} 2017, \aj, 153, 43,
  \dodoi{10.3847/1538-3881/153/1/43}

\bibitem[{{Service} {et~al.}(2016){Service}, {Lu}, {Campbell}, {Sitarski},
  {Ghez}, \& {Anderson}}]{Service2016}
{Service}, M., {Lu}, J.~R., {Campbell}, R., {et~al.} 2016, \pasp, 128, 095004,
  \dodoi{10.1088/1538-3873/128/967/095004}

\bibitem[{{Soummer} {et~al.}(2011){Soummer}, {Hagan}, {Pueyo}, {Thormann},
  {Rajan}, \& {Marois}}]{Soummer2011}
{Soummer}, R., {Hagan}, J.~B., {Pueyo}, L., {et~al.} 2011, \apj, 741, 55,
  \dodoi{10.1088/0004-637X/741/1/55}

\bibitem[{{Soummer} {et~al.}(2012){Soummer}, {Pueyo}, \&
  {Larkin}}]{Soummer2012}
{Soummer}, R., {Pueyo}, L., \& {Larkin}, J. 2012, \apjl, 755, L28,
  \dodoi{10.1088/2041-8205/755/2/L28}

\bibitem[{Tallis {et~al.}(2018)Tallis, Bailey, Macintosh, Chilcote, Poyneer,
  Ruffio, Hayward, \& Savransky}]{Tallis2018}
Tallis, M., Bailey, V.~P., Macintosh, B., {et~al.} 2018, Proc. SPIE, 10703,
  1070356, \dodoi{10.1117/12.2319615}

\bibitem[{{The Astropy Collaboration} {et~al.}(2018){The Astropy
  Collaboration}, {Price-Whelan}, {Sip{\H o}cz}, {G{\"u}nther}, {Lim},
  {Crawford}, {Conseil}, {Shupe}, {Craig}, {Dencheva}, {Ginsburg},
  {VanderPlas}, {Bradley}, {P{\'e}rez-Su{\'a}rez}, {de Val-Borro}, {Aldcroft},
  {Cruz}, {Robitaille}, {Tollerud}, {Ardelean}, {Babej}, {Bachetti}, {Bakanov},
  {Bamford}, {Barentsen}, {Barmby}, {Baumbach}, {Berry}, {Biscani}, {Boquien},
  {Bostroem}, {Bouma}, {Brammer}, {Bray}, {Breytenbach}, {Buddelmeijer},
  {Burke}, {Calderone}, {Cano Rodr{\'{\i}}guez}, {Cara}, {Cardoso},
  {Cheedella}, {Copin}, {Crichton}, {D{\'A}vella}, {Deil}, {Depagne},
  {Dietrich}, {Donath}, {Droettboom}, {Earl}, {Erben}, {Fabbro}, {Ferreira},
  {Finethy}, {Fox}, {Garrison}, {Gibbons}, {Goldstein}, {Gommers}, {Greco},
  {Greenfield}, {Groener}, {Grollier}, {Hagen}, {Hirst}, {Homeier}, {Horton},
  {Hosseinzadeh}, {Hu}, {Hunkeler}, {Ivezi{\'c}}, {Jain}, {Jenness}, {Kanarek},
  {Kendrew}, {Kern}, {Kerzendorf}, {Khvalko}, {King}, {Kirkby}, {Kulkarni},
  {Kumar}, {Lee}, {Lenz}, {Littlefair}, {Ma}, {Macleod}, {Mastropietro},
  {McCully}, {Montagnac}, {Morris}, {Mueller}, {Mumford}, {Muna}, {Murphy},
  {Nelson}, {Nguyen}, {Ninan}, {N{\"o}the}, {Ogaz}, {Oh}, {Parejko}, {Parley},
  {Pascual}, {Patil}, {Patil}, {Plunkett}, {Prochaska}, {Rastogi}, {Reddy
  Janga}, {Sabater}, {Sakurikar}, {Seifert}, {Sherbert}, {Sherwood-Taylor},
  {Shih}, {Sick}, {Silbiger}, {Singanamalla}, {Singer}, {Sladen}, {Sooley},
  {Sornarajah}, {Streicher}, {Teuben}, {Thomas}, {Tremblay}, {Turner},
  {Terr{\'o}n}, {van Kerkwijk}, {de la Vega}, {Watkins}, {Weaver}, {Whitmore},
  {Woillez}, \& {Zabalza}}]{Astropy2018}
{The Astropy Collaboration}, {Price-Whelan}, A.~M., {Sip{\H o}cz}, B.~M.,
  {et~al.} 2018, ArXiv e-prints.
\newblock \doarXiv{1801.02634}

\bibitem[{{van der Walt} {et~al.}(2014){van der Walt}, {S}ch\"onberger,
  {Nunez-Iglesias}, {B}oulogne, {W}arner, {Y}ager, {G}ouillart, {Y}u, \& the
  scikit-image contributors}]{scikitimage}
{van der Walt}, S., {S}ch\"onberger, J.~L., {Nunez-Iglesias}, J., {et~al.}
  2014, PeerJ, 2, e453, \dodoi{10.7717/peerj.453}

\bibitem[{{Vargas Catal{\'a}n} {et~al.}(2016){Vargas Catal{\'a}n}, {Huby},
  {Forsberg}, {Jolivet}, {Baudoz}, {Carlomagno}, {Delacroix}, {Habraken},
  {Mawet}, {Surdej}, {Absil}, \& {Karlsson}}]{Catalan2016}
{Vargas Catal{\'a}n}, E., {Huby}, E., {Forsberg}, P., {et~al.} 2016, \aap, 595,
  A127, \dodoi{10.1051/0004-6361/201628739}

\bibitem[{{Vigan} {et~al.}(2016){Vigan}, {Bonnefoy}, {Ginski}, {Beust},
  {Galicher}, {Janson}, {Baudino}, {Buenzli}, {Hagelberg}, {D'Orazi},
  {Desidera}, {Maire}, {Gratton}, {Sauvage}, {Chauvin}, {Thalmann}, {Malo},
  {Salter}, {Zurlo}, {Antichi}, {Baruffolo}, {Baudoz}, {Blanchard},
  {Boccaletti}, {Beuzit}, {Carle}, {Claudi}, {Costille}, {Delboulb{\'e}},
  {Dohlen}, {Dominik}, {Feldt}, {Fusco}, {Gluck}, {Girard}, {Giro}, {Gry},
  {Henning}, {Hubin}, {Hugot}, {Jaquet}, {Kasper}, {Lagrange}, {Langlois}, {Le
  Mignant}, {Llored}, {Madec}, {Martinez}, {Mawet}, {Mesa}, {Milli},
  {Mouillet}, {Moulin}, {Moutou}, {Orign{\'e}}, {Pavlov}, {Perret}, {Petit},
  {Pragt}, {Puget}, {Rabou}, {Rochat}, {Roelfsema}, {Salasnich}, {Schmid},
  {Sevin}, {Siebenmorgen}, {Smette}, {Stadler}, {Suarez}, {Turatto}, {Udry},
  {Vakili}, {Wahhaj}, {Weber}, \& {Wildi}}]{Vigan2016}
{Vigan}, A., {Bonnefoy}, M., {Ginski}, C., {et~al.} 2016, \aap, 587,
  \dodoi{10.1051/0004-6361/201526465}

\bibitem[{{Wizinowich} {et~al.}(2000){Wizinowich}, {Acton}, {Shelton},
  {Stomski}, {Gathright}, {Ho}, {Lupton}, {Tsubota}, {Lai}, {Max}, {Brase},
  {An}, {Avicola}, {Olivier}, {Gavel}, {Macintosh}, {Ghez}, \&
  {Larkin}}]{Wizinowich2000}
{Wizinowich}, P., {Acton}, D.~S., {Shelton}, C., {et~al.} 2000, \pasp, 112,
  315, \dodoi{10.1086/316543}

\bibitem[{{Wright} {et~al.}(2010){Wright}, {Eisenhardt}, {Mainzer}, {Ressler},
  {Cutri}, {Jarrett}, {Kirkpatrick}, {Padgett}, {McMillan}, {Skrutskie},
  {Stanford}, {Cohen}, {Walker}, {Mather}, {Leisawitz}, {Gautier}, {McLean},
  {Benford}, {Lonsdale}, {Blain}, {Mendez}, {Irace}, {Duval}, {Liu}, {Royer},
  {Heinrichsen}, {Howard}, {Shannon}, {Kendall}, {Walsh}, {Larsen}, {Cardon},
  {Schick}, {Schwalm}, {Abid}, {Fabinsky}, {Naes}, \& {Tsai}}]{Wright2010}
{Wright}, E.~L., {Eisenhardt}, P.~R.~M., {Mainzer}, A.~K., {et~al.} 2010, \aj,
  140, 1868, \dodoi{10.1088/0004-6256/140/6/1868}

\bibitem[{{Zacharias} {et~al.}(2012){Zacharias}, {Finch}, {Girard}, {Henden},
  {Bartlett}, {Monet}, \& {Zacharias}}]{Zacharias2012}
{Zacharias}, N., {Finch}, C.~T., {Girard}, T.~M., {et~al.} 2012, VizieR Online
  Data Catalog, 1322

\bibitem[{Zacharias {et~al.}(2013)Zacharias, Finch, Girard, Henden, Bartlett,
  Monet, \& Zacharias}]{Zacharias2013}
Zacharias, N., Finch, C.~T., Girard, T.~M., {et~al.} 2013, \apj, 145, 44,
  \dodoi{10.1088/0004-6256/145/2/44}

\end{thebibliography}
\bibliographystyle{aasjournal}

\end{document}